\documentclass[twocolumn]{aastex63}
\pdfoutput=1 
\usepackage{amsmath,amstext}
\usepackage[T1]{fontenc}
\usepackage{apjfonts} 
\usepackage[figure,figure*]{hypcap}
\usepackage{longtable}
\usepackage{threeparttable}
\usepackage{booktabs}
\usepackage{tabularx}

\usepackage{amssymb}
\usepackage{fontawesome}
\usepackage{gensymb}
\usepackage{mathrsfs}
\usepackage{upgreek}
\usepackage{hyperref}
\usepackage{float}

\renewcommand{\d}{\ensuremath{\rm d}}

\graphicspath{{./}{figures/}}

\begin{document}

\title{Search for High-Ionization Nebular Emission (SHINE). I.
A Systematically Selected [\ion{Ne}{5}] Sample at $z>3$ with
\textit{JWST}/NIRSpec PRISM}

\author[0000-0003-2277-2354]{Shobita Satyapal}
\affiliation{George Mason University, Department of Physics and Astronomy, MS3F3, 4400 University Drive, Fairfax, VA 22030, USA}
\author[0000-0003-3152-4328]{Sara Doan}
\affiliation{George Mason University, Department of Physics and Astronomy, MS3F3, 4400 University Drive, Fairfax, VA 22030, USA}
\author[0000-0001-7144-7182]{D. Schaerer}
\affiliation{Observatoire de Genève, Université de Genève, Chemin Pegasi 51, 1290 Versoix, Switzerland}
\affiliation{CNRS, IRAP, 14 Avenue E. Belin, 31400 Toulouse, France}
\author[0000-0001-8442-1846]{R. Marques-Chaves}
\affiliation{Department of Astronomy, University of Geneva, Chemin Pegasi 51, 1290 Versoix, Switzerland}
\author{A. Sawarkar}
\affiliation{Department of Astronomy, University of Geneva, Chemin Pegasi 51, 1290 Versoix, Switzerland}  
\correspondingauthor{Shobita Satyapal}
\email{ssatyapa@gmu.edu}
\author[0000-0003-3937-562X]{William Matzko}
\affiliation{George Mason University, Department of Physics and Astronomy, MS3F3, 4400 University Drive, Fairfax, VA 22030, USA}
\author[0000-0002-2121-8426]{Camilo Vazquez}
\affiliation{George Mason University, Department of Physics and Astronomy, MS3F3, 4400 University Drive, Fairfax, VA 22030, USA}
\author[0000-0002-3028-4759]{C. Daoutis}
\affiliation{Department of Astronomy, University of Geneva, Chemin Pegasi 51, 1290 Versoix, Switzerland}
\author[0000-0002-3897-6856]{D. Korber}
\affiliation{Department of Astronomy, University of Geneva, Chemin Pegasi 51, 1290 Versoix, Switzerland}
\author[0009-0002-8287-7406]{I. Morel}
\affiliation{Department of Astronomy, University of Geneva, Chemin Pegasi 51, 1290 Versoix, Switzerland}
\author[0000-0003-1207-5344]{Mengyuan Xiao}
\affiliation{Department of Astronomy, University of Geneva, Chemin Pegasi 51, 1290 Versoix, Switzerland}

\begin{abstract}

We conduct the first systematic Search for High-Ionization Nebular Emission (SHINE) using [\ion{Ne}{5}]~$\lambda3426$ in \textit{JWST}/NIRSpec PRISM spectroscopy at $z>3$. From more than 9,000 galaxies, we identify 25 [\ion{Ne}{5}] emitters spanning $z=3.059$--$9.444$ and $\log_{10}(L_{\rm [Ne\,V]}/\mathrm{erg\,s^{-1}})=40.70$--$42.86$. Their [\ion{Ne}{5}] luminosities overlap with those of local [\ion{Ne}{5}]-selected active galactic nuclei (AGNs) and exceed those of local metal-poor [\ion{Ne}{5}] emitters, although such low-luminosity systems would fall below our sensitivity. The population is diverse, spanning compact and extended morphologies and a broad range of continuum properties and stellar masses, including very low-mass hosts. It includes sources with broad Balmer emission as well as others whose higher-resolution spectra do not require a broad component. The [\ion{Ne}{5}] emitters overlap only partly with conventional AGN diagnostics: most do not satisfy conservative high-redshift AGN criteria based on strong rest-optical narrow-line ratios, and only a minority have adopted broad-line classifications. [\ion{Ne}{5}] upper limits for independently selected broad-line AGNs and little red dots are too shallow to establish a population-wide [\ion{Ne}{5}] deficit. Six sources have secure \textit{Chandra} counterparts, showing that X-ray weakness is not universal among luminous [\ion{Ne}{5}] emitters. Strong [\ion{Ne}{5}]/[\ion{Ne}{3}] emission is associated with redder ultraviolet slopes and stronger Balmer breaks, but not with UV luminosity, possibly linking strong high-ionization emission to recent changes in star formation. The observed incidence of luminous [\ion{Ne}{5}] emission shows no significant evolution over $z>3$, despite an increasing robust [\ion{Ne}{3}] detection fraction. The high [\ion{Ne}{5}] luminosities and hard line ratios favor black-hole accretion as the dominant power source, establishing [\ion{Ne}{5}] emission as a complementary probe of early black-hole growth.

\end{abstract}


\section{Introduction}
\label{sec:introduction}

The \textit{James Webb Space Telescope} (\textit{JWST}) has
revolutionized our view of black hole growth in the early Universe,
uncovering an unexpectedly abundant population of faint broad-line
AGNs at $z\gtrsim4$, together with an overlapping population of
compact red sources commonly referred to as Little Red Dots (LRDs)
\citep[e.g.,][]{Kocevski2023,Harikane2023,Matthee2024,Greene2024,Maiolino2024,Kokorev2024,Labbe2025,Kocevski2025,Akins2025,Hviding2025}.
These sources extend to substantially lower luminosities than the
UV-luminous quasars that dominated the pre-\textit{JWST} view of
early black hole growth, with number densities significantly exceeding
extrapolations of the quasar luminosity function
\citep[e.g.,][]{Harikane2023,Matthee2024,Kokorev2024,Akins2025}.
Yet despite several years of intensive study, their physical nature
remains uncertain. While exceptions have been reported \citep[e.g.,][]{Tang2025HighIonization,Gloudemans2025,Delvecchio2025,Barro2026} high-redshift AGNs uncovered by JWST generally differ from low-redshift counterparts in that they are often extremely compact, show red optical continua, are weak or undetected in X-rays, lack hot/cold dust emission, exhibit little variability, and display prominent broad Balmer lines alongside weak or absent high-ionization lines.
\citep[e.g.,][]{Ananna2024,Yue2024,Maiolino2025,Casey2025,Setton2025,KokuboHarikane2025,Zhang2025,Burke2026,Lambrides2026,Wang2026,Zucchi2026}. Proposed explanations include intrinsically X-ray-weak
or softer ionizing continua associated with high- or super-Eddington
accretion, weak X-ray coronae, heavy absorption, and accreting black
holes embedded in dense gaseous envelopes that absorb and reprocess
radiation from the central source
\citep[e.g.,][]{PacucciNarayan2024,InayoshiMaiolino2025,
Naidu2025,BegelmanDexter2025,Madau2026,Rusakov2026}. 
These discoveries suggest that \textit{JWST} is revealing modes of
black hole growth that were largely inaccessible to previous surveys.

Through the past several years, much of the rapidly expanding high-redshift AGN
population has been identified through broad Balmer lines or
photometric LRD selection. Broad-line searches by definition select
systems in which a broad-line region is present and directly visible,
and therefore probe only the Type~1 portion of the accreting
population. In the local Universe, obscured and narrow-line AGNs
constitute a large fraction of the active-galaxy population, showing
that black hole growth commonly occurs in physical configurations
that would not be recovered through Type~1 selection alone
\citep[e.g.,][]{MaiolinoRieke1995,HickoxAlexander2018}.
At high redshift, however, identifying the analogous population
through conventional narrow-line diagnostics is especially difficult.
The low metallicities and high ionization parameters characteristic
of early galaxies cause standard rest-optical diagnostic ratios to
overlap substantially between stellar and accreting-black-hole
photoionization models
\citep[e.g.,][]{Groves2006,Cann2019,Richardson2022,Cleri2025}. Recent \textit{JWST} studies have begun to recover candidate
narrow-line AGNs using combinations of rest-frame UV diagnostics,
auroral-line ratios, and high-ionization transitions
\citep[e.g.,][]{Scholtz2025,Mazzolari2024,Mazzolari2025}.
These searches demonstrate that a substantial narrow-line population
may be present at high redshift, but also emphasize that no single
rest-optical diagnostic provides both a complete and uncontaminated
selection of Type~2 AGNs under the physical conditions characteristic
of early galaxies
\citep[e.g.,][]{Scholtz2025,Mazzolari2025, Juodzbalis2026}.
The resulting incompleteness affects more than just estimates of AGN
abundance: selection through broad lines or LRD colors may preferentially
miss obscured, embedded, or otherwise distinct phases of black hole
activity and their connection to the evolving host galaxy. This
problem may be particularly acute for lower-mass black holes, for
which broad-line signatures become increasingly difficult to identify
and whose low-mass, metal-poor hosts occupy the regime in which
conventional narrow-line diagnostics provide the least discrimination
between stellar and accretion-powered ionization
\citep[e.g.,][]{ReinesComastri2016,Cann2019,Richardson2022,Nakajima2022}.
A substantial component of early black hole growth may therefore
remain poorly represented by the selection methods currently driving
the high-redshift AGN population.

Very-high-ionization nebular emission provides a complementary route
to identifying hard radiation from accreting black holes without
requiring either a directly visible broad-line region or placement in
the AGN region of conventional narrow-line diagnostic diagrams.
Emission from ions with ionization potentials above $\sim54$~eV
requires photons that are difficult for even extreme stellar populations to
produce in sufficient numbers, thereby offering additional
insight into the nature of the ionizing source
\citep[e.g.,][]{Satyapal2021,Cleri2025}.
Among these diagnostics, [\ion{Ne}{5}]~$\lambda3426$ is particularly
effective. The production of Ne$^{4+}$ requires photons with energies
above 97.1~eV, probing a significantly higher-energy regime of the ionizing continuum than those probed by the standard lower ionization optical lines \citep{Cleri2023Ne53}. Because neon is a noble gas, it does not deplete onto dust grains, leaving its gas-phase abundance unaffected by the presence of dust in the ionized gas \citep{McKaig2024}. Furthermore, [\ion{Ne}{5}]~$\lambda3426$ is typically the most luminous optical coronal line and among the most frequently detected in low-redshift spectroscopic samples \citep[e.g.,][]{Reefe2022,Negus2023, Doan2025, MatthewsAcuna2026}.
Coronal lines have long been used to identify highly ionized gas
associated with AGNs. Indeed, searches for [\ion{Ne}{5}] and other
coronal lines in the local Universe have uncovered accreting black
hole candidates missed by standard optical narrow-line diagnostics
\citep[e.g.,][]{Satyapal2008,Satyapal2009,Molina2021,
Reefe2022,Negus2023,MatthewsAcuna2026}. [\ion{Ne}{5}] has also proven effective for identifying obscured and
otherwise elusive AGNs
\citep[e.g.,][]{Gilli2010,Feuillet2024,Reiss2025,Peca2025}.
High-ionization lines can also offer valuable constraints on
lower-mass black holes. Under standard thin-disk assumptions, the
characteristic disk temperature increases toward lower black hole
mass at a fixed Eddington ratio and spin; consequently, photoionization models
predict prominent very-high-ionization emission from accreting
intermediate-mass black holes
\citep[e.g.,][]{KubotaDone2018,Cann2018}.

[\ion{Ne}{5}] is not, however, an unambiguous signature of accretion
in every galaxy. Optical [\ion{Ne}{5}]~$\lambda3426$
emission has been detected in a number of metal-poor star-forming
galaxies
\citep[e.g.,][]{Izotov2004,ThuanIzotov2005,Izotov2012,Izotov2021}.
Recent \textit{JWST}/MIRI spectroscopy has extended these results to
the mid-infrared, revealing [\ion{Ne}{5}]~$14.3\,\mu$m emission in
the extremely metal-poor galaxies SBS~0335--052E and I~Zw~18, as
well as in the metal-poor dwarf CGCG~007--025
\citep{Mingozzi2025,ArroyoPolonio2025,Hunt2025,
del-Valle-Espinosa2026}.
These observations demonstrate that very hard ionizing radiation can
be present in intensely star-forming, metal-poor environments, while
also emphasizing the difficulty of uniquely identifying its physical
source. Radiative shocks, unusually hard stellar populations,
X-ray binaries or ultraluminous X-ray sources, and accretion onto
lower-mass black holes have all been considered, with their relative
importance likely differing among individual systems
\citep[e.g.,][]{Izotov2012,Kehrig2018,Mingozzi2025,
ArroyoPolonio2025,Hunt2025,del-Valle-Espinosa2026}.
The interpretation of [\ion{Ne}{5}] therefore requires its luminosity
and excitation to be considered together with a broader
spectroscopic and multiwavelength analysis.

The apparent weakness of high-ionization emission in many
\textit{JWST}-selected AGNs makes such lines valuable probes of accretion physics
at high redshift. Several studies of broad-line AGNs and Little Red Dots (LRDs) have
reported weak or undetected high-ionization lines and X-ray emission
relative to expectations for conventional lower-redshift AGNs
\citep[e.g.,][]{Ananna2024,Maiolino2025,Wang2026,Lambrides2026,Zucchi2026}. In a sample of 851 $z>4$ galaxies targeted with NIRSpec grating spectroscopy,
\citet{Tang2025HighIonization} identified only a small fraction of candidate
very-high-ionization line emitters, indicating that such emission is either
intrinsically rare or typically faint at high redshift. Conversely, high-ionization emission is clearly
detected in a subset of high-redshift galaxies and AGNs
\citep[e.g.,][]{Tang2025HighIonization,Chisholm2024,
Scholtz2025,Valentino2026,Curti2025}, demonstrating that the
production of such lines is not universally suppressed.

Most existing studies, however, rely on samples selected
by broad lines, LRD colors, or other galaxy properties, only subsequently
evaluating whether high-ionization emission is present. Building on lower-redshift [\ion{Ne}{5}] surveys \citep[e.g.,][]{Gilli2010,Cleri2023CLEAR} and recent NIRSpec grating searches \citep{Tang2025HighIonization}, the growing \textit{JWST}/NIRSpec PRISM archive enables a complementary strategy at $z>3$: selecting target populations directly by [\ion{Ne}{5}] emission.

We refer to this survey as the \textbf{Search for High-Ionization Nebular Emission (SHINE)}.
SHINE is designed to identify high-redshift sources using high-ionization 
nebular emission, leveraging lines across a broad range of ionization energies to 
constrain the hard ionizing radiation field and surrounding gas. In this initial 
paper, we conduct a systematic search for [\ion{Ne}{5}]~$\lambda3426$ in public 
\textit{JWST}/NIRSpec PRISM spectroscopy at $z>3$ to construct the SHINE-[\ion{Ne}{5}] sample. We evaluate the source properties recovered when selecting purely by 
[\ion{Ne}{5}] emission, characterizing their continuum, host-galaxy, morphological, 
and emission-line features. Furthermore, we quantify the overlap between 
[\ion{Ne}{5}]-selected sources and existing broad-line, narrow-line, LRD, and 
X-ray-selected populations, examining line excitation and observed incidence across redshift. This selection strategy offers a complementary probe of 
early black hole growth independent of broad-line visibility, LRD colors, or 
conventional optical diagnostics, potentially revealing physical conditions that are poorly 
represented in current high-redshift AGN samples.

Throughout this work, we adopt a flat $\Lambda$CDM cosmology
corresponding to the \texttt{Planck18} cosmology implemented in
\texttt{Astropy}, with
$H_{0}=67.4$~km~s$^{-1}$~Mpc$^{-1}$,
$\Omega_{\rm m}=0.315$, and $\Omega_{\Lambda}=0.685$.

\section{JWST/NIRSpec Data and Construction of the SHINE-[NeV] Sample}
\label{sec:sample_selection}

\subsection{DJA parent sample}
\label{sec:dja_parent}

We searched for [\ion{Ne}{5}] emission using a new tool called PRISMATIC, which provides automatic fits of numerous spectral features (emission lines and others) in NIRSpec PRISM spectra and an efficient search and visualization tool (see Sawarkar et al., in preparation).
We used the public JWST/NIRSpec
spectroscopic products available through the Dawn JWST Archive
(DJA; \citealt{Brammer2023,Heintz2025,Valentino2025}) as of
2026 May 21. We restricted the analysis to NIRSpec PRISM/CLEAR
spectra assigned a secure spectroscopic-redshift quality grade of 3
in the DJA. When more than one pipeline reduction of a given
observation was available, we adopted the latest reduction available
on that date.

Many galaxies in the archive were observed more than once, either
through repeated visits or through overlapping JWST programs. For galaxies with more than one available PRISM observation, we
selected as the representative spectrum the observation with the
lowest local flux uncertainty at the expected observed wavelength of
[\ion{Ne}{5}]~$\lambda3426$. Each physical galaxy was therefore counted
only once, and multiple spectra of the same galaxy were not treated as
independent objects. After de-duplication, the parent catalog contained
9002 unique galaxies, with one representative PRISM spectrum adopted
for the initial [\ion{Ne}{5}] search.

The NIRSpec PRISM/CLEAR configuration provides broad, nearly
continuous observed-frame wavelength coverage from approximately
$0.6$ to $5.3~\mu{\rm m}$, with a strongly wavelength-dependent
spectral resolution \citep{Jakobsen2022,Ferruit2022}. The galaxies in
our parent catalog span the exact adopted spectroscopic-redshift range
$z=2.989$--14.462. Across this interval,
[\ion{Ne}{5}]~$\lambda3426$ is observed from $1.367$ to
$5.297~\mu{\rm m}$.
We evaluated the PRISM resolving power at the expected observed
wavelength of [\ion{Ne}{5}]~$\lambda3426$ for every galaxy in the parent
catalog using the wavelength-dependent PRISM dispersion relation. The
resulting resolving powers span $R\simeq31$--323, with a median value
of $R\simeq43$. These resolving powers correspond to instrumental
velocity scales of approximately
$c/R\simeq930$--$9550~{\rm km~s^{-1}}$. The appearance, blending, and
measured significance of the [\ion{Ne}{5}] doublet therefore vary
substantially with redshift across the parent sample.

The DJA also combines observations obtained by multiple JWST programs
with different exposure times, target-selection functions, observing
strategies, and source-specific data quality. Consequently, the
sensitivity to [\ion{Ne}{5}] emission is non-uniform across the parent
catalog and cannot be inferred from the wavelength-dependent spectral
resolution alone. 

\subsection{Automated candidate preselection}
\label{sec:automated_selection}

We used the automated PRISMATIC emission-line fits to construct an initial
list of candidate [\ion{Ne}{5}] emitters. The automated measurements were
used only for candidate preselection and were not adopted as the final
[\ion{Ne}{5}] measurements. We selected sources for which the automated
fit to [\ion{Ne}{5}]~$\lambda3426$ returned a nominal signal-to-noise
ratio of $\mathrm{S/N}>4$. This selection produced 187 unique galaxies
for visual inspection.

We adopted the $\mathrm{S/N}>4$ preselection following an initial
examination of candidates selected at automated
$\mathrm{S/N}>3$. Visual inspection showed that many of the
lower-significance candidates were associated with noise
fluctuations, spectral artifacts, or inaccurate continuum placement,
rather than with a convincing emission feature. Raising the automated
threshold reduced the number of obvious false positives and produced a
tractable candidate set for systematic visual inspection. The
automated threshold should therefore be understood as a practical
preselection criterion rather than as the final definition of an
[\ion{Ne}{5}] detection.

The automated measurements were not optimized specifically for weak
[\ion{Ne}{5}] emission. In particular, the automated fitting procedure did
not simultaneously model both members of the
[\ion{Ne}{5}]~$\lambda\lambda3346,3426$ doublet, and the continuum was
constrained over a comparatively broad wavelength interval. For weak
features, the resulting line flux and significance were sensitive to
the shape and placement of the fitted continuum. These limitations
motivated both visual assessment of the candidates and a subsequent
more detailed and rigorous fitting procedure for the visually retained
sources.

Because only objects passing the automated $\mathrm{S/N}>4$
preselection were inspected systematically, the sample is not expected
to contain every possible [\ion{Ne}{5}] emitter in the 9002-galaxy parent
catalog. Extending the search to lower automated significance would
require the visual inspection and detailed fitting of a substantially
larger number of spectra.

\subsection{Visual assessment of the candidates}
\label{sec:visual_vetting}

We visually examined the one- and two-dimensional spectra of each of
the 187 candidates to determine whether an emission feature was
present near the expected position of the
[\ion{Ne}{5}]~$\lambda3426$ line. A source was retained for detailed spectral
modeling when the feature was clearly apparent by eye and its
centroid was located within
\begin{equation}
\left|\Delta v\right| < 5000~{\rm km~s^{-1}}
\end{equation}
of the wavelength predicted from the adopted spectroscopic redshift.
This deliberately broad interval was used to reject clearly unrelated
features while accommodating uncertainties in the automated redshift,
continuum fit, and centroid of a weak feature in the low-resolution
PRISM spectrum. It was not adopted as a physical constraint on the
velocity of the [\ion{Ne}{5}]-emitting gas.

Twenty-eight galaxies passed the visual assessment and were carried
forward to a more detailed and rigorous spectral-fitting procedure.
Visual retention alone did not establish membership in the final
sample. Each of the 28 objects was subsequently required to satisfy
the same quantitative detection criterion based on the integrated
[\ion{Ne}{5}]~$\lambda3426$ line flux and its uncertainty.

\subsection{Detailed BADASS fitting and the SHINE-[NeV] sample}
\label{sec:badass_selection}

We reanalysed the 28 visually retained spectra using the Bayesian AGN
Decomposition Analysis for SDSS Spectra code
(BADASS; \citealt{Sexton2021}), following the implementation described
in Appendix~\ref{app:badass}. Relative to the automated catalog
measurements, this more detailed and rigorous fitting procedure used a
narrower local wavelength interval to constrain the continuum and
simultaneously included both members of the
[\ion{Ne}{5}]~$\lambda\lambda3346,3426$ doublet. This local fitting
strategy was adopted to reduce the sensitivity of the weak-line
measurements to continuum structure far from the [\ion{Ne}{5}] features
and to provide a more consistent estimate of the integrated line-flux
uncertainty.

The [\ion{Ne}{5}]~$\lambda3426$ fluxes obtained from the automated
catalog and from the final BADASS fits were generally consistent. The
inferred line significances varied somewhat between the two
procedures, however, because the BADASS analysis used a different
continuum window, a more robust local continuum treatment, and
simultaneous modeling of the [\ion{Ne}{5}] doublet. We therefore used
the BADASS integrated fluxes and uncertainties, rather than the
automated catalog values, to define the SHINE-[NeV] sample and to
calculate the [\ion{Ne}{5}] luminosities used throughout this work.

A galaxy was included in the SHINE-[NeV] sample only if it belonged to
the visually retained set and the BADASS fit returned a valid
integrated [\ion{Ne}{5}]~$\lambda3426$ flux, $F_{3426}$, and
integrated-flux uncertainty, $\sigma_{F_{3426}}$, satisfying
\begin{equation}
\frac{F_{3426}}{\sigma_{F_{3426}}} > 3.
\label{eq:nev_detection}
\end{equation}
We adopted the integrated [\ion{Ne}{5}]~$\lambda3426$ flux
significance defined in Equation~\ref{eq:nev_detection} as the
quantitative criterion for inclusion in the SHINE-[NeV] sample. Twenty-five of the 28 visually retained galaxies satisfied this
requirement and constitute the SHINE-[NeV] sample used
throughout this work. 

The significance assigned to a weak emission feature can depend on the
continuum interval, continuum parameterization, masking, noise model,
and treatment of neighboring features. We therefore applied a single
fitting procedure and a common integrated-flux requirement to all 28
visually retained objects. Sources not satisfying
Equation~\ref{eq:nev_detection} are not treated as [\ion{Ne}{5}]
detections in the subsequent statistical analyses.

Our uniform PRISM selection does not recover every previously published
[\ion{Ne}{5}] source in the parent catalog. 
Six previously reported sources in the parent catalog were not among
the 28 candidates retained for detailed fitting: five with secure
published [\ion{Ne}{5}] detections and one with a tentative detection.
Four of the five secure sources have [\ion{Ne}{5}] detections
established with medium- or high-resolution NIRSpec spectroscopy:
DJA-329 and DD-111 with G140 grating observations
\citep{Valentino2026}, GN~42437 with G235H/F170LP
\citep{Chisholm2024}, and JADES-NS-GS-10013609 with
G235M/F170LP \citep{Scholtz2025}. UNCOVER~45924 was reported as a
secure PRISM detection \citep{Treiber2025}, while GS-81034 shows a
tentative feature in the higher resolution G235M/F170LP spectrum
\citep{Tang2025HighIonization}. The SHINE-[NeV] sample should
therefore be regarded as a uniformly selected PRISM sample rather than
a complete inventory of all previously reported high-redshift
[\ion{Ne}{5}] emitters.

Table~\ref{tab:nev_sample} lists the ICRS coordinates, adopted spectroscopic redshift, exact PRISM/CLEAR
spectrum used in the analysis, and the apparent BADASS
[\ion{Ne}{5}]~$\lambda3426$ luminosity and its uncertainty for the full 25 galaxies in the SHINE-[NeV] sample. We list references for sources with previously published [\ion{Ne}{5}] emission among our final sample. Three sources lying in a
potential lensing field are flagged in Table~\ref{tab:nev_sample}. The magnifications are small.
NeV-01: 1.93
NeV-02: 1.94
NeV-03: 1.41
with typical 1-sigma (68\% CL) uncertainties of ~0.02-0.03.

\begin{deluxetable*}{lrrrlrc}
\tabletypesize{\scriptsize}
\tablewidth{0pt}
\tablecaption{The Adopted High-redshift [\ion{Ne}{5}] Emitter Sample
\label{tab:nev_sample}}
\tablehead{
\colhead{ID} &
\colhead{R.A.} &
\colhead{Decl.} &
\colhead{$z_{\rm spec}$} &
\colhead{Adopted PRISM analysis spectrum} &
\colhead{$\log L_{3426}^{\rm app}$} &
\colhead{Lens field?} \\
\colhead{} &
\colhead{(deg)} &
\colhead{(deg)} &
\colhead{} &
\colhead{} &
\colhead{($\mathrm{erg\,s^{-1}}$)} &
\colhead{}
}
\startdata
NeV-01\tablenotemark{a} & 3.566920 & -30.347271 & 3.46578 & \parbox[t]{0.39\textwidth}{\raggedright\ttfamily\path{uncover-v4_prism-clear_2561_45092.spec.fits}} & $42.36\pm0.04$ & Y \\
NeV-02 & 3.617833 & -30.404140 & 6.32973 & \parbox[t]{0.39\textwidth}{\raggedright\ttfamily\path{uncover-61-v4_prism-clear_2561_23409.spec.fits}} & $41.76\pm0.10$ & Y \\
NeV-03\tablenotemark{b} & 3.636960 & -30.406361 & 8.51786 & \parbox[t]{0.39\textwidth}{\raggedright\ttfamily\path{uncover-v4_prism-clear_2561_10646.spec.fits}} & $41.58\pm0.11$ & Y \\
NeV-04 & 34.301147 & -5.287817 & 4.81900 & \parbox[t]{0.39\textwidth}{\raggedright\ttfamily\path{rubies-uds43-v4_prism-clear_4233_19736.spec.fits}} & $42.14\pm0.10$ & N \\
NeV-05 & 34.322542 & -5.171391 & 3.94516 & \parbox[t]{0.39\textwidth}{\raggedright\ttfamily\path{gto-wide-uds13-v4_prism-clear_1215_3757.spec.fits}} & $41.85\pm0.06$ & N \\
NeV-06 & 34.391716 & -5.115013 & 6.23485 & \parbox[t]{0.39\textwidth}{\raggedright\ttfamily\path{mom-uds01-v4_prism-clear_5224_164378.spec.fits}} & $41.23\pm0.06$ & N \\
NeV-07 & 53.063600 & -27.848729 & 4.21594 & \parbox[t]{0.39\textwidth}{\raggedright\ttfamily\path{gds-looser-03-v4_prism-clear_5997_186744.spec.fits}} & $40.83\pm0.08$ & N \\
NeV-08 & 53.072848 & -27.908504 & 3.47808 & \parbox[t]{0.39\textwidth}{\raggedright\ttfamily\path{jades-gds06-v4_prism-clear_1286_158273.spec.fits}} & $41.40\pm0.07$ & N \\
NeV-09 & 53.086456 & -27.841707 & 3.47966 & \parbox[t]{0.39\textwidth}{\raggedright\ttfamily\path{gds-looser-03-v4_prism-clear_5997_189618.spec.fits}} & $40.70\pm0.06$ & N \\
NeV-10\tablenotemark{c} & 53.112434 & -27.774626 & 9.44391 & \parbox[t]{0.39\textwidth}{\raggedright\ttfamily\path{gds-udeep-v4_prism-clear_3215_265801.spec.fits}} & $40.86\pm0.13$ & N \\
NeV-11 & 53.121759 & -27.797633 & 5.93873 & \parbox[t]{0.39\textwidth}{\raggedright\ttfamily\path{gds-deep-v4_prism-clear_1210_13176.spec.fits}} & $40.81\pm0.06$ & N \\
NeV-12 & 53.124447 & -27.851707 & 3.72169 & \parbox[t]{0.39\textwidth}{\raggedright\ttfamily\path{jades-gds-w05-v3_prism-clear_b147.spec.fits}} & $42.86\pm0.03$ & N \\
NeV-13 & 53.129908 & -27.752004 & 3.44356 & \parbox[t]{0.39\textwidth}{\raggedright\ttfamily\path{jades-gds08-v4_prism-clear_1286_60148771.spec.fits}} & $41.19\pm0.05$ & N \\
NeV-14 & 53.139351 & -27.874542 & 3.47129 & \parbox[t]{0.39\textwidth}{\raggedright\ttfamily\path{jades-gds-w07-v3_prism-clear_b62.spec.fits}} & $41.86\pm0.04$ & N \\
NeV-15 & 69.965843 & -52.738358 & 3.34862 & \parbox[t]{0.39\textwidth}{\raggedright\ttfamily\path{borg-0440m5244-v4_prism-clear_1747_384.spec.fits}} & $41.04\pm0.12$ & N \\
NeV-16 & 150.073944 & 2.380797 & 4.54263 & \parbox[t]{0.39\textwidth}{\raggedright\ttfamily\path{glazebrook-cos-obs3-v4_prism-clear_2565_19702.spec.fits}} & $41.52\pm0.08$ & N \\
NeV-17 & 150.096344 & 2.425364 & 3.12566 & \parbox[t]{0.39\textwidth}{\raggedright\ttfamily\path{mom-cos05-v4_prism-clear_5224_318265.spec.fits}} & $41.33\pm0.08$ & N \\
NeV-18 & 150.139328 & 2.214673 & 5.50900 & \parbox[t]{0.39\textwidth}{\raggedright\ttfamily\path{capers-cos01-v4_prism-clear_6368_49388.spec.fits}} & $41.34\pm0.08$ & N \\
NeV-19 & 189.139572 & 62.238373 & 3.42241 & \parbox[t]{0.39\textwidth}{\raggedright\ttfamily\path{goodsn-wide1-v4_prism-clear_1211_7185.spec.fits}} & $41.56\pm0.05$ & N \\
NeV-20 & 214.931473 & 52.935528 & 3.05913 & \parbox[t]{0.39\textwidth}{\raggedright\ttfamily\path{ceers-ddt-v4_prism-clear_2750_1636.spec.fits}} & $41.25\pm0.04$ & N \\
NeV-21 & 214.931625 & 52.908684 & 3.43484 & \parbox[t]{0.39\textwidth}{\raggedright\ttfamily\path{ceers-ddt-v4_prism-clear_2750_1034.spec.fits}} & $42.61\pm0.04$ & N \\
NeV-22 & 214.944199 & 52.967579 & 4.47484 & \parbox[t]{0.39\textwidth}{\raggedright\ttfamily\path{capers-egs47-v4_prism-clear_6368_1832.spec.fits}} & $41.46\pm0.08$ & N \\
NeV-23 & 214.999176 & 52.973301 & 4.63750 & \parbox[t]{0.39\textwidth}{\raggedright\ttfamily\path{capers-egs47-v4_prism-clear_6368_223045.spec.fits}} & $41.52\pm0.06$ & N \\
NeV-24 & 215.031403 & 52.908157 & 3.46129 & \parbox[t]{0.39\textwidth}{\raggedright\ttfamily\path{rubies-egs52-v4_prism-clear_4233_19509.spec.fits}} & $41.52\pm0.07$ & N \\
NeV-25 & 215.132927 & 53.096295 & 3.40087 & \parbox[t]{0.39\textwidth}{\raggedright\ttfamily\path{gto-wide-egs1-v4_prism-clear_1213_4369.spec.fits}} & $42.17\pm0.04$ & N \\
\enddata

\tablenotetext{a}{Published as UNCOVER~45092, with a secure prior
[\ion{Ne}{5}] detection reported by \citet{Treiber2025}.}

\tablenotetext{b}{Corresponds to UNCOVER~10646 component~2
(LRD alias UNCOVER~9858). \citet{Treiber2025} discussed a possible
[\ion{Ne}{5}] feature but did not claim a secure detection.}

\tablenotetext{c}{Published as JADES-GS-z9-0 (JADES~10058975), with
a tentative prior [\ion{Ne}{5}] detection reported by
\citet{Curti2025}.}

\tablecomments{
The table lists the 25 adopted high-redshift [\ion{Ne}{5}] emitters
in order of increasing right ascension and then declination.
Coordinates are ICRS decimal degrees. The listed spectrum is the
exact DJA PRISM/CLEAR spectrum adopted for the manuscript analysis.
The [\ion{Ne}{5}]~$\lambda3426$ luminosities were calculated from the
integrated line fluxes obtained with the detailed BADASS fitting
procedure described in Section~\ref{sec:badass_selection}. Values are
reported as $\log L_{3426}^{\rm app}\pm\sigma_{\log L}$ and rounded to
two decimal places in the printed table; full numerical precision is
retained in the machine-readable table. The quoted uncertainty
reflects the line-flux uncertainty. Redshift and cosmological-parameter
uncertainties are not included. The luminosities are apparent and have
not been corrected for gravitational magnification. ``Lens field?''
identifies sources lying in a potential lensing field. The magnifications are small ( NeV-01: 1.93 NeV-02: 1.94 NeV-03: 1.41644) with typical 1-sigma (68\% CL) uncertainties of 0.02-0.03  Full literature metadata, internal
identifiers, and fitting-provenance fields are retained in the
machine-readable table.
}
\end{deluxetable*}

The additional emission-line measurements used elsewhere in this
paper were adopted directly from the DJA automated emission-line
catalog. These lines were generally substantially stronger and more
clearly detected than [\ion{Ne}{5}]~$\lambda3426$, and their measurements
were therefore less sensitive to the local continuum choices that
motivated our dedicated reanalysis of [\ion{Ne}{5}]. We did not refit
these other emission lines. Unless stated otherwise, only the
[\ion{Ne}{5}]~$\lambda3426$ measurements were replaced by the results of
the BADASS analysis.

When possible, we derived stellar-mass for the
SHINE-[NeV] sample from SED fitting based on the DAWN JWST
Archive products. Following the methodology described by \cite{Xiao2026} (see also: \citealt{Morel2026}), the fits combine JWST/NIRCam photometry
with NIRSpec/PRISM spectroscopy using \texttt{Bagpipes}
\citep{Carnall2018}, with the spectroscopic redshift fixed.
Residual spectrophotometric differences between the PRISM
spectra and the imaging are corrected using a polynomial
calibration of up to second order. The fits adopt a
non-parametric star-formation history, BPASS stellar-population
models, variable metallicity (from 0.01 to 2.5\,$Z_{\odot}$) and ionization parameter (from $\log U=-4$ to $-1$), and the \citet{Salim2018} dust
attenuation curve. We adopt the median of the posterior
stellar-mass distribution as the fiducial stellar mass and use
the 16th and 84th percentiles to characterize the formal fitting
uncertainty.
Stellar masses were thus derived for the majority of the sources (20 out of 25).

\section{Comparison Samples and Classification Framework}
\label{sec:comparison_samples}

We compare the adopted [\ion{Ne}{5}] sample with several independently
assembled high-redshift populations selected to represent complementary
methods for identifying accretion activity. These include published
broad-line AGN (BLAGN), galaxies selected using narrow-line AGN
diagnostics, and literature-selected little red dots (LRDs). We also
include two low-redshift [\ion{Ne}{5}] samples---the Coronal Line
Activity Spectroscopic Survey (CLASS) and a compilation of local
metal-poor [\ion{Ne}{5}] emitters---to examine how the high-redshift
population compares with [\ion{Ne}{5}]-emitting galaxies in the local
Universe.

The comparison samples are defined by different observational criteria
and are not mutually exclusive. A galaxy may, for example, be both an
LRD and a BLAGN, or both a [\ion{Ne}{5}] emitter and a source selected
by a narrow-line AGN diagnostic. We retain these overlapping
classifications throughout the analysis. For figures in which each
galaxy is shown only once, galaxies in the adopted [\ion{Ne}{5}] sample
are shown as [\ion{Ne}{5}] emitters; among the remaining sources,
published BLAGN are shown as BLAGN rather than a second time as LRDs.
This convention is used only to avoid plotting the same galaxy more
than once and does not change the classifications that apply to each
source.

The classifications of sources and the emission-line measurements used in
our diagnostic analyses are treated separately. Published
classifications are adopted from the observations on which they were
originally based, which may include NIRSpec PRISM or medium- and
high-resolution spectroscopy. For uniformity, however, the
emission-line fluxes and upper limits used for the high-redshift
comparison samples are taken from the automated PRISMATIC measurements
from the NIRSpec/PRISM spectra rather than from heterogeneous literature
line measurements. Thus, for example, a BLAGN classification may be
based on higher-resolution spectroscopy, while the
emission-line ratios used for that source in our diagnostic diagrams
are derived from its PRISMATIC line fluxes and upper limits. The
treatment of the adopted [\ion{Ne}{5}] sample, including the use of the
integrated BADASS [\ion{Ne}{5}] measurements, is described in
Section~\ref{sec:sample_selection}.

The samples defined below form the comparison populations used
throughout the paper. The number of galaxies entering a particular
analysis can be smaller because of source-specific wavelength coverage,
measurement quality, line strength, failed or unconstrained fits, or
the absence of a valid upper limit. The eligibility criteria, treatment
of detections and limits, and final sample sizes for each diagnostic
are described in Section~\ref{sec:nev_line_diagnostics}.

\subsection{Broad-line AGN}
\label{sec:comparison_blagn}

The BLAGN comparison sample was assembled from published JWST sources
classified as broad-line or Type~1 AGN. Specifically, we include the
published BLAGN samples from
\citet{Harikane2023,Greene2024,Matthee2024,PerezGonzalez2024,
Kocevski2025,Taylor2025,Hviding2025,Juodzbalis2026}.
Galaxies already included in our adopted [\ion{Ne}{5}] sample were
excluded from the comparison sample. After matching and
de-duplication, the BLAGN comparison sample contains 139 unique
galaxies.

The literature classifications are based on heterogeneous observations
and analysis methods, including different NIRSpec configurations,
spectral resolutions, fitting procedures, broad-line width thresholds,
and significance requirements. We therefore adopt the published BLAGN
classification for each source rather than attempting to reclassify all
of the literature samples using a single broad-line threshold. The
BLAGN comparison sample should consequently be regarded as a
compilation of published broad-line AGN rather than as a uniformly
selected sample. As described above, these published classifications
are treated independently of the emission-line measurements used in
our diagnostic comparisons, which are taken uniformly from the
PRISMATIC measurements of the NIRSpec/PRISM spectra.

BLAGN and LRD classifications are both retained when they apply to the
same galaxy. In figures in which each galaxy is shown only once, such
sources are shown as BLAGN and are not plotted a second time as LRDs.

\subsection{Galaxies selected using narrow-line AGN diagnostics}
\label{sec:comparison_nlagn}

We construct a second high-redshift comparison sample from galaxies
satisfying the adopted secure S2--VO87 AGN criterion using the
PRISMATIC emission-line measurements from the JWST/NIRSpec PRISM
spectra. We exclude galaxies in the adopted [\ion{Ne}{5}] sample,
sources with a published BLAGN classification, and sources with a
positive PRISM broad-component flag. After de-duplication, the
narrow-line-diagnostic comparison sample contains 353 galaxies.

We refer to these objects as galaxies selected using narrow-line AGN
diagnostics rather than as confirmed narrow-line AGN. Excluding known
or flagged broad-line sources prevents the same galaxy from being
included in both the BLAGN and narrow-line-diagnostic comparison
samples, but the absence of a broad-line classification does not
demonstrate that broad Balmer emission is absent. For many sources,
broad lines may be unassessable because of incomplete wavelength
coverage, limited spectral resolution, blending, or inadequate
sensitivity. These limitations are particularly relevant for PRISM
spectroscopy because its spectral resolution varies strongly with
wavelength.

The interpretation of conventional optical diagnostic diagrams also
becomes less straightforward at high redshift and low metallicity.
Hard stellar ionizing spectra and the evolving nebular conditions of
high-redshift galaxies can cause star-forming galaxies and AGN to
overlap in classical line-ratio diagrams
\citep{Scholtz2025}. We therefore treat this as an observationally
defined narrow-line-diagnostic comparison sample rather than as a
complete census of systems demonstrated to host exclusively
narrow-line AGN.

\subsection{Little Red Dots}
\label{sec:comparison_lrds}

The LRD comparison sample was assembled from published photometric and
spectroscopic catalogs. Specifically, we include the multi-field
compilation of \citet{Kocevski2025}, the GOODS/JADES sample of
\citet{Rinaldi2025}, the SMILES/JADES sample of
\citet{PerezGonzalez2024}, the COSMOS-Web catalog of
\citet{Akins2025}, and compact red sources identified in UNCOVER
spectroscopy by \citet{Greene2024}.

The LRD and BLAGN samples are not mutually exclusive, and some LRDs
also belong to the adopted [\ion{Ne}{5}] sample. For figures in which
each galaxy is shown only once, galaxies in the adopted
[\ion{Ne}{5}] sample and galaxies with a published BLAGN
classification are not plotted a second time as LRDs. After catalog
matching, de-duplication, and these exclusions, 116 galaxies are shown
as LRDs in these figures. Galaxies identified as LRDs in the literature
are still considered LRDs even when they are displayed under another
classification in such figures.

The contributing studies use different combinations of red optical
colors, continuum slopes, compactness, morphology, and
spectral-energy-distribution criteria. The resulting LRD sample is
therefore a heterogeneous literature compilation rather than a
uniformly selected physical population. Differences among the adopted
LRD definitions can affect the sample's redshift, luminosity, and
emission-line distributions.

Published BLAGN are not shown a second time as LRDs in figures
requiring each galaxy to appear only once. This is done solely to avoid
duplicate points and does not imply that the remaining LRDs lack
broad-line regions. For many LRDs, the available spectroscopy does not
provide the wavelength coverage, spectral resolution, or sensitivity
needed to determine whether broad Balmer emission is present.

\subsection{Local [\ion{Ne}{5}] comparison samples}
\label{sec:comparison_local_nev}

The preceding samples provide high-redshift comparisons selected
through broad permitted lines, narrow-line ratios, or compact red
continua. We also consider two low-redshift samples selected through
[\ion{Ne}{5}] emission itself. These samples provide a local reference
for galaxies known to produce the same coronal line used to define the
JWST sample.

\subsubsection{The Coronal Line Activity Spectroscopic Survey}
\label{sec:comparison_class}

The first low-redshift comparison is drawn from the Coronal Line
Activity Spectroscopic Survey (CLASS; \citealt{Reefe2022}), which
searched Sloan Digital Sky Survey spectra for a broad range of optical
coronal lines. We use the adopted subset of 61 CLASS galaxies with
reported [\ion{Ne}{5}]~$\lambda3426$ emission.

The CLASS sample provides a local comparison selected through the same
high-ionization transition as our JWST sample. We combine the
published CLASS [\ion{Ne}{5}] and [\ion{Ne}{3}] measurements with
optical emission-line measurements from the MPA--JHU catalogs where
required. For analyses involving nebular [\ion{He}{2}], we use the
subset matched to the strong-nebular-[\ion{He}{2}] catalog of
\citet{ShiraziBrinchmann2012}. The measurement and coverage
requirements used to define the CLASS subset entering each diagnostic
are given in Section~\ref{sec:nev_line_diagnostics}.

\subsubsection{Local metal-poor [\ion{Ne}{5}] emitters}
\label{sec:comparison_metalpoor}

The second low-redshift comparison contains 15 compact star-forming or
blue compact dwarf galaxies with published [\ion{Ne}{5}] emission.
The adopted line intensities were compiled from
\citet{Izotov2004,ThuanIzotov2005,Izotov2012,Izotov2021,Berg2021}.
These studies generally report extinction-corrected emission-line
intensities normalized to $I(\mathrm{H}\beta)=100$, which we preserve
in the line-ratio comparisons.

We include this sample to investigate whether the high-redshift
[\ion{Ne}{5}] emitters occupy similar ionization-diagnostic space to
low-metallicity galaxies with exceptionally hard radiation fields.
The local metal-poor galaxies are not treated as a pure
stellar-ionization control. The literature discusses several possible
contributors to their [\ion{Ne}{5}] emission, including radiative
shocks, accretion onto compact objects or massive black holes, and
unusually hard stellar ionizing spectra
\citep{Izotov2004,ThuanIzotov2005,Izotov2012,Izotov2021,Berg2021,Hatano2024}.
They therefore provide a physically informative comparison for
low-metallicity, hard-ionization conditions without constituting an
unambiguous population free of nonstellar ionization.

All 15 published galaxies are retained in the parent comparison
catalog. Requirements involving the availability and quality of
additional emission lines are applied separately for each diagnostic
in Section~\ref{sec:nev_line_diagnostics} and do not redefine membership
in the underlying literature sample.

\section{Results}
\label{sec:results}
\subsection{Observed Properties of the High-Redshift [\ion{Ne}{5}] Sample}
\label{sec:observed_nev_properties}

\subsubsection{Redshift and [\ion{Ne}{5}] Luminosity Distribution}
\label{sec:nev_redshift_luminosity}

\begin{figure*}
    \centering
    \includegraphics[width=\textwidth]
        {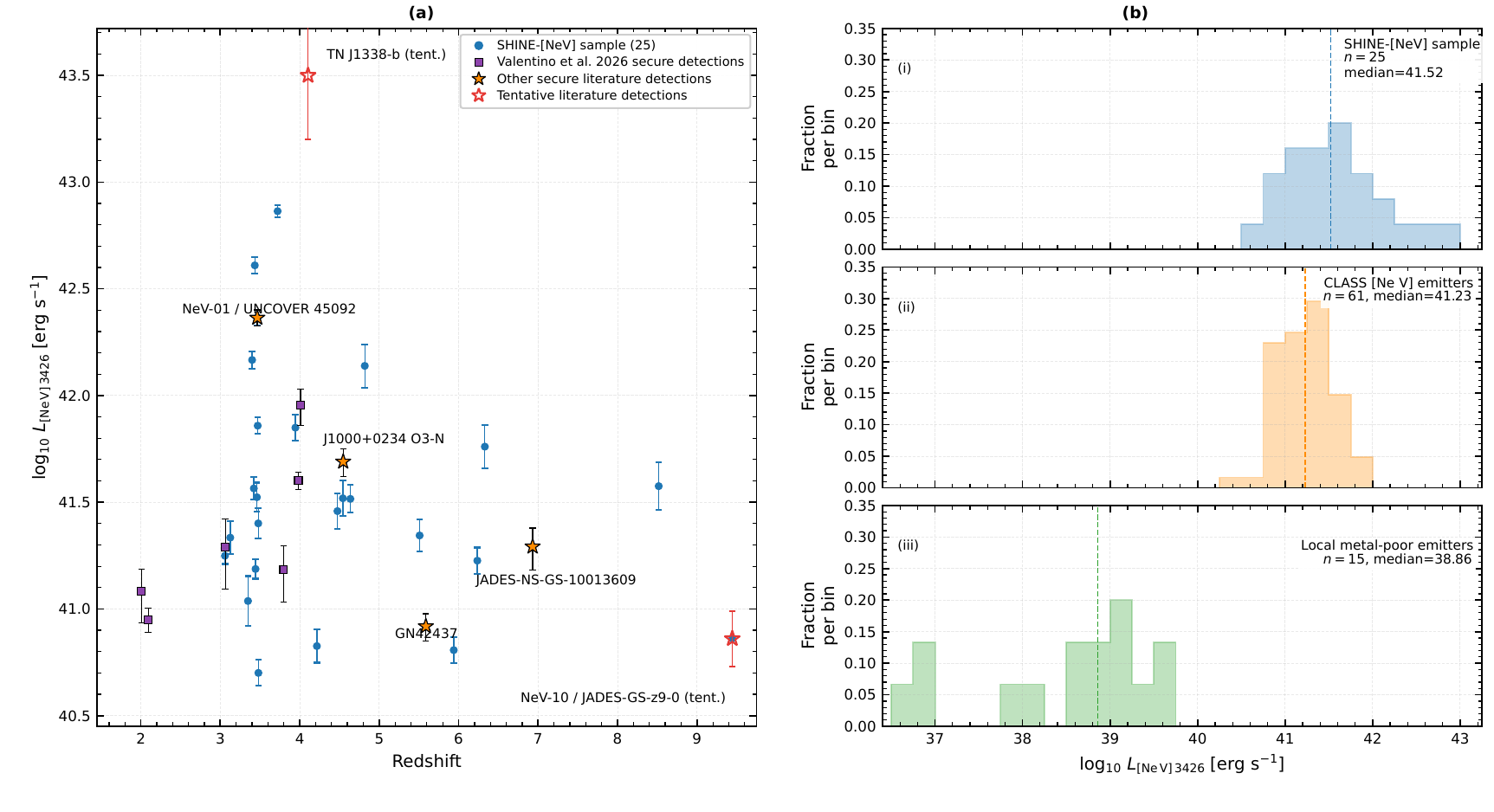}
    \caption{[\ion{Ne}{5}]~$\lambda3426$ luminosities of the
    SHINE-[NeV] and comparison samples.
    \textbf{(a)} Apparent line luminosity versus spectroscopic
    redshift for the 25 SHINE-[NeV] galaxies, with selected published
    high-redshift measurements shown for comparison.
    \textbf{(b)} Luminosity distributions for SHINE-[NeV], 61 CLASS
    emitters, and 15 local metal-poor emitters; dashed lines denote
    sample medians.}
    \label{fig:nev_luminosity_properties}
\end{figure*}

In Figure~\ref{fig:nev_luminosity_properties}(a), we show the apparent
[\ion{Ne}{5}]~$\lambda3426$ luminosities for the 25 galaxies in the
SHINE-[NeV] sample. The sample spans $z=3.059$--$9.444$ (median
$z=3.722$), with apparent line luminosities of
$\log_{10}(L_{3426}^{\rm app}/\mathrm{erg\,s^{-1}})=40.70$--$42.86$
(median 41.52). All values reported are apparent luminosities. The
UNCOVER DR4 lens models\footnote{\url{https://jwst-uncover.github.io/\#releases}}
give magnifications of $\mu=1.93$, 1.94, and 1.41 for NeV-01, NeV-02,
and NeV-03, respectively. These correspond to modest luminosity
corrections of $\simeq0.15$--$0.29$ dex and do not alter our
conclusions.

The highest-redshift source, NeV-10 ($z=9.444$), is JADES-GS-z9-0, in
which \citet{Curti2025} tentatively identified a $\sim2.9\sigma$
[\ion{Ne}{5}]~$\lambda3426$ feature. We retain NeV-10 in SHINE-[NeV]
because it was flagged through our visual inspection and satisfied
our uniform integrated-flux signal-to-noise threshold of
$F_{3426}/\sigma_{F_{3426}}>3$ in \textsc{badass}.

In Figure~\ref{fig:nev_luminosity_properties}(b), we compare the
SHINE-[NeV] luminosities with those of the low-redshift CLASS
[\ion{Ne}{5}] emitters ($N=61$) and the local metal-poor comparison
sample ($N=15$; Section~\ref{sec:comparison_local_nev}). The
SHINE-[NeV] luminosities substantially overlap those of the CLASS
sample. Twenty of the 25 sources fall within the CLASS range of
$\log_{10}(L_{3426}/\mathrm{erg\,s^{-1}})=40.50$--$41.94$ (median
41.23), while five exceed the CLASS maximum \citep{Reefe2022}. We
note that 57 of the 60 CLASS objects with usable MPA--JHU
classifications satisfy both the \citet{Kewley2001} and
\citet{Kauffmann2003} BPT criteria. SHINE-[NeV] thus occupies the
[\ion{Ne}{5}] luminosity regime of a predominantly BPT-classified AGN
population.

In contrast, the local metal-poor sample spans
$\log_{10}(L_{3426}/\mathrm{erg\,s^{-1}})=36.52$--$39.73$ (median
38.86), approximately 2.7 dex below the SHINE-[NeV] median. Every
SHINE-[NeV] source is more luminous than the brightest local
metal-poor object. This separation is strongly driven by sensitivity
limits. Among the 9,002 parent-sample galaxies with usable
[\ion{Ne}{5}] coverage and valid flux uncertainties, the deepest
source-specific $4\sigma$ luminosity limit is
$\log_{10}(L_{3426,\mathrm{lim}}/\mathrm{erg\,s^{-1}})=40.52$, 0.79
dex above the maximum luminosity of the local sample. Thus, analogous
low-luminosity [\ion{Ne}{5}] emitters could exist at high redshift but
remain undetected in our PRISM-selected sample. The origin of
[\ion{Ne}{5}] in local metal-poor galaxies also remains uncertain,
with contributions from extreme stellar populations, shocks, X-ray
binaries, and accretion considered in the literature
\citep[e.g.,][]{Izotov2004,ThuanIzotov2005,Izotov2012,Kehrig2018,
Berg2021,Izotov2021,Mingozzi2025}.

The most luminous [\ion{Ne}{5}] feature reported in the literature, detected at $\log_{10}(L_{3426}/\mathrm{erg\,s^{-1}})\simeq43.5$ in the extreme radio-galaxy/protocluster environment of TN~J1338-b, is only a $\sim2\sigma$ detection and is thus classified as tentative \citep{Saxena2024}.
Because the samples in Figure~\ref{fig:nev_luminosity_properties} are
subject to distinct selection functions and depth limits, these
comparisons are descriptive rather than evidence of intrinsic
luminosity evolution. We examine the redshift dependence of
[\ion{Ne}{5}] incidence among galaxies with [\ion{Ne}{3}] detections
in Section~\ref{sec:nev_line_diagnostics}.

\subsubsection{Spectral and imaging diversity of the
\texorpdfstring{\ion{Ne}{5}}{[Ne V]} sample}
\label{sec:nev_spectral_imaging_diversity}

\begin{figure*}
\centering
\includegraphics[width=0.99\textwidth]
{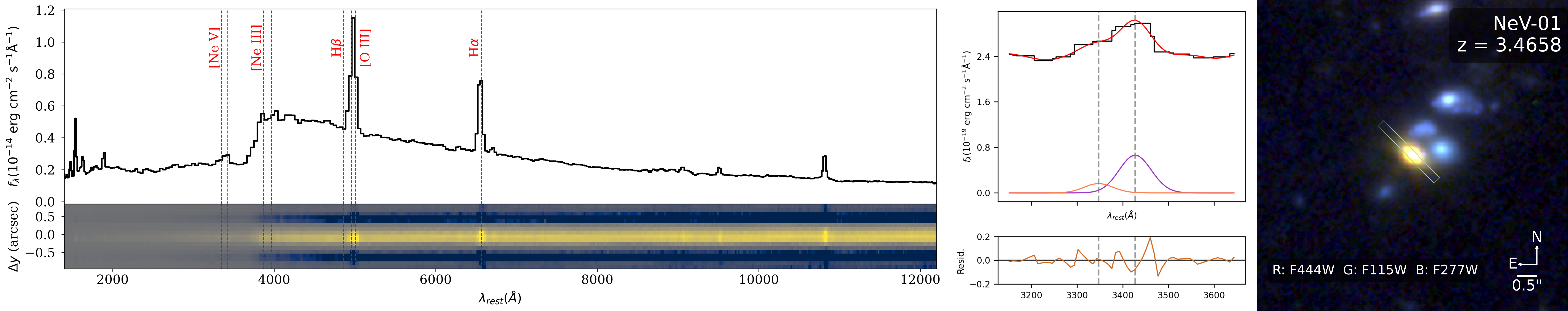}\par\smallskip
\includegraphics[width=0.99\textwidth]
{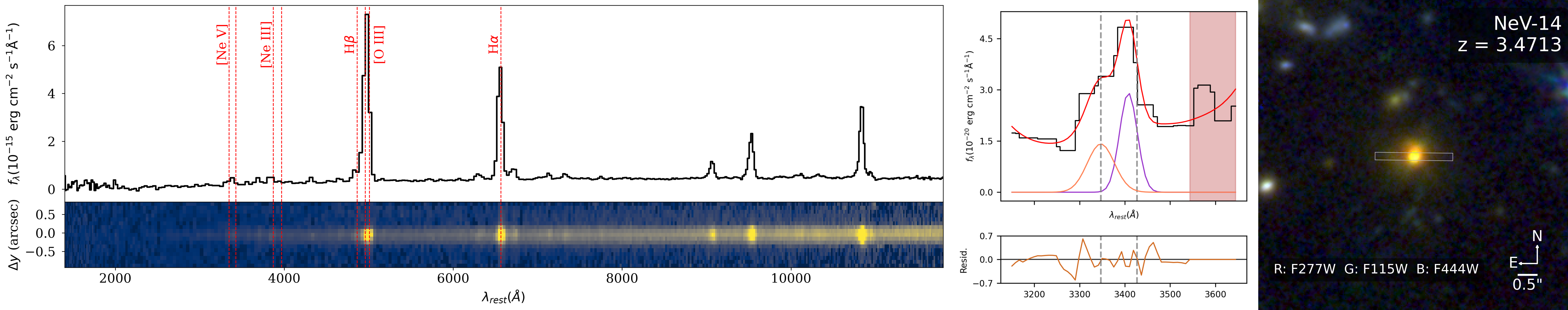}\par\smallskip
\includegraphics[width=0.99\textwidth]
{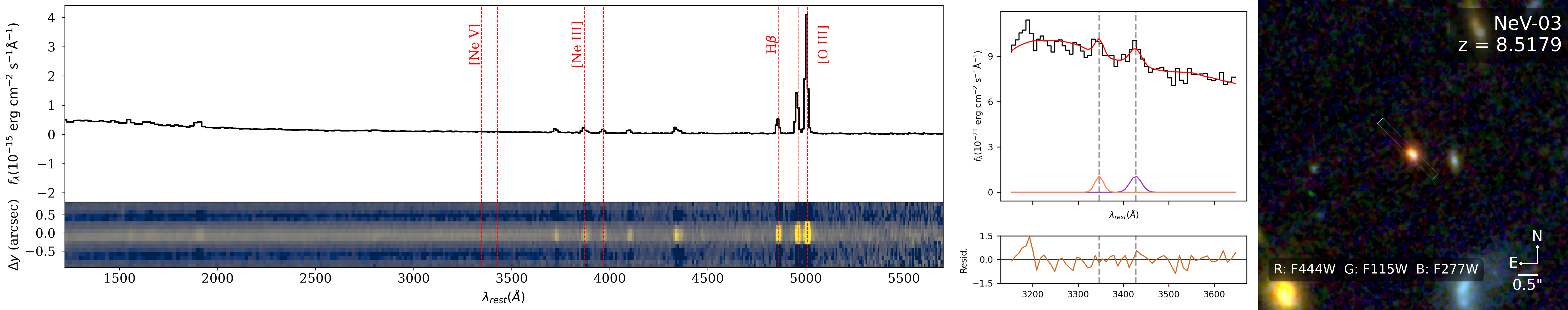}\par\smallskip
\includegraphics[width=0.99\textwidth]
{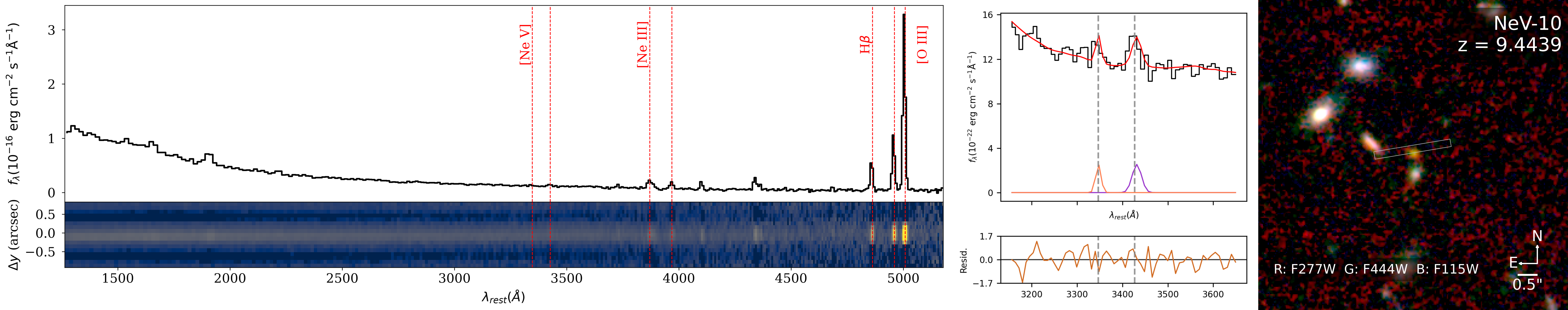}

```
\caption{Representative spectra and NIRCam cutouts for NeV-01,
NeV-14, NeV-03, and NeV-10, from top to bottom. The four sources
illustrate the range of redshifts, apparent [\ion{Ne}{5}]
luminosities, continuum shapes, rest-optical emission-line
strengths, and projected image structures in the sample. In each
row, the left panel shows the rest-frame PRISM one-dimensional
spectrum above the corresponding two-dimensional spectrum. Red
vertical markers identify selected emission lines. The central
panels show the BADASS fit to the
[\ion{Ne}{5}]~$\lambda\lambda3346,3426$ region and its residuals;
black shows the observed spectrum, red the total model, and the
colored curves the fitted line components. Gray dashed lines mark
the expected wavelengths of the [\ion{Ne}{5}] doublet. The right
panel shows the NIRCam RGB cutout, with F115W, F277W, and F444W
assigned to blue, green, and red, respectively. Orientation and
angular scale are indicated. The images provide visual context
and are not quantitative morphology classifications. The full
25-source atlas is presented in
Appendix~\ref{app:nev_atlas}.}
\label{fig:nev_representative_spectra}

\end{figure*}

Figure~\ref{fig:nev_representative_spectra} shows the PRISM spectra,
[\ion{Ne}{5}] fits, and NIRCam cutouts for four sources selected to
illustrate the diversity of the SHINE-[NeV] sample. They span
$z=3.466$--$9.444$ and
$\log_{10}(L_{3426}^{\rm app}/\mathrm{erg,s^{-1}})=40.86$--$42.36$.
They are illustrative examples rather than a statistically
representative subsample. We construct the RGB cutouts from DJA
imaging products \citep{Brammer2023,Heintz2025,Valentino2025}, with
F115W, F277W, and F444W assigned to blue, green, and red,
respectively. Because these filters probe different rest-frame wavelengths across the sample, we use the RGB images to illustrate spatial structure and internal color variations without interpreting their physical origin.

NeV-01 and NeV-14 lie at nearly the same redshift ($z=3.466$ and
$3.471$) but differ in continuum shape, relative rest-optical
emission-line strength, and NIRCam appearance. NeV-01 shows a
pronounced Balmer break that is not present at comparable strength in
the other three representative spectra. Medium-resolution NIRSpec
spectroscopy also reveals a broad H$\alpha$ component that is not
clearly identifiable in the PRISM spectrum. NeV-01 lies in the
Abell~2744/UNCOVER lensing field and has several nearby sources in
projection. NeV-14 shows stronger
rest-optical emission lines relative to its continuum and appears
more compact in the NIRCam image.

NeV-03 and NeV-10 are the two highest-redshift sources in the sample,
at $z=8.518$ and $z=9.444$, respectively. Both show prominent
rest-optical emission lines, but their continua and apparent
[\ion{Ne}{5}] luminosities differ. NeV-03 has
$\log_{10}(L_{3426}^{\rm app}/\mathrm{erg,s^{-1}})=41.58$, compared
with 40.86 for NeV-10. NeV-03 is also classified as an LRD in the
multi-field catalog of \citet{Kocevski2025}. Its NIRCam image shows a
compact central source with a nearby component, whereas NeV-10
appears more elongated and is accompanied by nearby sources.

The full atlas shows substantial variation in continuum shape,
rest-optical line strength, and projected image structure. We do not
use the NIRCam images as quantitative morphology classifications or
assume that nearby sources are physically associated. The remaining
21 SHINE-[NeV] sources are shown in
Appendix~\ref{app:nev_atlas}.

\subsection{Broad-Line Properties and Spectroscopic Classification}
\label{sec:nev_broadline_classification}

Three of the 25 SHINE-[NeV] sources are flagged as broad-line galaxies in the PRISMATIC catalog: NeV-04 and NeV-17 in H$\alpha$, and NeV-21 in both H$\alpha$ and H$\beta$. We treat these flags as qualitative indicators rather than uniform intrinsic line-width measurements, as the resolving power of the PRISM spectra varies strongly with wavelength \citep{Ferruit2022,Jakobsen2022}. The absence of a PRISMATIC flag does not rule out a broad-line region.

Medium- or high-resolution NIRSpec observations cover H$\alpha$ and/or H$\beta$ for eleven sources: ten in H$\alpha$, nine in H$\beta$, and eight in both. Publicly available spectra allowed us to inspect at least one Balmer line in eight sources; data for NeV-08, NeV-12, and NeV-14 remain proprietary. The medium- and high-resolution gratings provide nominal resolving powers of $R\sim1000$ and $R\sim2700$, respectively \citep{Ferruit2022,Jakobsen2022}. Because spectral quality and sensitivity vary across the sample, the lack of a required broad component in a given spectrum does not yield a uniform upper limit on broad-line width or flux.

\begin{figure*}
    \centering
    \includegraphics[width=0.96\textwidth]{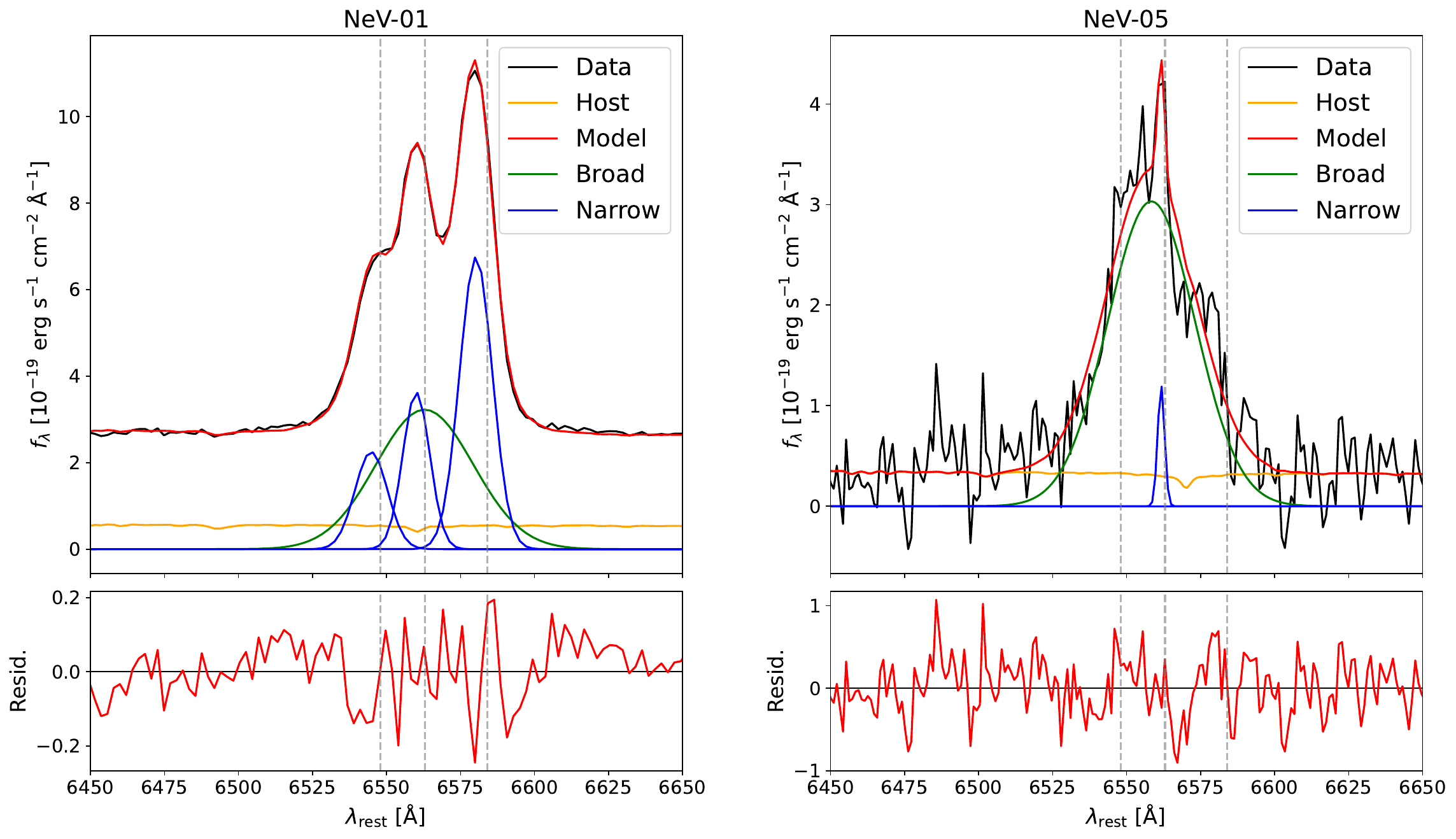}
    \caption{Higher-resolution H$\alpha$ and [\ion{N}{2}] fits for \textbf{(a)} NeV-05 (G395H/F290LP) and \textbf{(b)} NeV-01 (G235M/F170LP). Black curves show the observed spectra, orange the fitted continuum/host galaxy component, red the total model, green the broad component, and blue the narrow components. Lower panels display fit residuals. Vertical dashed lines denote the expected rest-frame wavelengths of H$\alpha$ and the neighboring [\ion{N}{2}]~$\lambda\lambda6548,6583$ doublet.}
    \label{fig:nev_new_broad_components_highres}
\end{figure*}

Higher-resolution spectra reveal two additional broad-line sources that are not flagged by PRISMATIC: NeV-01 exhibits broad H$\alpha$ in G235M/F170LP, while NeV-05 displays broad H$\alpha$ in G395H/F290LP and broad H$\beta$ in G235H/F170LP. Figure~\ref{fig:nev_new_broad_components_highres} presents the higher-resolution H$\alpha$ fits for these two additions.

Among the remaining six sources with public higher-resolution spectra, NeV-04 exhibits a blueshifted, asymmetric H$\alpha$ profile indicative of outflowing gas rather than a canonical broad-line region. Although we retain NeV-04 in our broad-line census based on its PRISMATIC classification, we do not count its higher-resolution spectrum as an independent broad-line detection. The five remaining inspected sources---NeV-10, NeV-11, NeV-23, NeV-24, and NeV-25---lack PRISMATIC broad-line flags and require no broad Balmer component in higher-resolution data.

Combining PRISMATIC classifications with higher-resolution spectral fits yields a final broad-line census of five sources: NeV-01, NeV-04, NeV-05, NeV-17, and NeV-21 (5/25, or 20\% of the SHINE-[NeV] sample). Higher-resolution data contribute NeV-01 and NeV-05, while cross-matching with literature broad-line AGN catalogs yields no further members. Table~\ref{tab:nev_broadline_census} summarizes source classifications and higher-resolution Balmer-line coverage.

\begin{deluxetable*}{llllll}

\tablecaption{Broad-line Classification and Higher-resolution Balmer-line Assessment of the SHINE-[NeV] Sample\label{tab:nev_broadline_census}}

\tabletypesize{\scriptsize}
\setlength{\tabcolsep}{3pt}
\tablewidth{0pt}

\tablehead{
\colhead{ID} &
\colhead{PRISM} &
\multicolumn{2}{c}{Higher-resolution Balmer-line status} &
\colhead{Higher-resolution} &
\colhead{Adopted} \\
\colhead{} &
\colhead{broad-line flag} &
\colhead{H$\alpha$} &
\colhead{H$\beta$} &
\colhead{configuration(s)} &
\colhead{classification}
}

\startdata
NeV-01 & No flag & Broad & Not public &
\shortstack[l]{H$\alpha$: G235M/F170LP;\\G395M/F290LP\\H$\beta$: G235M/F170LP} &
Broad (higher res.) \\
NeV-04 & H$\alpha$ & Asymmetric/outflow-like\tablenotemark{a} & No BLR comp.\tablenotemark{b} &
\shortstack[l]{H$\alpha$: G395M/F290LP\\H$\beta$: not listed} &
Broad (PRISM) \\
NeV-05 & No flag & Broad & Broad &
\shortstack[l]{H$\alpha$: G395H/F290LP\\H$\beta$: G235H/F170LP} &
Broad (higher res.) \\
NeV-08 & No flag & Not public & Not public &
\shortstack[l]{H$\alpha$: G235M/F170LP;\\G395H/F290LP; G395M/F290LP\\H$\beta$: G235M/F170LP} &
Unconstrained \\
NeV-10 & No flag & No coverage & No BLR comp. &
\shortstack[l]{H$\beta$: G395H/F290LP;\\G395M/F290LP} &
No BLR comp. (H$\beta$) \\
NeV-11 & No flag & No BLR comp. & No BLR comp. &
\shortstack[l]{H$\alpha$, H$\beta$: G395H/F290LP;\\G395M/F290LP} &
No BLR comp. \\
NeV-12 & No flag & Not public & Not public &
\shortstack[l]{H$\alpha$: G235H/F170LP;\\G395H/F290LP\\H$\beta$: G235H/F170LP} &
Unconstrained \\
NeV-14 & No flag & Not public & Not public &
\shortstack[l]{H$\alpha$: G235H/F170LP;\\G395H/F290LP\\H$\beta$: G235H/F170LP} &
Unconstrained \\
NeV-17 & H$\alpha$ & No coverage & No coverage & \ldots &
Broad (PRISM) \\
NeV-21 & H$\alpha$, H$\beta$ & No coverage & No coverage & \ldots &
Broad (PRISM) \\
NeV-23 & No flag & No BLR comp. & No BLR comp. &
\shortstack[l]{H$\alpha$: G395M/F290LP\\H$\beta$: G235M/F170LP} &
No BLR comp. \\
NeV-24 & No flag & No BLR comp. & No BLR comp.\tablenotemark{b} &
\shortstack[l]{H$\alpha$: G395M/F290LP\\H$\beta$: not listed} &
No BLR comp. \\
NeV-25 & No flag & No BLR comp. & No BLR comp.\tablenotemark{c} &
\shortstack[l]{H$\alpha$: G235H/F170LP;\\G395H/F290LP\\H$\beta$: G235H/F170LP} &
No BLR comp. adopted \\
\enddata

\tablenotetext{a}{The higher-resolution H$\alpha$ profile of NeV-04 is blueshifted and asymmetric and may be associated with outflowing gas rather than a conventional broad-line region. NeV-04 is retained in the broad-line census on the basis of its PRISM H$\alpha$ classification.}

\tablenotetext{b}{A higher-resolution inspection result is available for this line, but the corresponding grating/filter is not listed in the source metadata.}

\tablenotetext{c}{The H$\beta$ profile of NeV-25 includes a broad-like component, but its profile was interpreted as more consistent with non-BLR kinematics, such as an outflow, and it is therefore not adopted as a conventional broad-line detection.}

\tablecomments{The table includes all SHINE-[NeV] galaxies with nominal medium- or high-resolution Balmer-line coverage, together with NeV-17 and NeV-21, which are classified as broad-line sources from their PRISM spectra but lack corresponding higher-resolution Balmer-line coverage. The remaining 12 SHINE-[NeV] galaxies have neither a PRISM broad-line flag nor nominal medium- or high-resolution Balmer-line coverage. ``Broad'' indicates a broad Balmer component identified in the higher-resolution spectrum. ``No BLR comp.'' indicates that the inspected spectrum did not require a component interpreted as broad-line-region (BLR) emission. ``Not public'' indicates that higher-resolution spectroscopy with nominal coverage of the line exists but was not publicly available for inspection. ``No coverage'' indicates that no available medium- or high-resolution configuration nominally covers the corresponding Balmer line. The absence of a PRISM broad-line flag, or the lack of an inspectable higher-resolution spectrum, is not interpreted as evidence that a BLR is absent.}

\end{deluxetable*}

Because higher-resolution coverage and sensitivity remain non-uniform across the sample, this 20\% detection fraction represents a lower bound rather than a complete estimate of intrinsic broad-line incidence. Consequently, we refrain from classifying sources without detected broad components as Type~2 or narrow-line AGNs.

\subsection{Emission-line Excitation and Diagnostic Properties}
\label{sec:nev_line_diagnostics}

The classical BPT and S2--VO87 diagrams use strong rest-optical
emission-line ratios to distinguish AGNs from star-forming galaxies
\citep{Baldwin1981,VeilleuxOsterbrock1987,Kewley2001}. At
$z>3$, however, the lower gas metallicities and higher ionization
parameters of galaxies cause the predicted line ratios of AGNs and
star-forming systems to overlap
\citep{Feltre2016,Gutkin2016,NakajimaMaiolino2022,
Scholtz2025}. \citet{Scholtz2025} therefore defined conservative high-redshift
boundaries intended to identify sources beyond the region occupied
by the adopted star-forming models. Sources below these boundaries
may still host AGN. For example, GN~42437 exhibits low-ionization line ratios consistent with star formation despite prominent [\ion{Ne}{5}] emission and direct evidence for accretion \citep{Chisholm2024}.

High-ionization lines probe the ionizing spectrum at higher energies
than the standard optical diagnostics. We define
\begin{equation}
{\rm Ne53} \equiv
\log_{10}\left(
\frac{F_{3426}}
     {F_{3869}}
\right),
\label{eq:ne53}
\end{equation}
Because this ratio compares two ionization states of the same
element, it has been proposed as a relatively
metallicity-insensitive probe of the very-high-ionization spectrum
\citep{AbelSatyapal2008,Cleri2023}.

\begin{figure*}[t]
    \centering

    \begin{minipage}[t]{0.495\textwidth}
        \centering
        \textbf{(a)}\\[-0.5ex]
        \includegraphics[width=\linewidth]
        {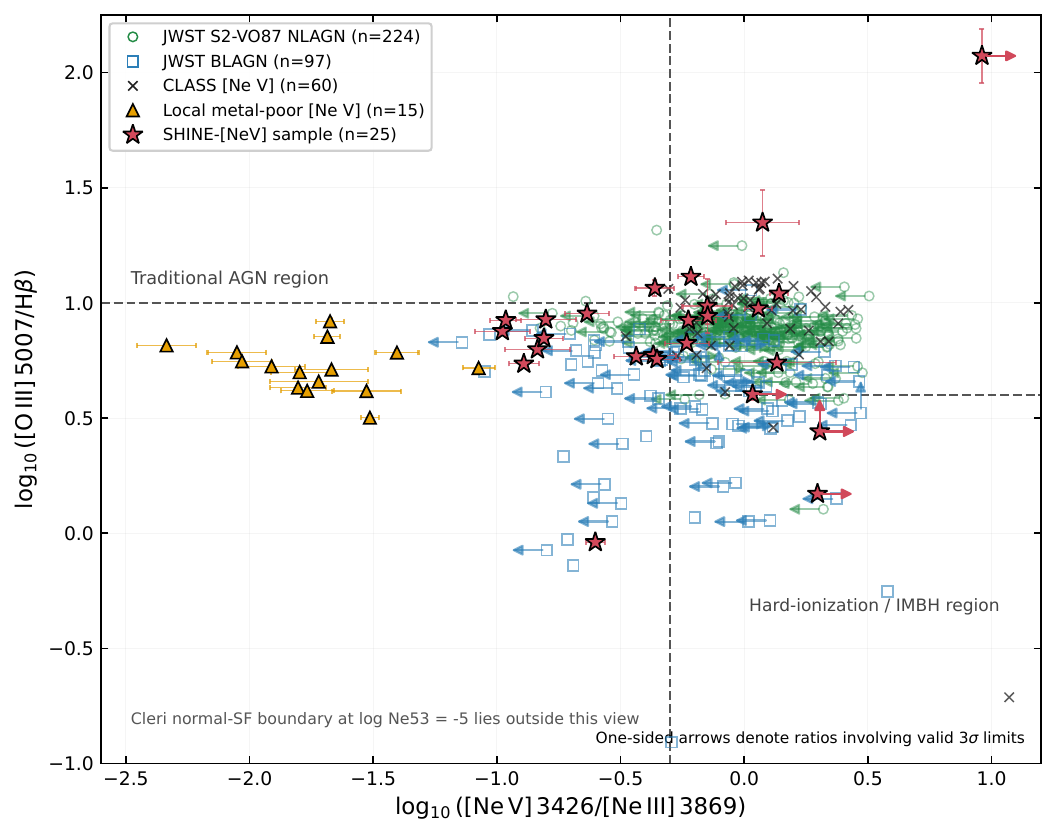}
    \end{minipage}
    \hfill
    \begin{minipage}[t]{0.495\textwidth}
        \centering
        \textbf{(b)}\\[-0.5ex]
        \includegraphics[width=\linewidth]
        {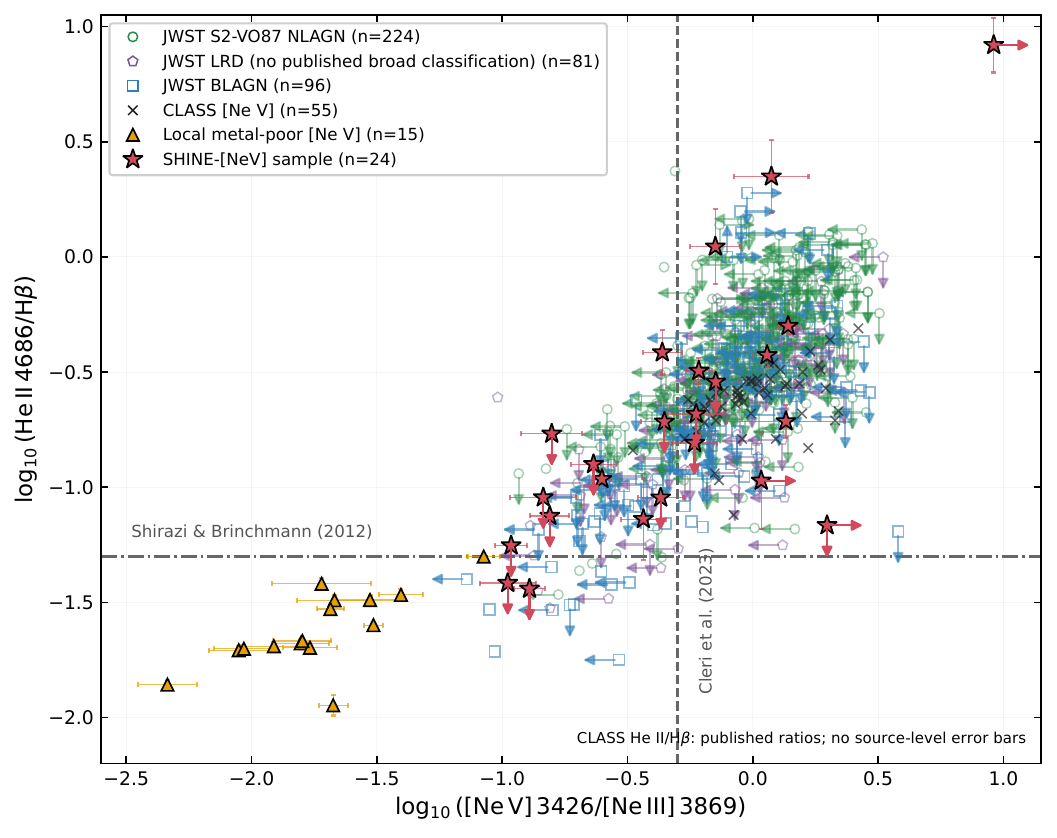}
    \end{minipage}

    \caption{High-ionization emission-line diagnostics for the
    SHINE-[NeV] sample.
    \textbf{(a)}
    $\log_{10}$([\ion{O}{3}]$~\lambda5007/\mathrm{H}\beta$) as a
    function of
    ${\rm Ne53}=
    \log_{10}$([\ion{Ne}{5}]$~\lambda3426/$
    [\ion{Ne}{3}]$~\lambda3869$).
    Red stars show the 25 SHINE-[NeV]
    galaxies. The comparison samples comprise 97
    published JWST broad-line AGNs, 224 galaxies selected using the
    S2--VO87 narrow-line diagnostic, 60 CLASS [\ion{Ne}{5}] emitters,
    and 15 local metal-poor [\ion{Ne}{5}] galaxies. The dashed lines
    show the reference divisions of \citet{Cleri2023}, including
    ${\rm Ne53}=-0.3$; the normal star-forming reference boundary
    lies outside the plotted range.
    \textbf{(b)}
    $\log_{10}(\mathrm{He\,II}~\lambda4686/\mathrm{H}\beta)$ as a
    function of Ne53. The 24 SHINE-[NeV] galaxies with a usable ratio
    are compared with 96 published JWST broad-line AGNs, 224
    galaxies selected using the S2--VO87 narrow-line diagnostic, 81
    literature LRDs without a published broad-line classification,
    55 CLASS [\ion{Ne}{5}] emitters, and 15 local metal-poor
    [\ion{Ne}{5}] galaxies. NeV-04 is omitted because both He~II
    and H$\beta$ are represented by limits. The vertical dashed line
    marks the \citet{Cleri2023} reference value
    ${\rm Ne53}=-0.3$. The horizontal dash-dotted line at
    $\log_{10}(\mathrm{He\,II}/\mathrm{H}\beta)=-1.3$ marks the
    low-metallicity He~II reference of
    \citet{ShiraziBrinchmann2012}, as adopted by
    \citet{Chisholm2024}. For the SHINE-[NeV] sample,
    [\ion{Ne}{5}] uses the adopted integrated BADASS flux and
    uncertainty, while the remaining JWST lines are taken from PRISMATIC. The CLASS ratios in
    panel (a) combine the CLASS [\ion{Ne}{5}] and
    [\ion{Ne}{3}] measurements with MPA--JHU
    [\ion{O}{3}] and H$\beta$ fluxes. The CLASS sample in panel
    (b) is restricted to sources matched to the
    \citet{ShiraziBrinchmann2012} catalog and is shown without
    He~II/H$\beta$ error bars because source-level uncertainties
    are not available for the published ratios. The local comparison
    uses published extinction-corrected line intensities. One-sided
    arrows indicate ratios involving valid $3\sigma$ limits. The
    reference divisions are shown for comparison and are not used to
    define the SHINE-[NeV] sample. Only one representative spectrum is
    included for the multiply imaged A2744-QSO1 system.}
    \label{fig:nev_highionization}
\end{figure*}

Figure~\ref{fig:nev_highionization}(a) compares Ne53 with
$\log_{10}$([\ion{O}{3}]$~\lambda5007/\mathrm{H}\beta$). The 25
SHINE-[NeV] galaxies span a broad region
of the diagram and overlap with the published JWST broad-line AGNs, LRDs,
galaxies selected using the S2--VO87 narrow-line diagnostic, and the
low-redshift CLASS [\ion{Ne}{5}] emitters. The local metal-poor [\ion{Ne}{5}] galaxies have lower measured
ratios. The local metal-poor
galaxies are not a purely stellar control sample, because shocks, accretion,
or unusually hard stellar spectra may contribute to their
high-ionization emission.

Figure~\ref{fig:nev_highionization}(b) compares Ne53 with optical
He~II~$\lambda4686$/H$\beta$. He~II is sensitive to photons above
the second ionization edge of helium and provides another measure of
the hardness of the ionizing spectrum
\citep{ShiraziBrinchmann2012,Chisholm2024}. Twenty-four of the 25
SHINE-[NeV] galaxies can be placed on this diagram. Among the 24 plotted galaxies, 11 have
measured He~II/H$\beta$ ratios and 13 have upper limits.

The SHINE-[NeV] sources overlap with the broad-line, S2--VO87-selected,
CLASS, LRD, and local metal-poor comparison samples. The large
number of upper limits prevents He~II/H$\beta$ from providing a
complete classification of the sample. A He~II detection provides
additional evidence for a hard ionizing spectrum, but its
nondetection does not argue against a secure [\ion{Ne}{5}]
detection. He~II~$\lambda4686$ is intrinsically faint, and even deep
JWST spectra often provide little constraint from individual upper
limits \citep{Scholtz2025}.

\begin{figure*}[t]
    \centering
    \includegraphics[width=0.78\textwidth]
    {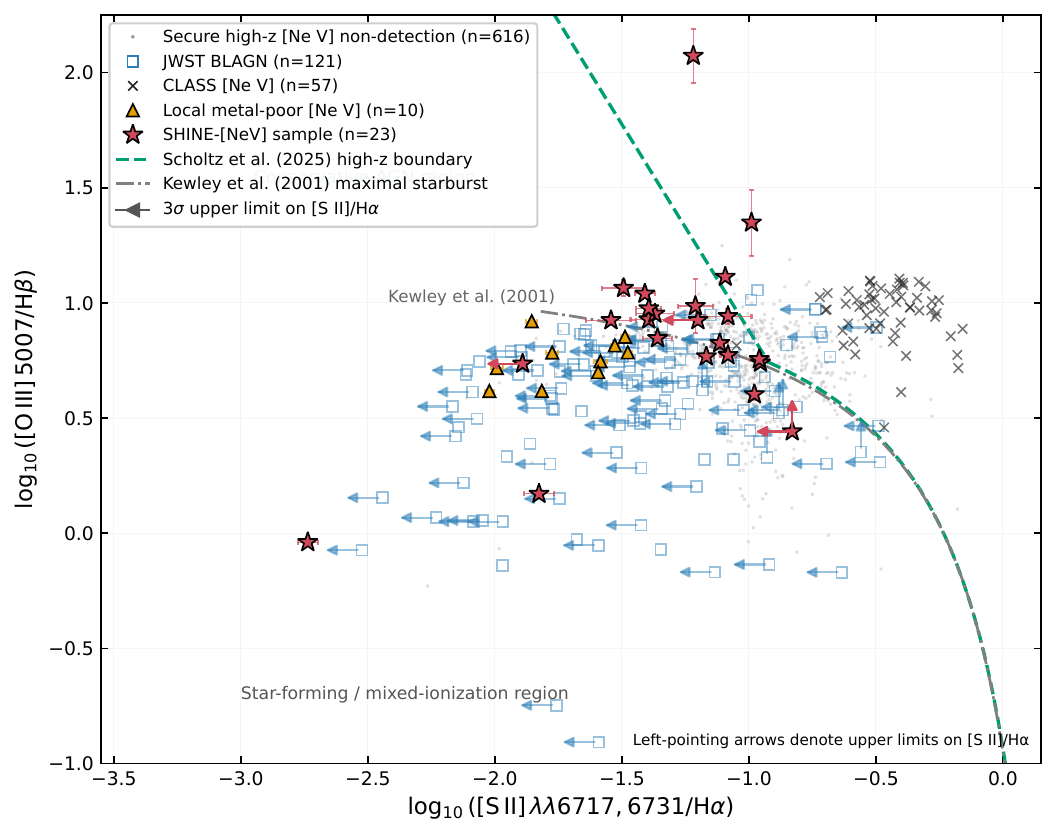}

    \caption{S2--VO87 diagram showing
    $\log_{10}$([\ion{O}{3}]$~\lambda5007/\mathrm{H}\beta$) as a
    function of
    $\log_{10}$([\ion{S}{2}]$~\lambda\lambda6717,6731/
    \mathrm{H}\alpha$). The plotted samples comprise 23 SHINE-[NeV]
    galaxies, 616 quality-qualified
    high-redshift [\ion{Ne}{5}] nondetections, 121 published JWST
    broad-line AGNs, 57 CLASS [\ion{Ne}{5}] emitters, and 10 local
    metal-poor [\ion{Ne}{5}] galaxies. The qualified nondetection
    sample consists of JWST sources outside the SHINE-[NeV] sample with nominal
    [\ion{Ne}{5}] coverage, a constrained [\ion{Ne}{5}] upper
    limit, complete S2--VO87 measurements, and $S/N>3$ in
    H$\alpha$, H$\beta$, [\ion{O}{3}], and [\ion{S}{2}]. It is
    shown as a diagnostic comparison population and not as an
    unbiased control sample. The green dashed curve is the
    conservative high-redshift boundary of \citet{Scholtz2025}, and
    the gray dash-dotted curve is the \citet{Kewley2001}
    maximal-starburst relation. Left-pointing arrows indicate
    $3\sigma$ upper limits on
    [\ion{S}{2}]~$\lambda\lambda6717,6731$/H$\alpha$. Two
    SHINE-[NeV] sources lack usable ratios and are not plotted. One of
    the 23 plotted sources has limits that cross the Scholtz et al.
    boundary and is not assigned to either side. Only one
    representative spectrum is included for the multiply imaged
    A2744-QSO1 system. Sources below the Scholtz et al. boundary are
    not necessarily free of AGN activity.}
    \label{fig:nev_s2vo87}
\end{figure*}

In Figure~\ref{fig:nev_s2vo87}, we show the location of the SHINE-[NeV] sample on the
S2--VO87 diagram,
[\ion{O}{3}]~$\lambda5007$/H$\beta$ versus
[\ion{S}{2}]~$\lambda\lambda6717,6731$/H$\alpha$
\citep{VeilleuxOsterbrock1987}. We show the
\citet{Kewley2001} maximal-starburst relation and the conservative
high-redshift boundary of \citet{Scholtz2025}. The latter was
defined to exclude the region reached by the adopted high-redshift
star-forming models and therefore provides a more restrictive AGN
selection.

Twenty-three of the 25 SHINE-[NeV] galaxies have line ratio or upper limits measurements that can be plotted on the S2--VO87 diagnostic diagram. Relative to the \citet{Kewley2001} relation,  14 of 21 ($67\%$) lie above the relation and 7 of 21 ($33\%$) lie below it. For the remaining two plotted galaxies, the allowed ranges defined by their upper limits cross the Kewley relation, preventing a secure classification relative to that boundary. Two additional galaxies lack usable S2--VO87 ratios and are not plotted. 

In contrast, only 3 of 22 classifiable SHINE-[NeV] sources ($14\%$)
lie above the more conservative \citet{Scholtz2025} boundary, while
19 of 22 ($86\%$) lie below it. Thus, most classifiable SHINE-[NeV] galaxies lie above the \citet{Kewley2001} relation, although 32\% of galaxies with constraining [\ion{Ne}{5}] nondetections also lie above this boundary. Only a small fraction of the SHINE-[NeV] sample satisfies the more conservative \citet{Scholtz2025} criterion. Sources below this boundary should not be interpreted as non-AGN. Indeed, \citet{Scholtz2025} emphasize the importance of high-ionization lines, including [\ion{Ne}{5}], for identifying AGN at high redshift.

The diagnostic methods identify overlapping but non-identical subsets
of the SHINE-[NeV] sample. Five of the 25 galaxies have adopted
broad-line classifications, and most sources lie above the \citet{Kewley2001} relation, whereas only a
small fraction lie above the more conservative \citet{Scholtz2025}
boundary. The SHINE-[NeV] galaxies also overlap with the comparison
populations in the high-ionization diagrams rather than occupying a
distinct region. [\ion{Ne}{5}] therefore provides complementary
information on the hard ionizing spectrum rather than reproducing any
single broad-line or rest-optical diagnostic selection.
Table~\ref{tab:nev_diagnostic_summary} summarizes the broad-line and
emission-line diagnostic classifications for each SHINE-[NeV] galaxy.

\begin{deluxetable*}{lcccc}

\tablecaption{Broad-line Status and Emission-line Diagnostic Regions of the
SHINE-[NeV] Sample
\label{tab:nev_diagnostic_summary}}

\tabletypesize{\scriptsize}
\setlength{\tabcolsep}{4pt}
\tablewidth{0pt}

\tablehead{
\colhead{ID} &
\colhead{Adopted broad-line} &
\colhead{\citet{Cleri2023}} &
\colhead{\citet{Kewley2001}} &
\colhead{\citet{Scholtz2025}} \\
\colhead{} &
\colhead{status} &
\colhead{region} &
\colhead{S2--VO87} &
\colhead{S2--VO87}
}

\startdata
NeV-01 & Broad (higher res.) & Traditional AGN & Above & Above \\
NeV-02 & Unconstrained & Traditional AGN & Unconstr. & Below \\
NeV-03 & Unconstrained & Composite & No usable ratio & No usable ratio \\
NeV-04 & Broad (PRISM) & AGN or hard-ion.\tablenotemark{a} & Unconstr. & Unconstr. \\
NeV-05 & Broad (higher res.) & Traditional AGN & Above & Below \\
NeV-06 & Unconstrained & Composite & Above & Below \\
NeV-07 & Unconstrained & Composite & Below & Below \\
NeV-08 & Unconstrained & Traditional AGN & Above & Above \\
NeV-09 & Unconstrained & Traditional AGN & Above & Below \\
NeV-10 & No BLR comp. (H$\beta$) & Composite & No usable ratio & No usable ratio \\
NeV-11 & No BLR comp. & Composite & Below & Below \\
NeV-12 & Unconstrained & Traditional AGN & Above & Below \\
NeV-13 & Unconstrained & Traditional AGN & Above & Below \\
NeV-14 & Unconstrained & Traditional AGN & Above & Above \\
NeV-15 & Unconstrained & Composite & Above & Below \\
NeV-16 & Unconstrained & Composite & Above & Below \\
NeV-17 & Broad (PRISM) & Hard-ion./IMBH\tablenotemark{b} & Below & Below \\
NeV-18 & Unconstrained & Composite & Below & Below \\
NeV-19 & Unconstrained & Traditional AGN & Above & Below \\
NeV-20 & Unconstrained & Traditional AGN & Below & Below \\
NeV-21 & Broad (PRISM) & Composite & Below & Below \\
NeV-22 & Unconstrained & Composite & Below & Below \\
NeV-23 & No BLR comp. & Traditional AGN & Above & Below \\
NeV-24 & No BLR comp. & Composite & Above & Below \\
NeV-25 & No BLR comp.\tablenotemark{c} & Traditional AGN & Above & Below \\
\enddata

\tablenotetext{a}{
For NeV-04, the upper limits place the source at
${\rm Ne53}>-0.3$ but do not determine whether
$\log_{10}$([\ion{O}{3}]$/\mathrm{H}\beta$) lies above or below the
\citet{Cleri2023} division at 0.6. The source therefore lies either in
the traditional-AGN region or in the hard-ionization/IMBH region.
}

\tablenotetext{b}{
The ``Hard-ion./IMBH'' label follows the region defined by
\citet{Cleri2023} at ${\rm Ne53}>-0.3$ and
$\log_{10}($[\ion{O}{3}]$/\mathrm{H}\beta)<0.6$. This label denotes the
location of the source in the Cleri et al. diagnostic diagram and is not a unique
AGN classification.
}

\tablenotetext{c}{
The higher-resolution H$\beta$ profile of NeV-25 contains a broad-like
component that was interpreted as more consistent with an outflow;
no conventional BLR component is adopted. See
Table~\ref{tab:nev_broadline_census} for details.
}

\tablecomments{
The broad-line column summarizes the adopted broad line classification from
Table~\ref{tab:nev_broadline_census}. ``Broad (PRISM)'' and
``Broad (higher res.)'' identify the observations supporting the adopted
broad-line classification. ``No BLR comp.'' indicates that the inspected
higher-resolution Balmer spectra did not require a component interpreted as
broad-line-region emission; when only one Balmer line could be inspected, that
line is specified. ``Unconstrained'' indicates that the available data do not
permit a secure broad-line assessment. It does not imply that broad Balmer
emission is absent. The Cleri et al. column reports the region containing each
source in the $\log_{10}$([\ion{O}{3}]$/\mathrm{H}\beta$)--Ne53 diagram; these
region labels describe diagnostic location rather than a unique physical
classification. ``Above'' and ``Below'' in the S2--VO87 columns indicate the
position relative to the \citet{Kewley2001} maximal-starburst relation and the
conservative high-redshift boundary of \citet{Scholtz2025}, respectively. A
source below either boundary is not classified here as a non-AGN.
``Unconstr.'' in an S2--VO87 column indicates that the measured upper limits
cross the relevant boundary. ``No usable ratio'' indicates that the source
lacks a valid S2--VO87 ratio and is not plotted in
Figure~\ref{fig:nev_s2vo87}. Emission line ratios and upper limits use the same PRISMATIC
catalog measurements; the SHINE-[NeV] [\ion{Ne}{5}] measurements use the adopted integrated
BADASS fluxes.
}

\end{deluxetable*}

\subsection{High-ionization Emission in Broad-line AGNs and LRDs}
\label{sec:nev_blagn_lrd_constraints}

Several studies have reported weak or absent high-ionization emission
lines in JWST BLAGNs and LRDs
\citep[e.g.,][]{Lambrides2026,Zucchi2026,Wang2026}, although
high-ionization lines have been detected in some systems
\citep[e.g.,][]{Tang2025HighIonization}. Using PRISMATIC measurements and upper
limits for our literature-assembled samples of BLAGNs and LRDs, we
systematically test whether the existing [\ion{Ne}{5}] nondetections
provide evidence that most members of these populations are genuinely
deficient in [\ion{Ne}{5}] relative to the SHINE sample.

To assess whether the [\ion{Ne}{5}] nondetections are sufficiently
constraining, we use Ne53 as defined in
Equation~\ref{eq:ne53}. Because [\ion{Ne}{5}] and [\ion{Ne}{3}]
arise from the same element and are close in wavelength, this ratio
reduces sensitivity to abundance, differential attenuation, and
absolute flux calibration. We restrict the comparison to galaxies
with robust [\ion{Ne}{3}] detections and finite $3\sigma$
[\ion{Ne}{5}] upper limits from PRISMATIC, and test whether those
limits exclude the Ne53 values observed in the SHINE-[NeV] sample.

\begin{figure*}
    \centering
    \includegraphics[width=\textwidth]
    {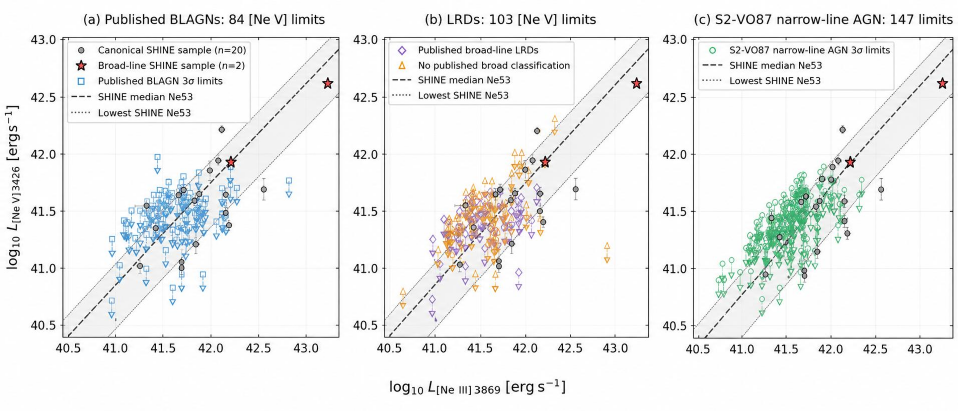}
    \caption{
    \textbf{Constraints on the [\ion{Ne}{5}] emission relative to the 
    [\ion{Ne}{3}] emission in JWST AGN comparison samples.}
    Apparent [\ion{Ne}{5}] luminosity versus apparent
    [\ion{Ne}{3}] luminosity for (a) 84 published broad-line AGNs,
    (b) 103 LRDs, and (c) 147 S2--VO87-selected AGNs with robust
    [\ion{Ne}{3}] detections and finite $3\sigma$
    [\ion{Ne}{5}] upper limits. The 20 SHINE galaxies with robust
    [\ion{Ne}{3}] detections are shown for reference in each panel;
    stars mark the two broad-line SHINE galaxies that satisfy this
    requirement. In panel (b), LRDs with and without a published
    broad-line classification are shown separately. The dashed line
    marks the median Ne53 ratio of the SHINE sample, and the dotted
    line marks the lowest observed SHINE ratio. Downward arrows denote
    $3\sigma$ [\ion{Ne}{5}] upper limits. Only SHINE galaxies are
    counted as [\ion{Ne}{5}] detections in this comparison; weaker or
    borderline emission may be present among sources classified here
    as non-detections. Luminosities are apparent and are not uniformly
    corrected for extinction or lensing. Class memberships overlap.
    }
    \label{fig:nev_neiii_blagn_lrd_constraints}
\end{figure*}

Figure~\ref{fig:nev_neiii_blagn_lrd_constraints} compares the apparent
[\ion{Ne}{5}] and [\ion{Ne}{3}] line luminosities of the SHINE sample with those of the published BLAGNs, LRDs, and S2--VO87-selected AGNs. The dashed and
dotted lines mark the median and lowest Ne53 values, respectively,
among the 20 SHINE galaxies with robust [\ion{Ne}{3}] detections.
These provide empirical reference points for determining whether the
upper limits exclude [\ion{Ne}{5}]/[\ion{Ne}{3}] ratios observed in
the SHINE sample.

As can be seen from Figure~\ref{fig:nev_neiii_blagn_lrd_constraints}, most of the BLAGN and LRD upper limits do not reach below the median
SHINE ratio. Of 84 published BLAGNs, only 33 (39\%) have $3\sigma$
limits below the SHINE median, and only 3 (4\%) fall below the lowest
ratio observed in SHINE. Similarly, 33 of 103 LRDs (32\%) have limits
below the SHINE median, and only 2 (2\%) fall below the lowest SHINE
value. For comparison, 23 of 147 (16\%) S2--VO87-selected AGNs have
limits below the SHINE median, and none falls below the lowest SHINE
ratio. Thus, most existing non-detections do not exclude
[\ion{Ne}{5}]/[\ion{Ne}{3}] ratios already observed in the SHINE
sample. The available PRISMATIC data therefore do not establish that
most BLAGNs or LRDs have weaker [\ion{Ne}{5}] relative to
[\ion{Ne}{3}] than the SHINE galaxies.

For this comparison, we count only the visually vetted and
quantitatively validated SHINE galaxies as [\ion{Ne}{5}] detections
and treat all other PRISM sources as non-detections. As a reminder, the SHINE search
systematically inspected sources passing an automated
$\mathrm{S/N}>4$ preselection and subsequently applied detailed
BADASS fitting. Sources outside the final sample did not all receive
the same level of inspection and may therefore include weaker or
borderline [\ion{Ne}{5}] emission. This analysis tests whether the
available upper limits exclude line ratios observed among the secure
SHINE detections; it does not measure the incidence of arbitrarily
weak [\ion{Ne}{5}] emission.

The SHINE sample itself provides direct counterexamples to a universal
absence of high-ionization emission in these populations. Broad Balmer
emission coexists with [\ion{Ne}{5}] in five SHINE galaxies
(Section~\ref{sec:nev_broadline_classification}), and one SHINE galaxy
is classified as an LRD. Thus, BLAGNs and LRDs can produce
[\ion{Ne}{5}], while the current upper limits are generally
insufficient to determine whether most members of these populations
have systematically weaker [\ion{Ne}{5}]/[\ion{Ne}{3}] ratios.
Deeper observations are required to establish such a population-level
difference.

\subsection{X-ray Properties of the [\ion{Ne}{5}] Emitters}
\label{sec:nev_xray}

JWST-selected AGNs, including both broad- and narrow-line systems, are
frequently undetected even in deep \textit{Chandra} imaging, and stacking
analyses have found X-ray emission below that expected for standard AGN
spectral energy distributions \citep{Maiolino2025,Mazzolari2025}.
Here we examine the X-ray properties of the [\ion{Ne}{5}]-selected
SHINE sample to explore whether a similar X-ray weakness is found.

Of the 25 SHINE-[NeV] galaxies, six have \textit{Chandra}
counterparts, 18 are undetected with valid local catalog detection
thresholds, and one, NeV-15, lacks a valid local detection-threshold
estimate. The fraction of SHINE-[NeV] detections is $6/25=24.0\%$, with a
68\% Wilson confidence interval of $16.6\%$--$33.4\%$. The detections
come from the 7~Ms CDF-S, 2~Ms CDF-N, AEGIS-XD, and X-UDS catalogs
\citep{Luo2017,Xue2016,Nandra2015,Kocevski2018}. Because these surveys
have different depths, this fraction is not a uniform-sensitivity measure of the X-ray detection rate.
For the non-detections, we use the local CSC~2.1 catalog detection
threshold at each source position; these values are not formal
statistical upper limits on the source flux. Details are given in
Appendix~\ref{app:xray_catalogs}.

\begin{deluxetable*}{lllcrrr}
\tabletypesize{\footnotesize}
\tablewidth{0pt}
\tablecaption{Secure Chandra Counterparts of the SHINE-[Ne~V] Sample
\label{tab:xray_secure_counterparts}}
\tablehead{
\colhead{NeV ID} &
\colhead{Balmer broad-line status} &
\colhead{X-ray catalog} &
\colhead{$\Delta\theta$ (\arcsec)} &
\colhead{$\log L_{2-10}$} &
\colhead{$\log L_{\rm [Ne\,V]}$} &
\colhead{$L_{2-10}/L_{\rm [Ne\,V]}$}
}
\startdata
NeV-05\tablenotemark{a} & Broad & X-UDS & 0.321 & 44.840 & 41.849 & 979.5 \\
NeV-12 & Undetermined & 7 Ms CDF-S & 0.263 & 44.190 & 42.864 & 21.2 \\
NeV-14 & Undetermined & 7 Ms CDF-S & 0.216 & 43.060 & 41.859 & 15.9 \\
NeV-19 & Undetermined & Improved 2 Ms CDF-N & 0.101 & 43.774 & 41.565 & 161.9 \\
NeV-21 & Broad & AEGIS-XD v4.2 & 0.113 & 44.666 & 42.610 & 113.7 \\
NeV-25 & No broad component required & AEGIS-XD v4.2 & 0.261 & 43.813 & 42.167 & 44.3 \\
\enddata
\tablecomments{
$\Delta\theta$ is the angular separation between the NIRSpec source position
and the coordinate used to identify the adopted counterpart. For the CDF-S
and AEGIS-XD matches, this is the cataloged multiwavelength counterpart
position; for the CDF-N and X-UDS matches, it is the X-ray centroid. The
luminosities are in units of $\mathrm{erg\,s^{-1}}$. The listed
$L_{2-10}$ values are observed-equivalent rest-frame 2--10~keV luminosities
calculated assuming $\Gamma=1.8$ and are not corrected for intrinsic
absorption. The [Ne~V] luminosities are the apparent BADASS values adopted in
Table~1 and are not uniformly corrected for extinction or gravitational
lensing. ``Undetermined'' means that the available spectroscopy is
insufficient to determine the presence or absence of broad Balmer emission.
``No broad component required'' means that the higher-resolution Balmer-line
data did not require a broad component;
it is not a uniform upper limit on broad-line width or strength.
}
\tablenotetext{a}{
The X-UDS counterpart is retained as secure, but the corresponding CSC entry
carries a confusion flag.
}
\end{deluxetable*}

We find that the X-ray detections are not restricted to galaxies with identified
broad Balmer lines. Two of the five sources with an adopted broad-line
classification are X-ray detected. Four additional X-ray detections
have no adopted broad-line classification. NeV-25 has higher-resolution
spectroscopy available that indicates that there is no broad component, whereas the
broad-line status of NeV-12, NeV-14, and NeV-19 remains undetermined.

\begin{figure}[t]
\centering
\includegraphics[width=\columnwidth]
{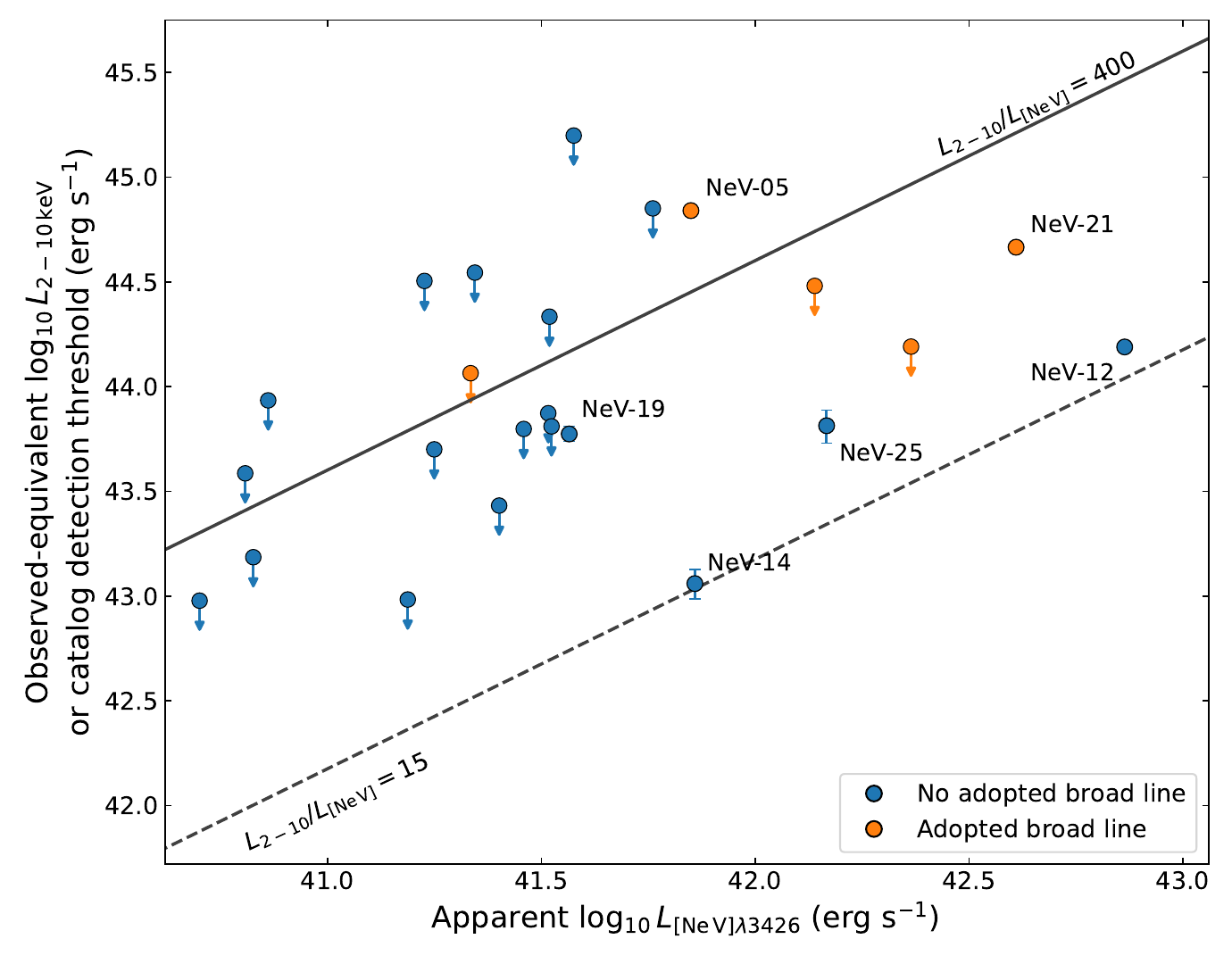}
\caption{
Observed-equivalent rest-frame 2--10~keV luminosity versus apparent
[\ion{Ne}{5}]~$\lambda3426$ luminosity for the SHINE-[NeV] sample.
Blue points have no adopted broad-line classification, while orange
points have an adopted broad-line classification; the sources indicated by the blue symbol should
not be interpreted as a uniformly confirmed narrow-line population.
Vertical error bars show catalog-based statistical uncertainties for
the six \textit{Chandra} detections; for several sources, the
uncertainties are smaller than the plotted symbols. Downward arrows mark the
local CSC~2.1 catalog detection thresholds for 18 nondetections; these
thresholds are not formal source-flux upper limits. NeV-15 is not shown
because no valid local X-ray detection threshold is available at its
position. The solid and dashed lines mark
$L_{2-10}/L_{\rm [Ne\,V]}=400$ and 15, respectively. In the local
calibration of \citet{Gilli2010}, the former is characteristic of
unobscured Seyferts, while ratios below the latter are associated
predominantly with Compton-thick sources. We show these values as
empirical local references rather than classification boundaries.
X-ray luminosities assume $\Gamma=1.8$ and are not corrected for
intrinsic absorption.
}
\label{fig:nev_xray_lx_lnev}
\end{figure}

Figure~\ref{fig:nev_xray_lx_lnev} compares the observed-equivalent
rest-frame 2--10~keV luminosity with the apparent [\ion{Ne}{5}]
luminosity. The six X-ray detections in our sample span
$L_{2-10}/L_{\rm [Ne\,V]}=15.9$--$979.5$. In local Seyferts,
\citet{Gilli2010} found that $L_{2-10}/L_{\rm [Ne\,V]}$ decreases
with increasing X-ray obscuration: $L_{2-10}/L_{\rm [Ne\,V]}\sim400$
is characteristic of unobscured systems, while sources with
$L_{2-10}/L_{\rm [Ne\,V]}<15$ were almost exclusively Compton thick.
We treat these values as empirical local reference points rather than formal classification boundaries. None of the six detections has $L_{2-10}/L_{\rm [Ne\,V]}<15$;
NeV-14 has the lowest ratio, $L_{2-10}/L_{\rm [Ne\,V]}\simeq15.9$.

\citet{Valentino2026} found that all four of their X-ray-detected
[\ion{Ne}{5}] emitters were broad-line AGNs, whereas the two
[\ion{Ne}{5}] emitters without broad lines were undetected in X-rays.
The SHINE-[NeV] sample shows that X-ray detection is not restricted to
broad-line sources. NeV-25 is securely detected at
$\log(L_{2-10}/{\rm erg\,s^{-1}})=43.81$, although its
higher-resolution Balmer-line spectrum does not require a broad
component. Three additional X-ray detections, NeV-12, NeV-14, and
NeV-19, have no identified broad lines, although their broad-line
status remains undetermined. NeV-12 reaches
$L_{2-10}>10^{44}\ {\rm erg\,s^{-1}}$, while NeV-19 has
$L_{2-10}/L_{\rm [Ne\,V]}\simeq162$.

Thus, the individually detected SHINE-[NeV] sources do not show the
pervasive X-ray faintness reported for other JWST-selected broad- and
narrow-line AGN samples, in which most sources remain undetected even
in deep \textit{Chandra} imaging and in stacked data
\citep{Maiolino2025,Mazzolari2025}. The SHINE-[NeV] detections instead
span both broad-line sources and at least one source, NeV-25, for which
higher-resolution spectroscopy does not require a broad Balmer
component.

The 18 SHINE-[NeV] nondetections are less constraining. None of their
local catalog detection thresholds reaches
$L_{2-10}/L_{\rm [Ne\,V]}<15$, while 10 reach below
$L_{2-10}/L_{\rm [Ne\,V]}=400$. We therefore cannot determine whether
the nondetected sources include a similarly X-ray-faint population.
The X-ray luminosities are not corrected for intrinsic absorption, and
the [\ion{Ne}{5}] luminosities are not uniformly corrected for
extinction or lensing. In addition, the \textit{Chandra} and JWST
observations are non-simultaneous, so AGN variability may contribute
to the scatter in $L_{2-10}/L_{\rm [Ne\,V]}$.
\subsection{Rest-frame UV and Optical Continuum Properties}
\label{sec:continuum_properties}

We characterize the rest-frame continuum using the absolute UV
magnitude measured near 1600~\AA, the UV continuum slope $\beta$,
defined by $f_{\lambda}\propto\lambda^{\beta}$ over
1350--3450~\AA, and the catalog Balmer-break strength,
\begin{equation}
B_{\rm Balmer} =
\frac{\langle f_{\nu}(4150\text{--}4350~\text{\AA})\rangle}
     {\langle f_{\nu}(3400\text{--}3600~\text{\AA})\rangle}.
\end{equation}
We require a mean continuum signal-to-noise ratio of at least three
over the UV fitting interval for $M_{\rm UV}$ and $\beta$, and in
both continuum windows for $B_{\rm Balmer}$. These criteria yield
reliable $M_{\rm UV}$ and $\beta$ measurements for 23 of the 25
SHINE-[NeV] galaxies and Balmer-break measurements for 24. The
sample spans a broad range of UV luminosity, continuum shape, and
Balmer-break strength:
$-24.18 \leq M_{\rm UV} \leq -18.42$,
$-2.45 \leq \beta \leq 0.15$, and
$0.84 \leq B_{\rm Balmer} \leq 3.30$, with medians of $-20.47$,
$-1.51$, and 1.22, respectively. Thus, SHINE-[NeV] includes both
UV-faint and UV-luminous galaxies, sources with very blue and
comparatively red UV continua, and galaxies with both weak and
strong Balmer breaks.

We next ask whether galaxies with prominent [\ion{Ne}{5}] emission
differ in their UV luminosities, UV slopes, or Balmer-break strengths
from galaxies with weaker very-high-ionization emission. A direct
comparison with all galaxies lacking a detected [\ion{Ne}{5}] line
would not provide a meaningful control sample. The spectra come from
programs with different depths, and a nondetection may reflect either
intrinsically weak [\ion{Ne}{5}] emission or insufficient
sensitivity. Such a comparison would therefore mix genuinely weak
emitters with galaxies in which comparable [\ion{Ne}{5}] emission
could have remained undetected.

We instead restrict the analysis to galaxies with robust
[\ion{Ne}{3}] detections and use
[\ion{Ne}{5}]/[\ion{Ne}{3}] as the comparison metric.
[\ion{Ne}{3}] provides a detected reference line against which the
[\ion{Ne}{5}] sensitivity can be evaluated for each galaxy.
Moreover, [\ion{Ne}{5}]~$\lambda3426$ and
[\ion{Ne}{3}]~$\lambda3869$ are close in wavelength, reducing the
effects of differential attenuation and wavelength-dependent flux
calibration.

We define strong-ratio sources by
\begin{equation}
\frac{F_{[\mathrm{Ne\,V}]\,\lambda3426}}
     {F_{[\mathrm{Ne\,III}]\,\lambda3869}}
\geq 0.33.
\label{eq:strong_neon_ratio}
\end{equation}
The threshold of 0.33 is the lowest measured
[\ion{Ne}{5}]/[\ion{Ne}{3}] ratio among BPT-selected AGNs in the
local CLASS comparison sample. We use it as an operational,
sample-dependent threshold rather than a universal AGN boundary.

The lower-ratio comparison consists of galaxies outside the
SHINE-[NeV] sample for which
\begin{equation}
\frac{3\,\sigma\!\left(F_{[\mathrm{Ne\,V}]\,\lambda3426}\right)}
     {F_{[\mathrm{Ne\,III}]\,\lambda3869}}
< 0.33.
\label{eq:lower_neon_ratio_limit}
\end{equation}
This criterion requires sufficient sensitivity to detect
[\ion{Ne}{5}] at greater than $3\sigma$ if the source had a ratio
of 0.33. The upper limit determines whether a galaxy enters the
comparison; we do not treat it as a measured line ratio. The
comparison galaxies are not assumed to be non-AGNs.

Twelve SHINE-[NeV] galaxies satisfy the strong-ratio criterion,
eight have measured ratios below 0.33, and five lack robust
[\ion{Ne}{3}] measurements. The lower-ratio comparison contains
195 galaxies. After applying the continuum signal-to-noise
requirements, the $M_{\rm UV}$ and $\beta$ comparisons contain
11 strong-ratio and 191 lower-ratio galaxies, while the
Balmer-break comparison contains 12 and 190, respectively.

\begin{figure*}
    \centering
    \includegraphics[width=\textwidth]
    {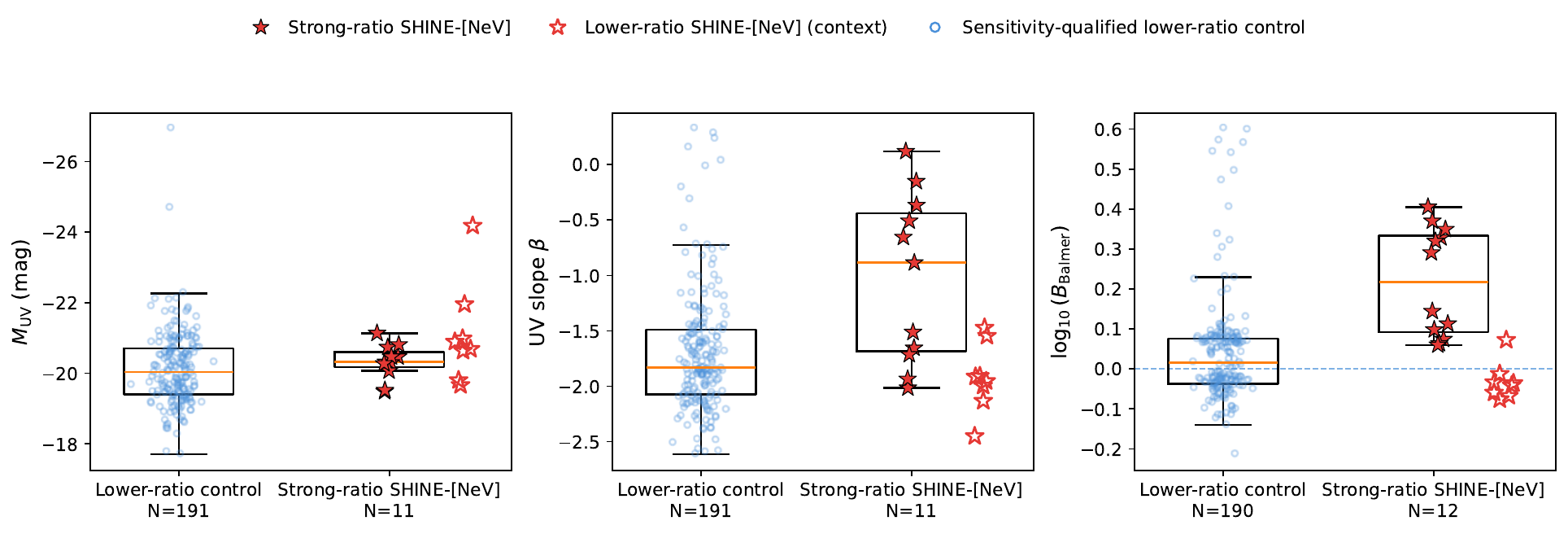}
    \caption{
    Rest-frame continuum properties of the strong
    [\ion{Ne}{5}]/[\ion{Ne}{3}] subsample and the lower-ratio
    comparison sample. Filled red stars show strong-ratio
    SHINE-[NeV] galaxies, open red stars show SHINE-[NeV]
    galaxies with measured ratios below 0.33, and open blue
    circles show the lower-ratio comparison. Box plots summarize
    the full distributions; statistical results in the text use
    the redshift-matched comparison. The dashed line in the right
    panel marks $B_{\rm Balmer}=1$.
    }
    \label{fig:ratio_continuum}
\end{figure*}

Because the strong-ratio and lower-ratio samples have different
redshift distributions, we use a redshift-matched comparison for
the statistical analysis. We match each strong-ratio source to the
three lower-ratio galaxies closest in redshift within the redshift
range shared by the two samples, and compare each continuum
property with the mean of the three matched galaxies. We assess whether the matched differences are consistent with zero
using a two-sided permutation (randomization) test and adjust the
resulting $p$-values for the three continuum properties tested.
Figure~\ref{fig:ratio_continuum} shows the full distributions,
whereas the statistical results below refer to the redshift-matched
comparison.

The strong-ratio and lower-ratio samples have similar UV
luminosities. Their full-sample medians are
$M_{\rm UV}=-20.32$ and $-20.04$, and the redshift-matched
difference is $\Delta M_{\rm UV}=-0.08$~mag
(adjusted $p=0.62$). In contrast, the strong-ratio sources have
redder UV slopes, with median $\beta=-0.89$ compared with $-1.83$
and a matched difference of $\Delta\beta=0.63$
(adjusted $p=0.038$). They also have stronger Balmer breaks:
the median ratios are 1.67 and 1.04, and the matched difference is
$\Delta\log_{10}(B_{\rm Balmer})=0.183$~dex
(adjusted $p=0.0088$), corresponding to a factor of approximately
1.5.

The redder UV slopes persist when we restrict the lower-ratio
comparison to BLAGNs and S2--VO87-selected AGNs, while the
Balmer-break difference becomes only suggestive. Because this
AGN-restricted comparison is small and the samples remain poorly
matched in redshift, we treat it only as a supporting test.

The continuum differences do not uniquely identify their physical
origin. Redder UV slopes may reflect dust or differences in the
stellar and AGN continua, while stronger Balmer breaks can arise
from composite stellar populations rather than a uniformly older
population. Similar strong Balmer breaks in high-redshift nitrogen
emitters have been interpreted as evidence for young line-emitting
regions superposed on older rest-optical stellar populations
\citep{Morel2026}. Stellar masses are not available for a
sufficiently large fraction of the lower-ratio comparison sample to
test whether differences in host stellar mass contribute to these
continuum trends.
\subsubsection{Low-Stellar-Mass [\ion{Ne}{5}] Emitters and the
Search for Intermediate-Mass Black Holes}
\label{sec:low_mass_nev}
Very-high-ionization emission lines can identify accreting black holes
in low-mass galaxies that are missed by standard AGN diagnostics.
Hotter accretion disks expected around lower-mass black holes may enhance the
high-ionization lines even when optical narrow-line ratios remain
consistent with star formation \citep{Cann2018,Cann2019}. In the local
CLASS survey, the highest-ionization coronal lines preferentially occur
in lower-mass galaxies, and most coronal-line emitters in dwarf galaxies
are not classified as AGNs using conventional optical narrow-line
diagnostics \citep{Reefe2022}. Very-high-ionization lines such as
[\ion{Ne}{5}] may therefore provide a complementary means of
identifying accretion in galaxies where broad-line or standard
narrow-line AGN signatures are weak or absent.
Stellar-mass estimates are available for 20 of the 25 SHINE-[NeV]
galaxies from the SED-fitting products described in the analysis
methods. These estimates span
$7.39 \leq \log(M_\ast/M_\odot) \leq 11.09$, with a median of
$\log(M_\ast/M_\odot)=9.29$. Six sources have
$\log(M_\ast/M_\odot)<9$, including one with
$\log(M_\ast/M_\odot)<8$. The SHINE-[NeV] sample therefore extends
into the low-stellar-mass galaxy population.

\begin{figure}
\centering
\includegraphics[width=\columnwidth]
{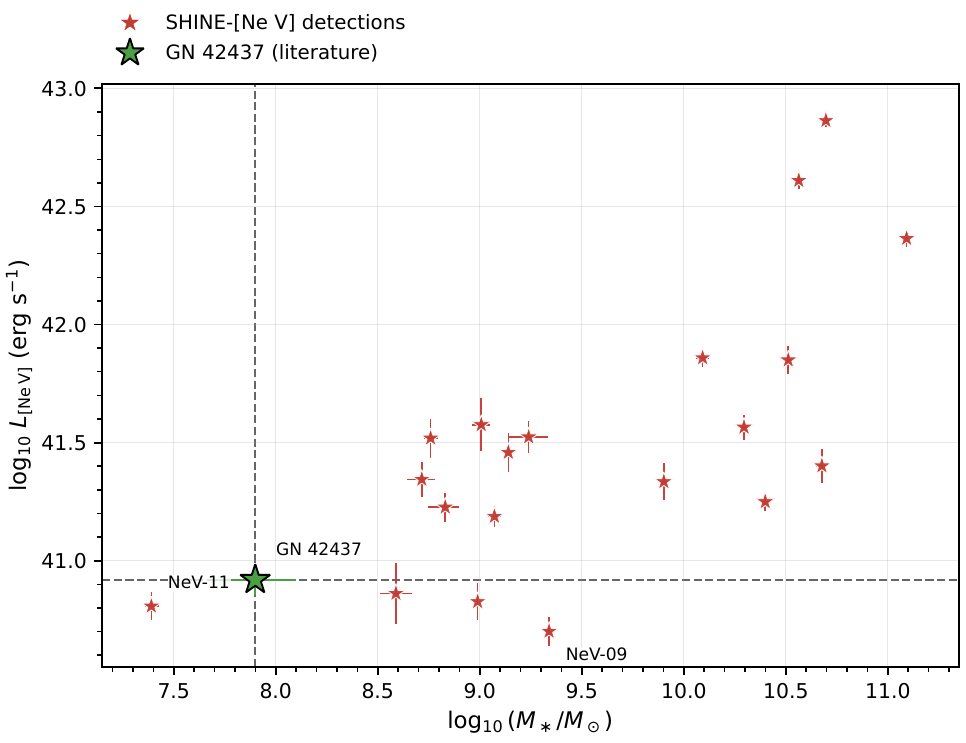}
\caption{
Apparent [\ion{Ne}{5}] luminosity as a function of stellar mass
for the 20 SHINE-[NeV] galaxies with available stellar-mass
estimates. Red stars show the SHINE-[NeV] sources using the
validated integrated BADASS [\ion{Ne}{5}] luminosities, and the
green star shows GN~42437 using its published stellar mass and
apparent [\ion{Ne}{5}] luminosity \citep{Chisholm2024}. Dashed
lines mark the GN~42437 stellar mass and [\ion{Ne}{5}] luminosity.
Horizontal error bars show the 16th--84th percentile intervals
from the current SED fits and do not include systematic modeling
uncertainties; vertical error bars show the statistical
[\ion{Ne}{5}] luminosity uncertainties. SHINE-[NeV] luminosities
are apparent values.
}
\label{fig:nev_mass_gn42437}
\end{figure}

GN~42437 illustrates how high-ionization emission lines can reveal accretion activity in low-mass hosts that traditional diagnostics miss. This compact galaxy at $z=5.587$
has a reported stellar mass of
$\log(M_\ast/M_\odot)=7.9\pm0.2$ and a significant narrow
[\ion{Ne}{5}] detection \citep{Chisholm2024}. Higher-resolution Balmer
profiles show no broad component, while conventional narrow-line optical diagnostic
diagrams place the source in the star-forming region. Through photoionization modeling of the line ratios, \citet{Chisholm2024} showed that pure stellar ionization fails to reproduce the observed spectrum, pointing instead to an accreting intermediate-mass black hole.
NeV-11 provides the closest SHINE-[NeV] comparison in stellar mass.
It lies at $z=5.939$ and has a current stellar-mass estimate of
$\log(M_\ast/M_\odot)=7.39^{+0.04}_{-0.03}$, 0.51~dex below the
published value for GN~42437. Its apparent [\ion{Ne}{5}] luminosity,
$\log(L{\rm [Ne,V]}/{\rm erg,s^{-1}})=40.81$, differs from that of
GN~42437 by only 0.11~dex. Inspection of the medium- and high-resolution NIRSpec spectra
reveal no broad component in either the H$\alpha$ or H$\beta$ profiles. NeV-11 also lies below the adopted S2--VO87 AGN
boundary. Thus, [\ion{Ne}{5}] identifies highly ionized gas in a low-stellar-mass galaxy that evades AGN classification via both broad Balmer emission and our adopted narrow-line diagnostics.
We note also that the mass of NeV-11 is probably fairly uncertain, since it shows a peculiar spectrum with a strong contribution from nebular continuum emission \citep[see][]{Katz2025}. This galaxy, identified previously as GS-NDG-9422 by \cite{Cameron2024}, may be dominated by a very young population with a top-heavy IMF \citep{Cameron2024,Katz2025}, who estimate a lower limit of $\log(M_\ast/M_\odot) \sim 6.2-6.6$. The quoted uncertainty from our SED fits is only formal, and the fits do not allow for IMF deviations. In any case, even if stellar masses for individual objects can be fairly uncertain, the relative ranking of all NeV emitters in $M_\ast/M_\odot$ should be meaningful, at least on average.
UV faintness does not provide an equivalent proxy for low stellar
mass. NeV-09 is the UV-faintest SHINE-[NeV] source, with
$M_{\rm UV}=-18.42$, and has an apparent [\ion{Ne}{5}] luminosity
comparable to GN~42437. Its current stellar-mass estimate is
$\log(M_\ast/M_\odot)=9.34$, more than an order of magnitude larger
than the published stellar mass of GN~42437. We therefore use stellar
mass, rather than $M_{\rm UV}$, to identify possible low-mass analogs
of GN~42437.
Figure~\ref{fig:nev_mass_gn42437} shows apparent [\ion{Ne}{5}]
luminosity as a function of stellar mass for the 20 SHINE-[NeV]
galaxies with available stellar-mass estimates. NeV-11 lies close to
GN~42437 in this plane, while the SHINE-[NeV] sample spans nearly four
orders of magnitude in stellar mass.
These measurements establish that SHINE-[NeV] includes
[\ion{Ne}{5}] emitters with very low inferred stellar masses. They do
not establish that NeV-11 or any other SHINE-[NeV] source hosts an
intermediate-mass black hole. No direct black-hole mass measurements
are available, and any black-hole mass inferred from [\ion{Ne}{5}]
depends on assumptions about the ionizing spectral energy distribution,
bolometric correction, and accretion rate. Stellar masses are also
available for only 20 of the 25 SHINE-[NeV] galaxies and not for a
uniformly selected parent sample. We therefore do not infer either the
incidence of [\ion{Ne}{5}] emission in low-mass galaxies or the
frequency of intermediate-mass black holes from the present sample.
\subsection{Incidence and Redshift Dependence of [\ion{Ne}{5}] Emission}
\label{sec:nev_incidence}

\begin{figure*}[t]
    \centering
    \includegraphics[width=\textwidth]
    {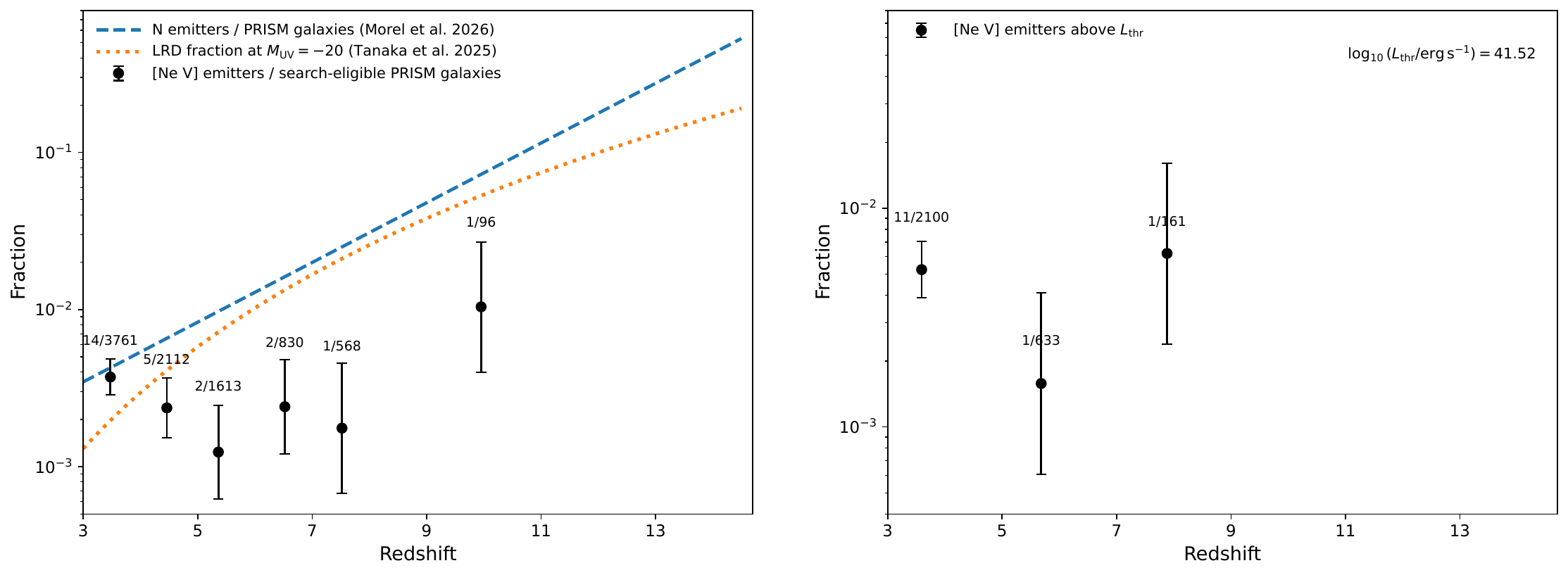}
    \caption{Observed [\ion{Ne}{5}] fractions as a function of
    redshift. \textit{Left:} black points show the number of visually
    validated and BADASS-confirmed [\ion{Ne}{5}] emitters divided by
    the number of unique search-eligible NIRSpec/PRISM galaxies in
    each redshift bin. The dashed curve shows the N-emitter fraction
    among PRISM galaxies reported by \citet{Morel2026}, and the dotted
    curve shows the LRD fraction at fixed $M_{\rm UV}=-20$ from
    \citet{Tanaka2025}. \textit{Right:} black points show the fraction
    of [\ion{Ne}{5}] emitters above the median apparent
    [\ion{Ne}{5}] luminosity threshold,
    $\log_{10}(L_{\rm thr}/{\rm erg\,s^{-1}})=41.52$, restricting the
    denominator to unique NIRSpec/PRISM galaxies capable of detecting
    a line at that luminosity. The numerator is the number of
    [\ion{Ne}{5}] emitters with
    $L_{\rm [Ne\,V]}\geq L_{\rm thr}$. The denominator includes all
    [\ion{Ne}{5}] emitters, whose line luminosities are measured, and
    nonemitters whose catalog $4\sigma$ luminosity limit is at or
    below $L_{\rm thr}$. Because fewer emitters lie above the
    threshold, the right panel uses three broader redshift bins:
    $3\leq z<5$, $5\leq z<7$, and $7\leq z<15$. In both panels,
    vertical error bars show 68\% Wilson score confidence intervals,
    and labels give the exact numerator and denominator. The N-emitter
    and LRD curves have different operational definitions and are
    shown for qualitative comparison. Apparent luminosities are
    computed using the adopted cosmology and are not corrected for
    lensing. The left panel traces the observed archive fraction and
    is not defined above a common luminosity threshold. The right
    panel repeats the calculation above a fixed luminosity threshold
    using only sufficiently sensitive spectra, but is not a
    completeness-corrected measurement of the intrinsic incidence.}
    \label{fig:nev_incidence_redshift}
\end{figure*}

To compare the [\ion{Ne}{5}] emitters with the underlying galaxy
population and provide a first estimate of their relative incidence,
we used the full unique-galaxy PRISMATIC parent catalog described in
Section~\ref{sec:sample_selection}. We restricted the denominator to sources at $z\geq3$ for which
[\ion{Ne}{5}]~$\lambda3426$ falls within the adopted PRISM wavelength
range and for which the catalog provides a finite, positive line-flux
uncertainty. This yields 8980 galaxies included in the
[\ion{Ne}{5}] incidence calculation.  The resulting fraction of [\ion{Ne}{5}]
emitters as a function of redshift is shown in the left panel of
Figure~\ref{fig:nev_incidence_redshift}. Error bars throughout this
section show 68\% Wilson score confidence intervals.

The observed [\ion{Ne}{5}] fraction displays no systematic trend with redshift and remains consistent with a constant value within uncertainties. At $3\leq z<5$, we identify 19 [\ion{Ne}{5}] emitters among
5873 galaxies in the parent sample ($0.324\%$), compared with
$6/3107=0.193\%$ at $z\geq5$. Although the value of the fraction is lower in the
higher-redshift subsample, the difference is not statistically
significant according to Fisher's exact test ($p=0.30$). We therefore
find evidence for neither an increase nor a decrease in the observed
incidence of [\ion{Ne}{5}] emission.

For comparison, the left panel of
Figure~\ref{fig:nev_incidence_redshift} also shows the increasing
N-emitter fraction reported by \citet{Morel2026} and the increasing
LRD fraction at $M_{\rm UV}=-20$ derived by \citet{Tanaka2025}.
Unlike these populations, the [\ion{Ne}{5}] emitters show no evidence
for a similarly pronounced increase toward high redshift. However,
these curves have different operational definitions: the N-emitter
fraction is measured relative to the available PRISM galaxy
population, whereas the LRD fraction is derived from the ratio of the
LRD and total-galaxy luminosity functions at fixed UV magnitude. The
comparison is therefore qualitative.

The raw [\ion{Ne}{5}] fraction is not measured above a common
line-flux or luminosity threshold, and its redshift dependence may be
affected by variations in spectral depth and line sensitivity. We
therefore repeated the calculation above using the median apparent
[\ion{Ne}{5}] luminosity of the emitter sample,
$\log_{10}(L_{\rm thr}/{\rm erg\,s^{-1}})=41.52$, restricting the
denominator to spectra capable of detecting a line at that luminosity.
The numerator includes the 13 [\ion{Ne}{5}] emitters with
$L_{\rm [Ne\,V]}\geq L_{\rm thr}$. The denominator includes all 25
[\ion{Ne}{5}] emitters, whose line luminosities are measured, together
with nonemitters whose catalog $4\sigma$ [\ion{Ne}{5}] luminosity
limit is at or below $L_{\rm thr}$. Because the number of emitters
above the threshold is small, we show this luminosity-thresholded
fraction in three broad redshift intervals in the right panel of
Figure~\ref{fig:nev_incidence_redshift}. The fractions are
$11/2100=0.524\%$ at $3\leq z<5$, $1/633=0.158\%$ at
$5\leq z<7$, and $1/161=0.621\%$ at $7\leq z<15$. Combining the
two higher-redshift bins gives $2/794=0.252\%$, compared with
$11/2100=0.524\%$ at $3\leq z<5$; this difference is also not
statistically significant ($p=0.53$). Restricting the denominator to
sufficiently sensitive spectra reduces the influence of nonuniform
line-detection limits, but does not provide a full completeness
correction.

To examine whether the absence of an increasing [\ion{Ne}{5}]
fraction could simply reflect a general decline in the detectability
of rest-optical emission lines, we also measured the fraction of
galaxies with robust [\ion{Ne}{3}]~$\lambda3869$ emission
(Figure~\ref{fig:neiii_incidence_diagnostics}). The [\ion{Ne}{3}]
line lies close in wavelength to [\ion{Ne}{5}]~$\lambda3426$ and is
therefore affected by broadly similar wavelength-dependent PRISM
coverage and sensitivity, although its intrinsic strength and
astrophysical origin differ. We define robust [\ion{Ne}{3}] emitters
as sources with
$F_{\rm [Ne\,III]}/\sigma_{F_{\rm [Ne\,III]}}>4$. Their fraction
increases from $783/5871=13.3\%$ at $3\leq z<5$ to
$762/3097=24.6\%$ at $z\geq5$. Thus, the absence of a corresponding
increase in the [\ion{Ne}{5}] fraction is not accompanied by a general
decline in the detection of this nearby rest-optical line.

We also calculated the fraction of robust [\ion{Ne}{3}] emitters that
are [\ion{Ne}{5}] emitters. This conditional fraction decreases from
$13/783=1.66\%$ at $3\leq z<5$ to $6/762=0.79\%$ at $z\geq5$,
although the difference is not statistically significant ($p=0.17$).
This calculation includes 19 of the 25 SHINE-[NeV] galaxies because
six do not satisfy the adopted robust [\ion{Ne}{3}] criterion. The
plotted quantity is a conditional population fraction rather than the
source-level [\ion{Ne}{5}]/[\ion{Ne}{3}] line ratio. These
[\ion{Ne}{3}] calculations are treated as selection diagnostics rather
than direct measurements of intrinsic evolution. The raw
[\ion{Ne}{3}] fraction is not defined above a common line-flux,
luminosity, or equivalent-width threshold and may therefore reflect
both astrophysical evolution and changes in survey targeting and
sensitivity. Moreover, because the [\ion{Ne}{3}]-selected population
itself evolves with redshift, the conditional [\ion{Ne}{5}] fraction
does not provide a fixed parent population against which to measure
AGN evolution.

Taken together, the raw and luminosity-thresholded calculations
provide no statistically significant evidence for either an increase
or a decrease in the observed incidence of [\ion{Ne}{5}] emission
with redshift. The data therefore do not show the pronounced rise
reported for N-emitters and LRDs, although differences in selection
and denominator definitions limit the comparison to a qualitative
one. The increasing robust [\ion{Ne}{3}] fraction further indicates
that the absence of an [\ion{Ne}{5}] increase is not simply associated
with a general decline in nearby rest-optical line detections.
Nevertheless, the current sample size and heterogeneous PRISM
sensitivity remain insufficient to determine whether the intrinsic
incidence of [\ion{Ne}{5}]-selected AGN evolves over this redshift
range.

\section{Discussion}
\label{sec:discussion}

The systematic detection of [\ion{Ne}{5}] in 25 galaxies at
$3.06<z<9.44$ provides a view of high-redshift black-hole
accretion complementary to the broad-line, narrow-line, X-ray, and
photometric AGN populations that have been the focus of many recent
JWST studies. The SHINE-[NeV] galaxies do not define a homogeneous
class of AGN: they span a broad range of spectroscopic, continuum,
X-ray, stellar-mass, and morphological properties. All show luminous
[\ion{Ne}{5}] emission, which requires photons with energies above
97.1~eV. Although [\ion{Ne}{5}]~$\lambda3426$ has a high critical
density for an optical forbidden line,
$n_{\rm crit}\simeq1.8\times10^7~{\rm cm^{-3}}$ at
$T_e=1.5\times10^4$~K \citep{HoldenTadhunter2023}, this density is
well below the $n_{\rm H}\gtrsim10^{9}$--$10^{10}~{\rm cm^{-3}}$
characteristic of the classical broad-line region, where forbidden
emission is collisionally suppressed
\citep[e.g.,][]{Peterson2006,Netzer2013}. Detectable
[\ion{Ne}{5}] therefore requires the hard ionizing radiation to
reach lower-density gas outside the classical broad-line region.
This emission may arise in a coronal-line region or the inner
narrow-line region.

In the following sections, we discuss the implications of this
population for AGN selection at high redshift, the origin and escape
of the hard ionizing radiation field, the connection between
high-ionization emission and the host-galaxy continuum, and the
incidence of luminous [\ion{Ne}{5}] emission over the redshift
range probed by the current NIRSpec archive.

\subsection{[\ion{Ne}{5}] as a Complementary View of the
High-redshift AGN Population}
\label{sec:discussion_selection}

The location of the SHINE-[NeV] galaxies on the S2--VO87 diagram
depends strongly on the adopted AGN boundary. Among sources that can
be securely classified relative to the \citet{Kewley2001}
maximal-starburst relation, 67\% lie above it. However, 32\% of
galaxies with constraining [\ion{Ne}{5}] nondetections and
high-significance measurements of all four S2--VO87 lines also lie
above this relation, showing that this region is not restricted to
[\ion{Ne}{5}] emitters. In contrast, only 14\% of the securely
classifiable SHINE-[NeV] galaxies lie above the more conservative
high-redshift boundary of \citet{Scholtz2025}. This low fraction
should not be interpreted as evidence against AGN activity in the
remaining sources. \citet{Scholtz2025} designed this boundary as a
conservative AGN selection and themselves use high-ionization lines,
including [\ion{Ne}{5}], to identify AGN that are not cleanly
separated by conventional optical diagnostics. Sources below the
Scholtz et al. boundary therefore cannot be classified as non-AGN on
the basis of S2--VO87 alone.

The SHINE-[NeV] sample also spans different broad-line
classifications. Five of the 25 galaxies have an adopted broad-line
classification. For the remaining sources, no broad-line
classification is adopted; some higher-resolution spectra do not
require a component associated with the broad-line region, while
other sources remain unassessable with the available data. The
coexistence of broad Balmer emission and [\ion{Ne}{5}] in several
galaxies shows that a visible broad-line region and luminous forbidden
high-ionization emission can occur simultaneously. Most SHINE-[NeV]
galaxies, however, do not have an adopted broad-line classification.
Broad-line and [\ion{Ne}{5}] selection therefore identify overlapping
but non-identical populations.

The diversity of the SHINE-[NeV] sample is also apparent in the
NIRCam imaging. Only one source is identified as an LRD in our
adopted literature compilation, while the images span compact,
extended, elongated, structured, and apparently multi-component
systems. We have not performed a quantitative morphological analysis,
and projected companions cannot be assumed to lie at the same
redshift. The available imaging nevertheless shows that luminous
[\ion{Ne}{5}] emission is not restricted to a single obvious
host-galaxy morphology. This contrasts with several JWST-selected
AGN populations initially identified in part through compact
morphologies or unusual continuum colors
\citep{Matthee2024,Greene2024,Kocevski2025}.

The complementarity of [\ion{Ne}{5}] selection should not be confused
with completeness. In the local Universe, the CLASS survey searched
nearly one million SDSS galaxy spectra without an initial AGN
preselection and detected at least one optical coronal line in only
$\sim0.03\%$ of the sample \citep{Reefe2022}. Among dwarf-galaxy
coronal-line emitters, $\sim80\%$ lacked optical narrow-line ratios
indicative of nuclear activity \citep{Reefe2022}. Thus, coronal-line
selection can identify systems missed by standard optical narrow-line
criteria, but detectable coronal-line emission itself represents a
rare subset of the galaxy population.

Photoionization calculations provide a natural
physical explanation for at least part of this rarity. In particular, \citet{McKaig2024} demonstrated that dust strongly suppresses optical coronal-line emission via two mechanisms: refractory element depletion onto grains and direct absorption of the ionizing continuum. The latter mechanism affects non-refractory species like neon by absorbing ionizing photons before they can photoionize the gas that powers [\ion{Ne}{5}]. Strong coronal-line emission therefore
depends not only on the presence of an AGN and a sufficiently hard
ionizing continuum, but also on the dust content, density,
ionization parameter, and geometrical distribution of the gas
\citep[e.g.,][]{Ferguson1997,Cann2018,McKaig2024}. Recent photoionization calculations further predict that [\ion{Ne}{5}] can become stronger toward lower metallicity as the declining dust-to-gas ratio increases fraction of ionizing photons available to ionize the gas (Matzko et al., submitted).

The SHINE-[NeV] sample should therefore not be interpreted as a
complete census of high-redshift AGN. Instead, it selects systems in
which a sufficiently hard radiation field is both produced and able
to illuminate gas under conditions favorable for [\ion{Ne}{5}]
emission. This distinction is important when comparing the incidence
of [\ion{Ne}{5}] with the incidence of AGN selected using broad
lines, X-rays, or other diagnostics. An AGN can be present without a
detectable coronal-line region, and differences in [\ion{Ne}{5}]
incidence may reflect differences in circumnuclear gas and radiative
transfer as well as differences in the underlying black-hole
population.

The presence of [\ion{Ne}{5}] alone does not uniquely identify the
ionizing source, since sufficiently extreme stellar populations,
accreting stellar remnants, and fast radiative shocks can in principle
produce photons at these energies
\citep[e.g.,][]{ShiraziBrinchmann2012,Izotov2012,Olivier2022,
Garofali2023}. However, the hard-ionization line ratios in the SHINE-[NeV] sample are
difficult to reproduce with ordinary stellar photoionization. The
photoionization calculations of \citet{Cleri2023b} predict negligible
[\ion{Ne}{5}] from normal stellar populations using BPASS, with
$\log$([\ion{Ne}{5}]/[\ion{Ne}{3}])$\sim-5$ across a broad range of
metallicity and ionization parameter, far below the ratios measured
for the SHINE-[NeV] galaxies. The He~II/H$\beta$ measurements provide
complementary evidence for a hard ionizing spectrum relative to the
stellar-photoionization reference of
\citet{ShiraziBrinchmann2012}, although the available upper limits
prevent this diagnostic from classifying every source.

The local metal-poor [\ion{Ne}{5}] emitters illustrate a second
limitation of the diagnostic. These galaxies generally have much
lower [\ion{Ne}{5}] luminosities and substantially lower
[\ion{Ne}{5}]/[\ion{Ne}{3}] ratios than the SHINE-[NeV] sample.
However, the physical origin of their hard ionizing radiation remains
uncertain in several cases, and contributions from shocks, compact
accreting sources, or unusually hard stellar populations have been
considered
\citep{Izotov2004,ThuanIzotov2005,Izotov2012,Izotov2021,Berg2021}.
Recent mid-infrared observations of the extremely metal-poor galaxy
I~Zw~18 provide a particularly useful reminder that Ne$^{4+}$ can be
produced in a star-forming galaxy without an established massive AGN,
with spatially extended [\ion{Ne}{5}] emission detected on galactic
scales \citep{ArroyoPolonio2025,Hunt2025}.  The mechanism responsible
for such hard ionizing radiation, however, is not necessarily ordinary
stellar photoionization.  More generally, the origin of the hard
radiation in local metal-poor [\ion{Ne}{5}] emitters remains uncertain
in several cases, with unusually hard stellar populations, X-ray
binaries or ultraluminous X-ray sources, shocks, and accreting
intermediate-mass black holes all having been considered
\citep{Izotov2004,ThuanIzotov2005,Izotov2012,Izotov2021,
Berg2021,Garofali2023,Mingozzi2025,Hunt2025}.
The local metal-poor sample therefore provides an important comparison
with systems lacking an established classical AGN, but it should not
be interpreted as a control sample in which [\ion{Ne}{5}] is known to
be powered solely by normal massive stars.

The absolute luminosity scale of the [\ion{Ne}{5}] emission provides an important
additional constraint.  Known stellar-remnant and shock-powered
nebulae producing optical [\ion{Ne}{5}] are orders of magnitude
fainter than the SHINE-[NeV] galaxies.  For example, the entire
shock-excited LMC supernova remnant N49 radiates only
$1.2\times10^{36}$~erg~s$^{-1}$ in
[\ion{Ne}{5}]~$\lambda3426$ \citep{Rakowski2007}, while the
[\ion{Ne}{5}]-emitting nebula surrounding the ultraluminous X-ray
source NGC~5408~X-1 has
$L_{\rm [Ne\,V]}\simeq10^{36}$~erg~s$^{-1}$
\citep{KaaretCorbel2009}.  Even the faintest SHINE-[NeV] source is
therefore more than four orders of magnitude more luminous in the
same transition.  The local metal-poor [\ion{Ne}{5}] emitters
considered here are likewise substantially less luminous than
SHINE-[NeV]. Their luminosities lie below the
regime that can be probed for most galaxies in the current
high-redshift NIRSpec archive.  Our survey is consequently only sensitive
to a luminous [\ion{Ne}{5}] population and cannot determine the
abundance of much fainter [\ion{Ne}{5}] emitters analogous to the
local metal-poor systems.

 Fast radiative shocks provide another possible origin for
high-ionization emission. Published shock+precursor calculations
can reproduce some of the observed high-ionization line ratios
\citep[e.g.,][]{Allen2008,Izotov2012,Flury2025}.  However, agreement
in line-ratios does not by itself demonstrate that shocks can
supply the observed absolute line luminosities.  As an
order-of-magnitude energetic check, we normalized the public
MAPPINGS~V shock+precursor grids of \citet{Flury2025} to the
kinetic-energy flux through the shock.  Even the most
[\ion{Ne}{5}]-efficient model across the full grid converts only
$\simeq2\times10^{-3}$ of the incoming shock power into
[\ion{Ne}{5}]~$\lambda3426$.  The median SHINE-[NeV] luminosity would therefore require
$\gtrsim1.6\times10^{44}$~erg~s$^{-1}$ in radiative shocks even
under this deliberately [\ion{Ne}{5}]-efficient model, rising to
$\sim5\times10^{44}$~erg~s$^{-1}$ for the $300$--$500$~km~s$^{-1}$
shock velocities commonly invoked for [\ion{Ne}{5}]-emitting
star-forming systems. At the luminous end of the SHINE-[NeV] sample, the
corresponding requirements rise to $\gtrsim3.5\times10^{45}$ and
$\sim10^{46}$~erg~s$^{-1}$, respectively.

For comparison, stellar-population models predict a mechanical
energy injection from supernovae and stellar winds of order
$6.5\times10^{41}$~erg~s$^{-1}$ per
$M_\odot$~yr$^{-1}$ of continuous star formation
\citep{Leitherer1999,StricklandHeckman2009}. Even if this mechanical
power were converted into radiative shocks with 100\% efficiency,
the shock powers required at the luminous end of SHINE-[NeV] would
correspond to star-formation rates of $\sim5\times10^3$--$1.5\times
10^4~M_\odot$~yr$^{-1}$. As an empirical benchmark, the exceptionally
powerful $\sim300~M_\odot$~yr$^{-1}$ compact starburst Makani has an
asymptotic stellar mechanical power of only
$\sim8\times10^{43}$~erg~s$^{-1}$ based on Starburst99
\citep{Rupke2023}, approximately 40--125 times below the shock power
required for the most luminous SHINE-[NeV] sources.
These estimates are conservative because the
[\ion{Ne}{5}] luminosities are not corrected for internal extinction.
Thus, while shocks or stellar-remnant accretion may contribute in
individual systems, reproducing the characteristic SHINE-[NeV]
luminosities without massive-black-hole accretion requires
increasingly extreme energetic conditions.

The selection based on [\ion{Ne}{5}] may nevertheless be particularly useful at the
low-mass end of the high-redshift galaxy population. Accretion disks
around lower-mass black holes are expected, under standard
thin-disk assumptions, to peak at higher energies and can produce
enhanced high-ionization emission
\citep{Cann2018,Cann2019}. GN~42437 provides an important example:
high-resolution JWST spectroscopy revealed strong
[\ion{Ne}{5}] in a compact galaxy with
$\log(M_\ast/M_\odot)\simeq7.9$, despite narrow Balmer lines and
rest-optical diagnostic ratios that would not by themselves have
identified the source as an AGN \citep{Chisholm2024}.

NeV-11 provides a particularly interesting comparison. This source,
previously studied as GS-NDG-9422 (JADES 1210\_13176), has an
inferred stellar mass of only
$\log(M_\ast/M_\odot)=7.39$. Detailed analyses of its unusually
nebular-dominated spectrum have previously favored a stellar origin
for its ionizing radiation. \citet{Cameron2024} showed that, when
considering the ensemble of UV and optical emission-line ratios, the
source is consistent with ionization by a young, metal-poor stellar
population and found no strong evidence that an AGN is required.
Some individual diagnostics are more ambiguous, however: the source
lies near or overlaps the star-forming/AGN transition in parts of
diagnostic space and had previously been included in a narrow-line
AGN selection based on its He~II/H$\beta$ ratio
\citep{Cameron2024,Scholtz2025}. \citet{Katz2025} subsequently
showed that its continuum can be dominated by nebular emission,
with hot, massive stars providing a plausible explanation, and
described the existing nebular diagnostics as consistent with a
source that is not AGN dominated, or at most composite.

The secure [\ion{Ne}{5}] detection reported here adds a qualitatively
different constraint on the ionizing spectrum of this galaxy.
NeV-11 has
$\log(L_{\rm [Ne\,V]}/{\rm erg\,s^{-1}})=40.81$, demonstrating the
presence of a substantial radiation field above the 97.1~eV
ionization threshold in a remarkably low-mass host. Its available
higher-resolution Balmer spectra do not require an adopted broad
component, and the source lies below the S2--VO87 AGN boundary.
The combination of luminous [\ion{Ne}{5}] emission with otherwise
largely star-forming-like nebular diagnostics illustrates why
high-ionization selection can provide information that is missed by
conventional diagnostic diagrams.

We caution, however, against associating the low stellar mass of the
host directly with a low black-hole mass. Indeed,
\citet{Trakhtenbrot2025} argued that the similarly luminous
[\ion{Ne}{5}] emission in GN~42437 implies a luminous AGN and, under
an Eddington-limited interpretation of an empirical
$L_{\rm [Ne\,V]}$--$L_{\rm bol}$ relation, a black-hole mass well
above the conventional intermediate-mass regime. Even strongly
super-Eddington accretion substantially weakens, but does not remove,
this constraint. The applicability of locally calibrated
[\ion{Ne}{5}]-to-bolometric luminosity relations to low-metallicity
galaxies at high redshift remains uncertain, and we do not use such a
scaling to estimate a black-hole mass for NeV-11 here. The present
data therefore establish that very luminous high-ionization emission
can occur in an exceptionally low-mass galaxy, but they do not
establish that NeV-11 hosts an intermediate-mass black hole or
distinguish among black-hole seeding scenarios.

\subsection{The Production and Escape of Hard Ionizing Radiation
in JWST AGNs}
\label{sec:discussion_hard_sed}

A growing number of JWST studies have found that many high-redshift
broad-line AGN and LRDs differ substantially from classical
low-redshift Type~1 AGN. In addition to their unusual continuum
properties, many show weak X-ray emission and weak or absent
high-ionization UV lines
\citep{Maiolino2025,Zucchi2026,Lambrides2026}. These observations
have motivated several related explanations. One possibility is that
high accretion rates produce an intrinsically softer ionizing
spectral energy distribution, with a weak corona and a reduced
relative number of EUV and X-ray photons
\citep{PacucciNarayan2024,Lambrides2026}. Alternatively, dense gas
with a large covering factor may absorb or reprocess the ionizing
continuum before it reaches the line-emitting gas
\citep{Rusakov2026}. A third possibility is that the radiation field
is strongly anisotropic, so that different regions surrounding the
black hole are exposed to very different ionizing spectra
\citep{Madau2026}. These possibilities are not mutually exclusive.

The [\ion{Ne}{5}] detections provide a useful additional constraint
because they probe a different physical component from many of the
high-ionization lines reported thus far in broad-line AGN studies. C~IV and
N~V are permitted resonance transitions, while He~II is a
recombination line. They can therefore remain strong in gas at
densities at which forbidden transitions are collisionally
suppressed. In contrast, [\ion{Ne}{5}]~$\lambda3426$ requires both
photons above the 97.1 eV ionization threshold and gas with densities
low enough for the forbidden transition to be emitted efficiently.
A detection of [\ion{Ne}{5}] thus demonstrates that very hard
radiation has reached a lower-density coronal or narrow-line gas
phase; it does not simply demonstrate that hard photons are present
within the broad-line region
\citep{Gilli2010,Cann2018,Chisholm2024}.

This distinction is important when considering recent claims for
weak high-ionization emission in JWST broad-line AGN.
\citet{Zucchi2026} find a marked deficit of several high-ionization
emission lines in their high-redshift Type~1 AGN sample and show that
standard sub-Eddington SEDs have difficulty reproducing the observed
line spectrum. \citet{Lambrides2026} similarly find no X-ray
detections and weak high-ionization UV emission in their selected
sample of broad-line AGN and argue that accretion near or above the
Eddington limit can naturally soften the ionizing SED. These results
provide strong evidence that such conditions occur in at least some
JWST AGN. They do not, however, require that all high-redshift
broad-line AGN lack photons capable of producing [\ion{Ne}{5}].

Indeed, five SHINE-[NeV] galaxies have broad Balmer emission while
also showing secure [\ion{Ne}{5}] detections. These galaxies demonstrate
directly that broad-line emission and the escape of radiation above
97 eV to a lower-density gas phase can coexist. The comparison with
the larger BLAGN and LRD samples is also instructive. Although
[\ion{Ne}{5}] is generally not detected in those samples, most of
the available limits are not sufficiently deep to exclude line
ratios typical of SHINE-[NeV]. Only 6 of 84 BLAGN [\ion{Ne}{5}] upper limits and
4 of 103 LRD upper limits fall below the lower quartile of the
SHINE-[NeV] Ne53 distribution. The current data therefore do not
establish that BLAGN or LRDs as populations have systematically
weaker [\ion{Ne}{5}] relative to [\ion{Ne}{3}].

This sensitivity issue is important in interpreting the emerging
high-ionization-line literature. For example, \citet{Tang2025HighIonization}
identified narrow high-ionization emission in only a small fraction
of LRDs, but also emphasized that many individual spectra do not
provide stringent constraints on intrinsically weak lines. The
presence of narrow high-ionization emission in some LRDs led Tang et al. to suggest that dense neutral gas may not cover the central
source uniformly. Our results extend this argument specifically to
the much higher ionization potential traced by [\ion{Ne}{5}].
The SHINE-[NeV] detections are difficult to reconcile with a
universal geometry in which an optically thick envelope completely
encloses the ionizing source and prevents EUV photons from reaching
lower-density gas in all directions. They are fully consistent,
however, with clumpy or partially covering gas in which hard photons
escape along some sightlines.

A similar conclusion follows from models in which super-Eddington
accretion produces a geometrically thick accretion flow.
Super-Eddington accretion need not imply that the ionizing spectrum
is soft in every direction. Thick-disk models can produce strong
shadowing in equatorial directions while collimating radiation
through a polar funnel \citep{Madau2026}. In such a configuration,
the broad-line region could receive a softer or filtered radiation
field while gas along less obscured directions is exposed to a much
harder continuum. Weak broad C~IV or He~II emission and detectable
narrow [\ion{Ne}{5}] are therefore not necessarily contradictory.
The SHINE-[NeV] sample does not rule out super-Eddington accretion;
rather, it argues against treating a globally soft, isotropic
ionizing spectrum as a universal description of the high-redshift
AGN population.

The X-ray properties reinforce this picture of diversity. Six of
the 25 SHINE-[NeV] galaxies have secure Chandra counterparts. Their
observed-equivalent rest-frame $2$--$10$ keV luminosities span a
wide range relative to their [\ion{Ne}{5}] luminosities, and none of the secure
detections falls clearly below the empirical
$L_{\rm X}/L_{\rm [Ne\,V]}=15$ reference associated with the most
heavily obscured systems in the local calibration of
\citet{Gilli2010}. Thus, at least some luminous high-redshift
[\ion{Ne}{5}] emitters are not extremely X-ray faint relative to
their coronal-line emission. This is consistent with the existence
of X-ray-bright exceptions among other JWST AGN samples and with the
luminous [\ion{Ne}{5}]-detected systems reported by
\citet{Valentino2026}.

The absence of an X-ray counterpart for the remaining SHINE-[NeV]
galaxies cannot be interpreted in the same way. The available
constraints are heterogeneous, and for 18 galaxies the adopted
values are local catalog source-detection thresholds rather than
formal source-flux upper limits. None reaches below the local
$L_{\rm X}/L_{\rm [Ne\,V]}=15$ reference. We therefore cannot
determine from the current data what fraction of the SHINE-[NeV]
population is intrinsically X-ray weak or Compton thick. This
distinction is especially relevant in light of recent stacking
results. \citet{Maiolino2025} find widespread X-ray weakness among
both broad- and narrow-line JWST AGN, while \citet{Comastri2026}
report a hard-band stacked signal for Type~2 sources consistent with
heavy obscuration but no corresponding detection for the Type~1
sample. The mechanisms responsible for X-ray faintness may therefore
not be identical in every spectroscopic class.

These observations suggest that the emerging JWST
AGN population cannot be described by a single ionizing SED or
circumnuclear geometry. Some sources may be characterized by
intrinsically soft ionizing continua associated with high accretion
rates, some may be obscured by high-column-density gas, and some may
permit hard radiation to escape anisotropically through a clumpy or
funnel-like structure. The SHINE-[NeV] sample selects systems in which, regardless of the
underlying accretion state or geometry, hard EUV radiation has
reached lower-density gas capable of producing [\ion{Ne}{5}].
Because the sample is selected on the presence of [\ion{Ne}{5}], it
cannot establish how common this state is among all AGN. It does
establish that complete suppression of escaping hard ionizing
radiation is not a universal property of high-redshift AGN,
including broad-line systems.

\subsection{Strong High-ionization Emission and the Continuum
State of the Host}
\label{sec:discussion_continuum}

A second unexpected result of this work is the connection between
the relative strength of [\ion{Ne}{5}] and the continuum properties
of the host. The strong
[\ion{Ne}{5}]/[\ion{Ne}{3}] sources do not differ significantly in
$M_{\rm UV}$ from the sensitivity-qualified lower-ratio comparison
sample. They do, however, have redder UV slopes and stronger measured
Balmer breaks. In the redshift-matched comparison, the differences
are $\Delta\beta=0.63$ and
$\Delta\log_{10}(B_{\rm Balmer})=0.183$ dex, corresponding to a
Balmer break approximately 1.5 times stronger on average. The
association is therefore with continuum shape rather than simply
with UV luminosity.

Several aspects of the analysis argue against this trend being
produced by a simple [\ion{Ne}{5}] detectability bias. First,
[\ion{Ne}{5}]~$\lambda3426$ and [\ion{Ne}{3}]~$\lambda3869$ are
separated by only $\simeq440$~\AA\ in the rest frame. Their ratio is
therefore relatively insensitive to large-scale continuum shape,
differential attenuation, and changes in instrumental
sensitivity compared with ratios involving lines widely separated
in wavelength. Second, the lower-ratio comparison contains only
galaxies for which the source-specific [\ion{Ne}{5}] uncertainty is
small enough that a line at the adopted ratio threshold would have
been detected. The comparison is therefore not simply between line
detections and arbitrarily shallower nondetections. Finally, we find
no evidence that the continuum signal-to-noise relevant to the UV
slope or Balmer-break measurements systematically tracks the
[\ion{Ne}{5}]/[\ion{Ne}{3}] strength. These considerations make it unlikely that the
observed continuum differences are produced solely by the way the
[\ion{Ne}{5}] sample was detected.

The physical interpretation of the  association between the relative strength of [\ion{Ne}{5}] and the continuum properties
of the host is not straightforward. A Balmer break
is naturally produced by stellar populations after the contribution
of the youngest massive stars declines, and the combination of strong
emission lines with a measurable break can arise from composite
populations containing both a young line-emitting component and an
older stellar population. 

Such Balmer break objects have, e.g., been observed by \citet{Kuruvanthodi2024,Witten2025} and also found among some high-redshift nitrogen emitters \citep{Morel2026}. 

An interesting complementary result comes from \citet{Valentino2026},
who began with a sample of massive quenched galaxies and subsequently
searched their spectra for [\ion{Ne}{5}]. They found the strongest
[\ion{Ne}{5}] emission preferentially in the youngest
post-starburst systems, with weak or absent [\ion{Ne}{5}] in the
oldest quenched galaxies. Although their galaxies occupy a very
different stellar-mass and evolutionary regime from much of
SHINE-[NeV], both studies suggest that strong high-ionization
activity may preferentially occur during a transitional phase rather
than exclusively in either the youngest star-forming systems or the
oldest passive populations.

This interpretation is also consistent with previous evidence at
lower redshift that [\ion{Ne}{5}] activity can be associated with
galaxies in or shortly following a shutdown of star formation
\citep{Vergani2018,Barchiesi2024,Barchiesi2025}. In this picture,
the redder UV slopes and stronger Balmer breaks in the strong-Ne53
population could reflect an increasing contribution from an evolved
stellar population while substantial black-hole accretion remains
active. Such an interpretation is appealing, but the Balmer-break
measurement used here does not provide a unique age or star-formation
history, and a redder UV slope does not uniquely imply increased dust
attenuation. Stellar population modeling will therefore be required
to determine whether the strong-Ne53 galaxies have systematically
different recent star-formation histories.

An alternative interpretation has become particularly relevant with
the discovery of LRDs with very strong Balmer discontinuities. In
objects such as the ``Cliff,'' \citep{deGraaff2025} the observed break is too extreme to
be reproduced easily by conventional stellar populations, and dense
hydrogen gas surrounding an accreting black hole has been proposed
as the origin of both the break and other unusual spectral features
\citep{deGraaff2025}. Related dense-cocoon and ``black-hole star''
models can substantially reprocess the intrinsic AGN continuum and,
in some cases, alter the apparent widths and strengths of the Balmer
lines \citep{Rusakov2026}. The Balmer breaks in SHINE-[NeV] are
generally more moderate than the extreme cases that motivated these
models, and most of the SHINE-[NeV] sample is not selected as LRDs.

We therefore do not regard the continuum trend as evidence that the
strong-Ne53 galaxies are related to black-hole stars or contain dense cocoons.

There is nevertheless an interesting connection between the two
phenomena. If dense nuclear gas contributes significantly to the
Balmer break in some SHINE-[NeV] galaxies, the simultaneous presence
of strong [\ion{Ne}{5}] requires that this gas cannot prevent hard
ionizing photons from reaching lower-density material in every
direction. A nearly closed, optically thick envelope would naturally
suppress the formation of an external coronal-line region. A clumpy
or anisotropic envelope, in contrast, could both reprocess a large
fraction of the continuum and permit hard radiation to escape through
lower-column-density channels. The association between stronger
Balmer breaks and stronger Ne53 may therefore ultimately provide a
useful constraint on the geometry of the gas surrounding rapidly
growing black holes, but the current continuum measurements alone
cannot distinguish this possibility from a stellar-population
origin.

The diversity of the NIRCam morphologies provides an additional
reason for caution in assigning a single interpretation. Strong
[\ion{Ne}{5}] is present in both compact and visibly structured
systems rather than being restricted to the compact systems characteristic of the LRD
population. The simplest interpretation of the present data is
therefore that the continuum correlations identify a connection
between the conditions that favor strong high-ionization emission
and the state of the surrounding galaxy or nuclear environment,
without yet determining whether the relevant continuum is dominated
by stellar evolution, dust, dense nuclear gas, or some combination
of these components.

\subsection{Incidence, Cosmic Evolution, and Implications for
Early Black-hole Growth}
\label{sec:discussion_evolution}

The large parent PRISM sample also allows us to ask whether the
observed incidence of luminous [\ion{Ne}{5}] emission changes over
the redshift interval probed by this work. For both the raw and
luminosity-thresholded measurements presented in this paper, we find
no statistically significant difference between the lower- and
higher-redshift bins. The same is true when considering
[\ion{Ne}{5}] emission among galaxies with robust [\ion{Ne}{3}] detections.
Thus, within the range $z>3$ probed here, there is no evidence for a
strong rise or fall in the observed fraction of galaxies exhibiting
luminous [\ion{Ne}{5}] emission.

This result should not be interpreted as evidence that the intrinsic
AGN population does not evolve. The parent NIRSpec archive combines
programs with different target selections, exposure times, and
sensitivities, and [\ion{Ne}{5}] visibility itself depends on the
ionizing SED and physical state of the coronal-line gas. In addition,
our survey begins at $z\simeq3$ and therefore does not constrain
possible evolution between the local Universe, cosmic noon, and the
redshift range studied here. The appropriate conclusion is more
limited: over $z>3$, the current data provide no evidence for a
dramatic change in the observed incidence of luminous
[\ion{Ne}{5}] emitters with redshift.

The behavior of [\ion{Ne}{3}] provides an interesting contrast.
The raw robust-[\ion{Ne}{3}] detection fraction increases from
approximately 13\% at $3\leq z<5$ to approximately 25\% at
$z\geq5$. Although the raw [\ion{Ne}{3}] detection fraction is not measured
above a common intrinsic luminosity or equivalent-width threshold,
its increase toward higher redshift provides an important empirical
check on redshift-dependent sensitivity. In particular, the absence
of a corresponding increase in the [\ion{Ne}{5}] detection fraction
cannot be explained simply by a general loss of sensitivity to
rest-frame optical emission lines at higher redshift, since the
nearby [\ion{Ne}{3}]~$\lambda3869$ line is detected in an increasing
fraction of the parent sample. Instead, an increasing fraction
of the observed galaxies enters a strong [\ion{Ne}{3}]-emitting
population without a proportional increase in the fraction showing
luminous [\ion{Ne}{5}].

One possible interpretation is that increasingly bursty or
high-excitation phases become more common toward high redshift,
increasing the fraction of galaxies with strong intermediate-ionization
nebular emission. Related rapid evolution has been reported for other
unusual emission-line populations. For example,

the fraction of extreme emission lines galaxies increases with redshift \citep{Boyett2024}, and
\citet{Morel2026} find that the fraction of nitrogen emitters rises
strongly with redshift and discuss an increasing prevalence of young,
compact star-forming systems as one possible origin. The
[\ion{Ne}{3}] trend in our sample does not establish the same
physical mechanism. It does, however, suggest that whatever controls
the visibility of luminous [\ion{Ne}{5}] is not simply the increasing
frequency of strong emission-line phases in the galaxy population.
The production and escape of photons above 97 eV appears to impose
an additional requirement.

This point may also be relevant when comparing SHINE-[NeV] with the
LRD population. The abundance of LRDs evolves strongly over the
redshift interval probed by JWST, with a broad maximum around the
epoch $z\sim5$--6 under commonly adopted selections
\citep{Matthee2024,Kocevski2025,Tanaka2025}. We do not observe a
similarly dramatic evolution in luminous [\ion{Ne}{5}] incidence.
This difference does not imply that LRDs and [\ion{Ne}{5}] emitters
represent unrelated black-hole populations. Rather, the observability
of the two phenomena may depend on different combinations of
accretion rate, gas covering fraction, orientation, host-galaxy
contrast, and dust. An LRD-like continuum and a visible
[\ion{Ne}{5}]-emitting region need not occur at the same time or
along the same line of sight.

The detection of [\ion{Ne}{5}] in very low-mass galaxies adds a
further dimension to this picture. Traditional AGN searches become
increasingly difficult in low-mass hosts because star formation can
dominate the integrated optical spectrum, the broad-line region may
be weak or difficult to identify, and X-ray emission from stellar
populations becomes increasingly important
\citep{Cann2018,ReinesComastri2016}. High-ionization forbidden lines
therefore provide an attractive complementary method for identifying
accretion in this regime. GN~42437 demonstrated this potential at
$z=5.59$ \citep{Chisholm2024}, and the low inferred stellar mass of
NeV-11 extends the population of high-redshift [\ion{Ne}{5}]
emitters into an even lower host-mass regime.

These objects are potentially important for understanding early
black-hole growth, but their implications for black-hole seeding
should remain limited at present. Stellar mass is not a black-hole
mass measurement, and both virial and line-ratio estimates of
black-hole mass carry large model dependencies in the physical
regimes now being uncovered by JWST. Moreover, the current
[\ion{Ne}{5}] luminosity limits do not allow us to determine the
abundance of much fainter accreting black holes in low-mass galaxies.
The most robust conclusion is that [\ion{Ne}{5}] provides a means of
finding accretion activity in host galaxies where many conventional
AGN diagnostics fail. Uniform deeper spectroscopy, together with
higher-resolution line profiles and independent constraints on the
stellar and black-hole masses, will be required before this population
can be used to distinguish among black-hole seed models.

The results from this work suggest that high-redshift
AGN should not be viewed as a single spectroscopic or photometric
class. The SHINE-[NeV] galaxies span broad-line and non-broad-line
classifications, X-ray detections and nondetections, compact and
structured morphologies, and a wide range of stellar masses and
continuum properties. Their common feature is that hard ionizing
radiation is able to reach gas capable of producing luminous
[\ion{Ne}{5}]. The diversity within this population, together with
the weak constraints on [\ion{Ne}{5}] in many independently selected
BLAGN and LRDs, suggests that the emerging differences among JWST
AGN may reflect not only differences in black-hole accretion rate,
but also differences in how the ionizing radiation is generated,
filtered, and allowed to escape. A complete census of black-hole
growth in the early Universe will therefore require combining
[\ion{Ne}{5}] and other coronal lines with broad-line, narrow-line,
continuum, and multiwavelength AGN diagnostics rather than relying
on any single selection method.
\section{Conclusions}
\label{sec:conclusions}

We have presented the first systematic SHINE search for
[\ion{Ne}{5}]~$\lambda3426$ emission in public
\textit{JWST}/NIRSpec PRISM spectroscopy at $z>3$. 
Starting from our new PRISMATIC emission line flux catalog of more than 9000 PRISM spectra from the DJA archive \citep[see][Sawarkar et al., in prep.]{Heintz2025},  
we identify a SHINE-[NeV] sample of
25 visually vetted and quantitatively validated [\ion{Ne}{5}]
emitters and use their emission-line, continuum, broad-line,
X-ray, stellar-mass, and imaging properties to determine what
population is recovered when very-high-ionization emission defines the selection.

Our main conclusions are as follows:

\begin{enumerate}

\item The SHINE-[NeV] sample spans $z=3.059$--$9.444$ and more than
two orders of magnitude in apparent [\ion{Ne}{5}] luminosity,
$40.70 \leq \log(L_{\rm [Ne\,V]}/{\rm erg\,s^{-1}}) \leq 42.86$.
The galaxies also span a wide range of UV luminosities, UV slopes,
Balmer-break strengths, stellar masses, and projected morphologies.
The NIRCam imaging includes compact sources as well as extended,
elongated, and structured systems, demonstrating that luminous
[\ion{Ne}{5}] emission is not restricted to any specific morphological class.

\item The SHINE-[NeV] sample's [\ion{Ne}{5}] luminosities overlap with those of the
local CLASS [\ion{Ne}{5}] emitters: 20/25 SHINE-[NeV] galaxies lie
within the CLASS luminosity range and five are more luminous than
the most luminous CLASS source. Among the CLASS objects with usable
optical classifications, 57/60 are identified as AGNs from their BPT
ratios. In contrast, the 15 local metal-poor [\ion{Ne}{5}] emitters
span only
$36.52 \leq \log(L_{\rm [Ne\,V]}/{\rm erg\,s^{-1}}) \leq 39.73$,
and all are less luminous than the SHINE-[NeV] galaxies. The
luminosity overlap with the predominantly AGN-classified CLASS sample,
together with the separation from the local metal-poor population,
supports an AGN origin for the luminous [\ion{Ne}{5}] emission in
SHINE-[NeV]. However, the current PRISM data are not sensitive to
[\ion{Ne}{5}] luminosities characteristic of the local metal-poor
population, so a fainter high-redshift [\ion{Ne}{5}] population could
remain undetected.

\item {[}\ion{Ne}{5}] selection overlaps only partly with other AGN
diagnostics. Among SHINE-[NeV] galaxies securely classifiable in
S2--VO87, 67\% lie above the \citet{Kewley2001} maximal-starburst
relation, whereas only 14\% lie above the more conservative
high-redshift boundary of \citet{Scholtz2025}. However, 32\% of
galaxies with constraining [\ion{Ne}{5}] nondetections also lie above
the Kewley relation, showing that this region is not specific to
[\ion{Ne}{5}] emitters. Five of the 25 SHINE-[NeV] galaxies have
adopted broad-line classifications; for the remainder, broad emission
is either not required in the assessable spectra or cannot be
determined from the available data. Broad-line, narrow-line, and
[\ion{Ne}{5}] selection therefore provide complementary views of the
high-redshift AGN population. Stellar masses are available for 20
sources and span approximately
$\log(M_\ast/M_\odot)\sim7.4$--11, including one exceptionally
low-mass source at $\log(M_\ast/M_\odot)\sim7.4$. Luminous
very-high-ionization emission can therefore occur in low-mass hosts,
although low stellar mass alone does not imply the presence of an
intermediate-mass black hole.

\item The current data do not establish a population-wide
[\ion{Ne}{5}] deficit in independently selected broad-line AGNs or
LRDs. When the comparison is restricted to galaxies with robust
[\ion{Ne}{3}] detections and constraining [\ion{Ne}{5}]
upper limits, only 6 of 84 broad-line AGN limits and 4 of 103 LRD
limits reach below the lower quartile of the SHINE-[NeV] Ne53
distribution. 
Most available spectra are not deep enough to exclude

[\ion{Ne}{5}]/[\ion{Ne}{3}] ratios found within the SHINE-[NeV]
sample. In addition, broad Balmer emission and narrow
[\ion{Ne}{5}] coexist in five SHINE-[NeV] galaxies. These results
demonstrate that the escape of photons capable of producing
[\ion{Ne}{5}] is not universally suppressed in high-redshift
broad-line systems.

\item X-ray weakness is likewise not universal among luminous
high-redshift [\ion{Ne}{5}] emitters. Six of the 25 SHINE-[NeV]
galaxies have secure \textit{Chandra} counterparts, spanning
$L_{2-10}/L_{\rm [Ne\,V]}=15.9$--$979.5$, with none below the
empirical $L_{\rm X}/L_{\rm [Ne\,V]}=15$ reference associated with
the most heavily obscured local systems. The detected galaxies
therefore do not show uniformly low X-ray/[Ne V] ratios. For most
non-detections, however, the available constraints are catalog
detection thresholds rather than formal source-flux upper limits
and are not deep enough to determine the intrinsic X-ray-weak
fraction. The combined X-ray and high-ionization results argue
against a universal picture in which high-redshift AGNs lack both
observable X-rays and escaping very hard ionizing radiation.

\item The relative strength of [\ion{Ne}{5}] is associated with
continuum shape. Compared with a sensitivity-qualified,
redshift-matched lower-ratio population, the strong
[\ion{Ne}{5}]/[\ion{Ne}{3}] galaxies show no significant difference
in $M_{\rm UV}$ but have systematically redder UV slopes
($\Delta\beta=0.63$) and stronger Balmer breaks
($\Delta\log_{10}B_{\rm Balmer}=0.183$ dex). The comparison is
constructed so that galaxies in the lower-ratio sample have spectra
capable of excluding the adopted strong-ratio threshold, and the
continuum trends are not associated with systematically higher
continuum signal-to-noise. These differences therefore do not appear
to arise from a simple [\ion{Ne}{5}] detection bias. Their physical
origin remains uncertain.  One possibility is that strong [\ion{Ne}{5}] emission preferentially occurs
during or shortly after a decline in star formation, potentially in a
phase associated with AGN feedback. However, the present continuum
measurements do not establish that these galaxies are recently
quenched or that AGN feedback caused a decline in their star
formation. Differences in stellar populations, dust, dense
circumnuclear gas, or continuum reprocessing could also contribute.

\item We find no statistically significant evidence for evolution in
the observed incidence of luminous [\ion{Ne}{5}] emission over
$z>3$. The raw [\ion{Ne}{5}] fraction is $0.324\%$ at
$3\leq z<5$ and $0.193\%$ at $z\geq5$, and the
luminosity-thresholded calculation likewise shows no significant
difference between the lower- and higher-redshift samples. In
contrast, the robust [\ion{Ne}{3}] detection fraction increases from
approximately 13\% to 25\% over the same two redshift intervals.
Because [\ion{Ne}{3}] lies close in wavelength to [\ion{Ne}{5}],
this increase argues against a general loss of rest-optical
emission-line sensitivity as the origin of the different trends.
The current heterogeneous archive and small [\ion{Ne}{5}] sample do
not provide a completeness-corrected measurement of intrinsic
evolution, but they show no evidence for the pronounced rise toward
high redshift reported for some other JWST-selected populations.

\end{enumerate}

The SHINE-[NeV] sample therefore does not define a single new
spectroscopic or morphological class of high-redshift AGN. Rather,
it isolates systems in which photons above the 97.1~eV ionization
threshold reach gas capable of producing luminous forbidden
[\ion{Ne}{5}] emission. These systems span broad-line and
non-broad-line classifications, X-ray states, host morphologies,
stellar masses, and continuum properties. At the same time,
[\ion{Ne}{5}] selection is necessarily incomplete: an accreting
black hole can lack detectable coronal-line emission if the
ionizing continuum, dust content, gas density, or covering geometry
is unfavorable. A complete census of early black-hole growth will
therefore require combining high-ionization lines with broad-line,
narrow-line, continuum, and multiwavelength diagnostics.
\appendix

\section{Detailed Fitting of the
\texorpdfstring{[\ion{Ne}{5}]}{[Ne V]} Emission}
\label{app:badass}

As described in Section~\ref{sec:badass_selection}, we performed a
detailed reanalysis of the 28 visually validated
[\ion{Ne}{5}] candidates using the Bayesian AGN Decomposition
Analysis for SDSS Spectra code (BADASS; \citealt{Sexton2021}).
BADASS has been used in our previous systematic searches for optical
coronal-line emission \citep{Reefe2022,Doan2025}. The fitting
procedure first determines a maximum-likelihood solution and then
uses Markov Chain Monte Carlo sampling to determine the parameter
uncertainties and covariances. In the general BADASS framework,
emission lines are modeled simultaneously with the underlying
continuum, and individual coronal lines are represented by Gaussian
profiles with their line parameters allowed to vary
\citep{Sexton2021,Reefe2022,Doan2025}.

The weak [\ion{Ne}{5}] emission in the PRISM spectra required
particular care in defining the local continuum. The inferred properties of a weak
line can depend strongly on the wavelength interval used to
constrain the continuum and on the adopted continuum model
\citep{Reefe2022,Doan2025}. We found this effect to be important for
the present sample. Fits over broader wavelength intervals were more
sensitive to large-scale continuum curvature and to spectral
structure unrelated to [\ion{Ne}{5}], whereas very restricted
intervals provide less information with which to constrain the
continuum. For several of the weakest candidates, changes in the
fitting interval or continuum prescription produced appreciable
changes in the inferred [\ion{Ne}{5}] flux uncertainty and hence in
the formal line significance.

We therefore adopted a local fitting procedure centered on the
[\ion{Ne}{5}]~$\lambda\lambda3346,3426$ doublet. Both members of the
doublet were included simultaneously in the model, and the continuum
was constrained using the same local fitting prescription for every
visually retained candidate. This approach follows a general
strategy in which restricted
spectral regions are used to obtain a reliable estimate of the local
continuum rather than allowing distant portions of the spectrum to
drive the fit \citep{Reefe2022,Doan2025}. The final fitting regions for each
source were constrained and masked in order to disregard erroneous emission features that could potentially bias the underlying continuum, without needing to unnecessarily complicate
the model with an extra set of Gaussian fit parameters.

The line measurements used throughout this paper are the integrated
fluxes and uncertainties from these final local fits. In particular,
we define the [\ion{Ne}{5}] significance as
\begin{equation}
{\rm S/N}_{\rm int}
=
\frac{F_{3426}}{\sigma_{F_{3426}}},
\end{equation}
where $F_{3426}$ is the integrated
[\ion{Ne}{5}]~$\lambda3426$ flux and
$\sigma_{F_{3426}}$ is its integrated-flux uncertainty. This quantity
is distinct from a peak-amplitude signal-to-noise ratio. A source was
included in the SHINE-[NeV] sample only when this integrated
significance exceeded three, as defined in
Equation~\ref{eq:nev_detection}.

The dependence of the formal significance on the local continuum
treatment is particularly relevant near the adopted threshold. The
line flux itself was generally consistent between the automated
catalog measurements and the detailed BADASS fits, but the uncertainty
on the integrated flux could change as the continuum interval and
continuum model were changed. Consequently, a candidate with a
visually apparent feature could move above or below a nominal
$3\sigma$ threshold under different reasonable fitting choices. We
therefore do not interpret small differences in formal significance
between fitting prescriptions as evidence that the underlying
spectral feature has appeared or disappeared. Instead, the
$F_{3426}/\sigma_{F_{3426}}>3$ requirement defines a uniform
operational threshold based on the single fitting prescription
adopted for the SHINE-[NeV] sample.

Twenty-five of the 28 visually retained candidates satisfy this
criterion and comprise the SHINE-[NeV] sample. The spectra and final
[\ion{Ne}{5}] fits for these sources are shown in
Figure~\ref{fig:nev_representative_spectra} and
Appendix~\ref{app:nev_atlas}. These fits are also the source of the
[\ion{Ne}{5}] fluxes and apparent luminosities reported in
Table~\ref{tab:nev_sample}. Other emission lines used in this work
were adopted from the DJA emission-line catalog and were not refitted
with BADASS.

%

\subsection{Spectral and imaging atlas of the remaining
\texorpdfstring{\ion{Ne}{5}}{[Ne V]} emitters}
\label{app:nev_atlas}

\begin{figure*}
    \centering
    \includegraphics[width=0.99\textwidth]
    {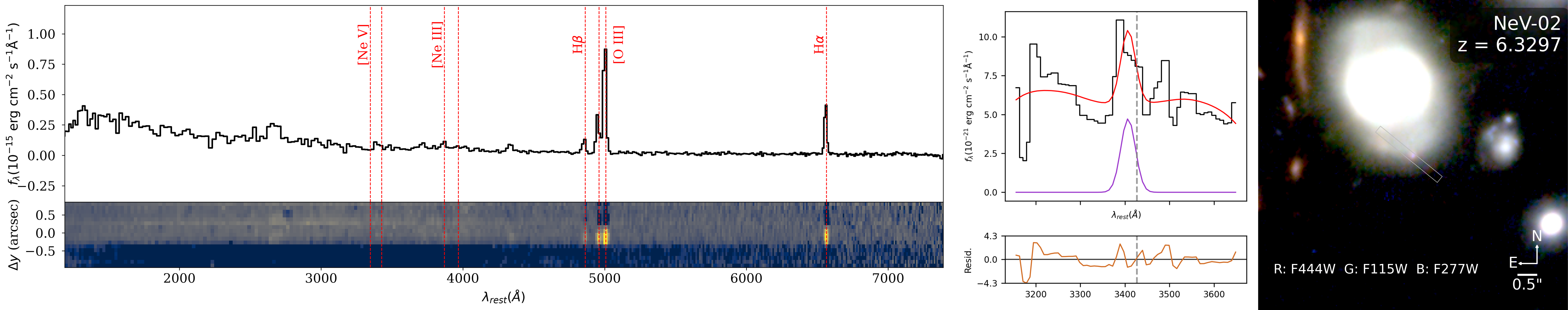}\par\smallskip
    \includegraphics[width=0.99\textwidth]
    {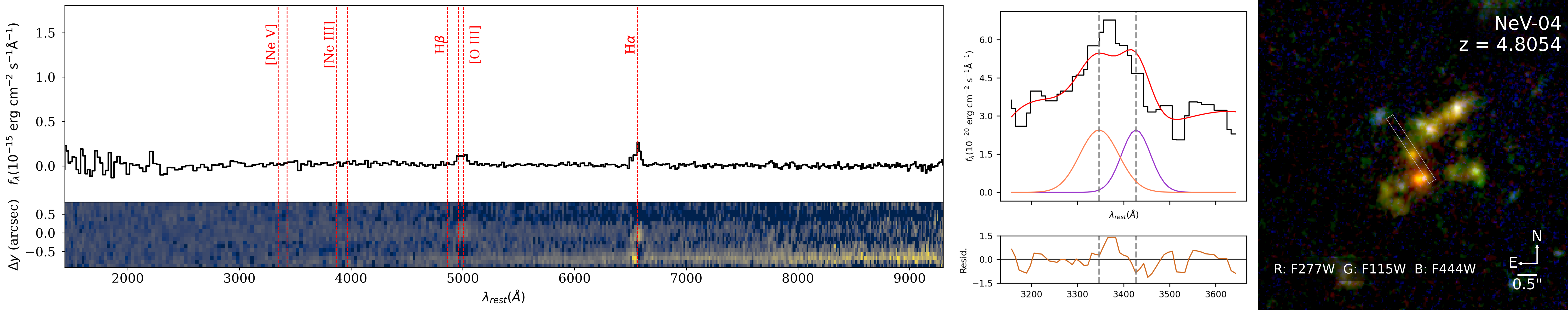}\par\smallskip
    \includegraphics[width=0.99\textwidth]
    {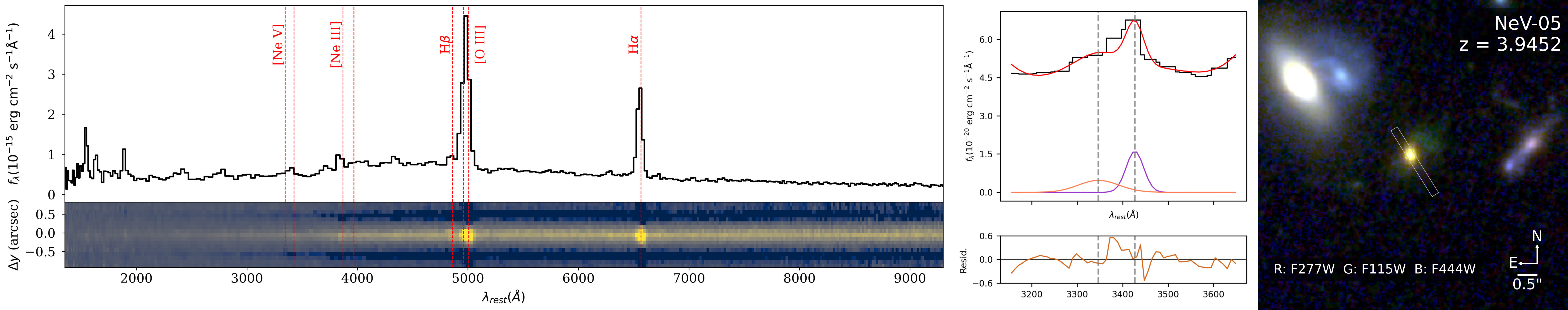}\par\smallskip
    \includegraphics[width=0.99\textwidth]
    {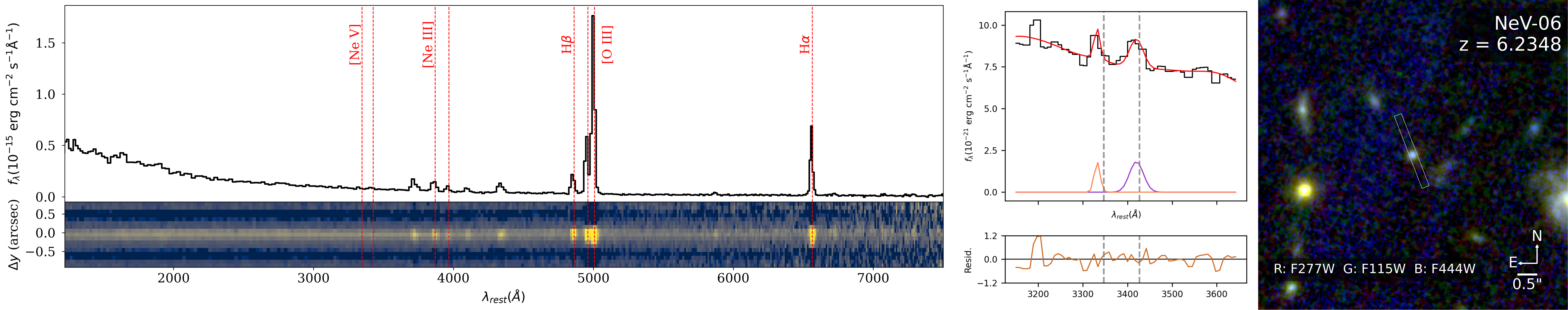}

   \caption{Spectral and imaging atlas for NeV-02, NeV-04, NeV-05,
and NeV-06, shown from top to bottom. In each row, the left panel
shows the full rest-frame PRISM one-dimensional spectrum in black
above the associated two-dimensional spectrum, with red vertical
markers identifying selected emission lines. The upper central
panel shows the detailed BADASS fit around the
[\ion{Ne}{5}]~$\lambda\lambda3346,3426$ doublet, and the lower
central panel shows the residuals. Black shows the observed
spectrum, red shows the total fitted model, the colored curves show
the fitted emission-line components, and gray dashed lines mark the
expected doublet wavelengths. The right panel shows the NIRCam RGB
cutout, with F115W ($1.154~\mu{\rm m}$), F277W
($2.776~\mu{\rm m}$), and F444W ($4.402~\mu{\rm m}$) assigned to
blue, green, and red, respectively. The plotting conventions are the
same as in Figure~\ref{fig:nev_representative_spectra}.}
    \label{fig:nev_atlas_1}
\end{figure*}

\begin{figure*}
    \centering
    \includegraphics[width=0.99\textwidth]
    {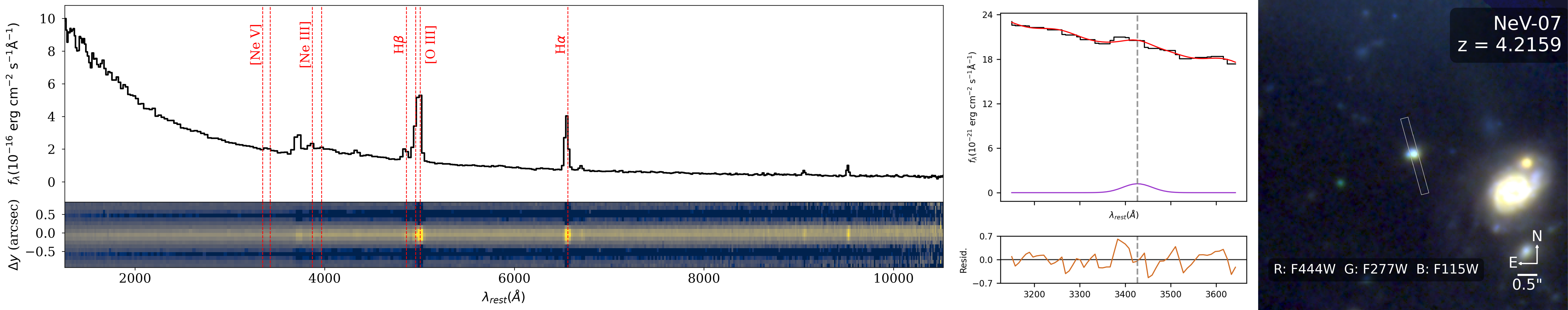}\par\smallskip
    \includegraphics[width=0.99\textwidth]
    {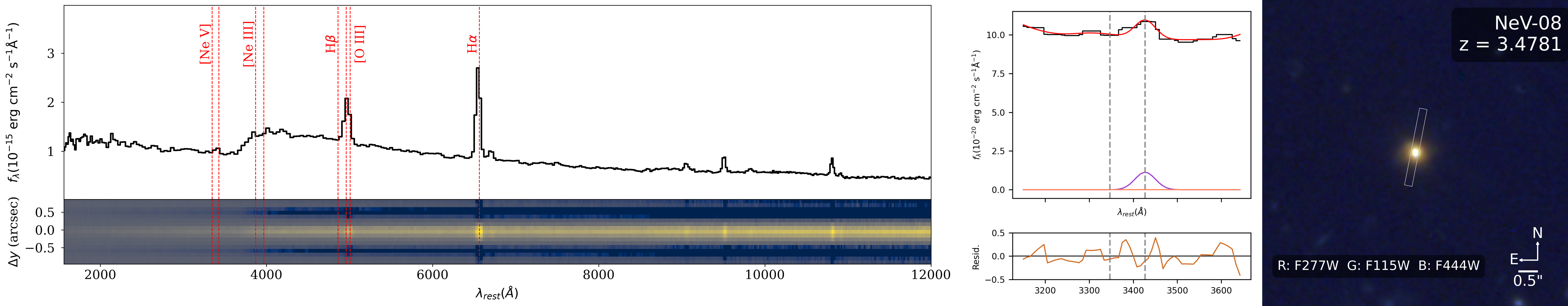}\par\smallskip
    \includegraphics[width=0.99\textwidth]
    {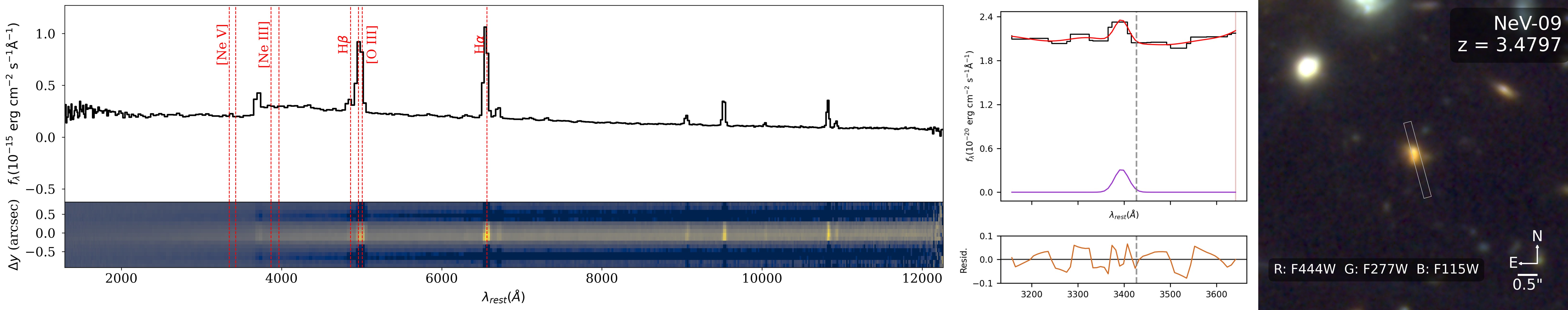}\par\smallskip
    \includegraphics[width=0.99\textwidth]
    {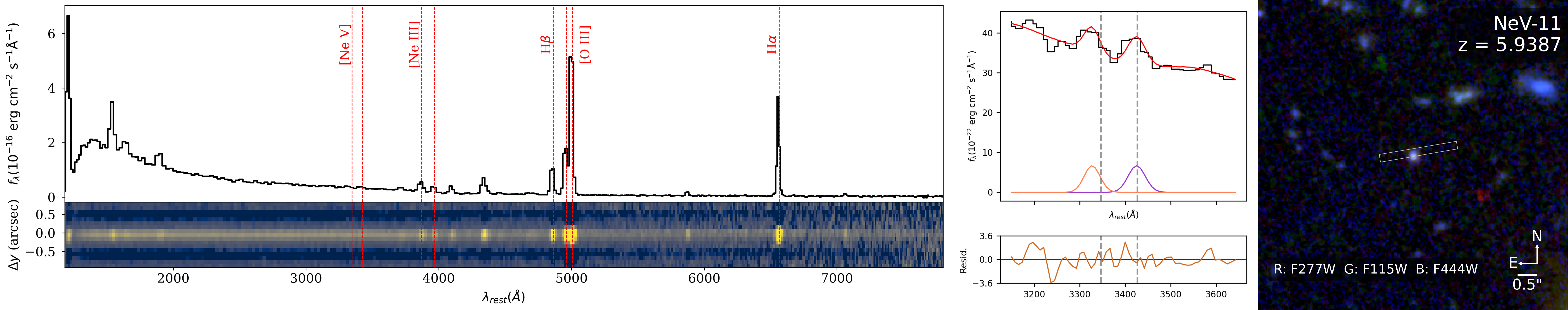}

    \caption{Spectral and imaging atlas for NeV-07, NeV-08, NeV-09,
    and NeV-11, shown from top to bottom. The plotting conventions are
    the same as in Figures~\ref{fig:nev_representative_spectra} and~
    \ref{fig:nev_atlas_1}.}
    \label{fig:nev_atlas_2}
\end{figure*}

\begin{figure*}
    \centering
    \includegraphics[width=0.99\textwidth]
    {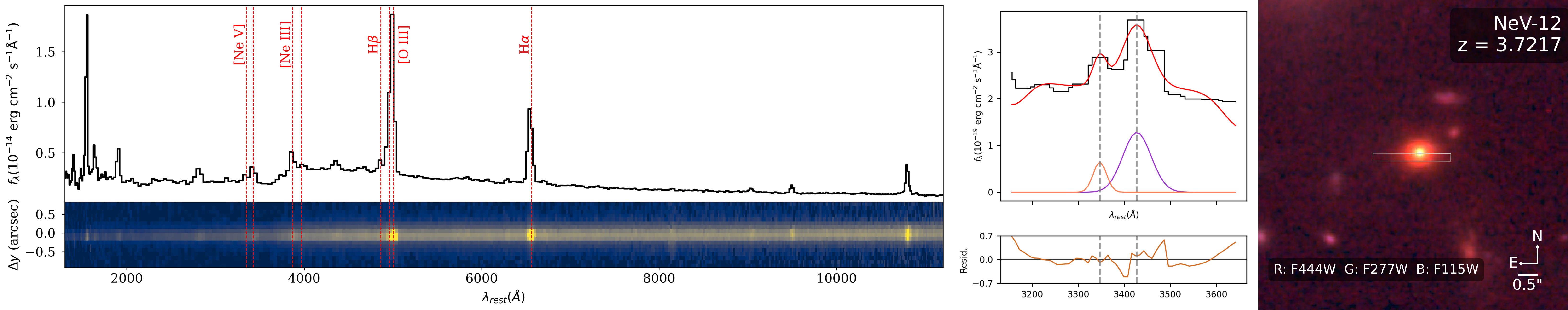}\par\smallskip
    \includegraphics[width=0.99\textwidth]
    {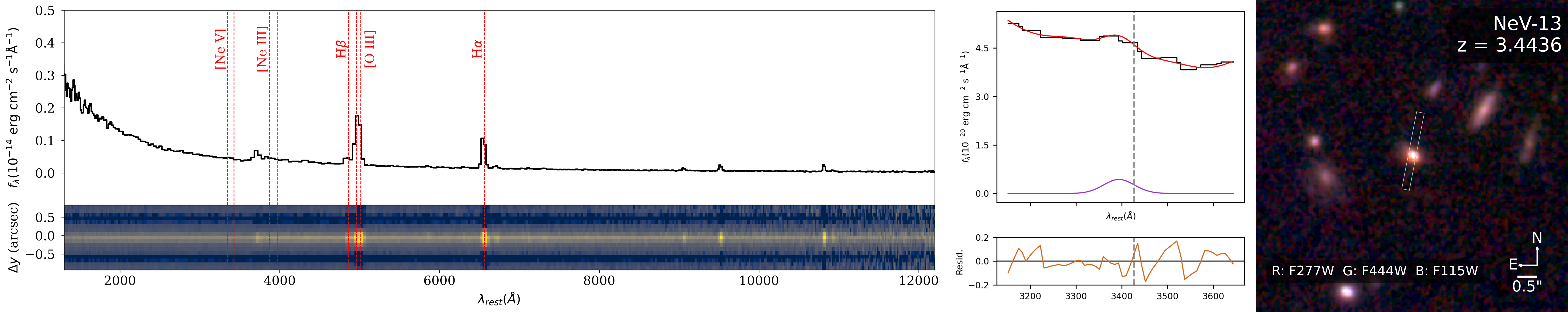}\par\smallskip
    \includegraphics[width=0.99\textwidth]
    {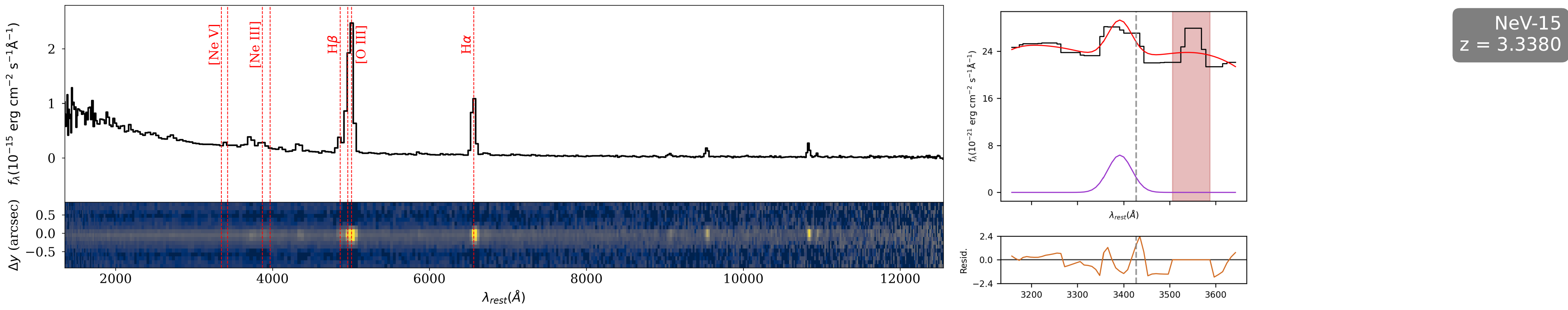}\par\smallskip
    \includegraphics[width=0.99\textwidth]
    {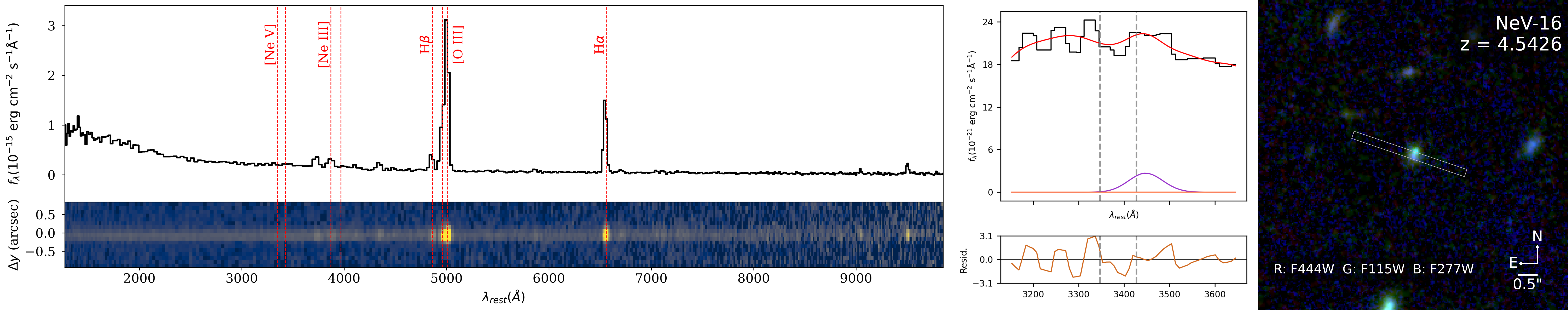}

    \caption{Spectral and imaging atlas for NeV-12, NeV-13, NeV-15,
    and NeV-16, shown from top to bottom. The plotting conventions are
    the same as in Figures~\ref{fig:nev_representative_spectra} and~
    \ref{fig:nev_atlas_1}.}
    \label{fig:nev_atlas_3}
\end{figure*}

\begin{figure*}
    \centering
    \includegraphics[width=0.99\textwidth]
    {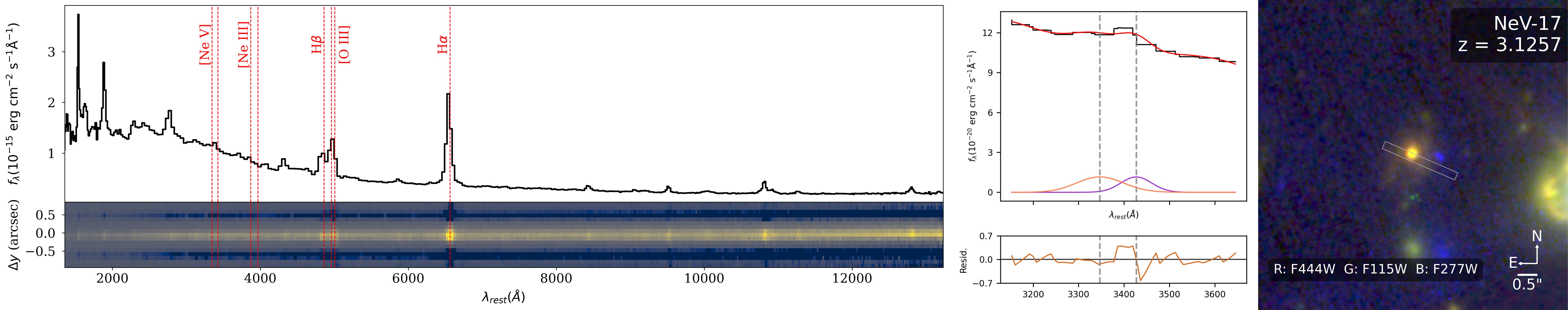}\par\smallskip
    \includegraphics[width=0.99\textwidth]
    {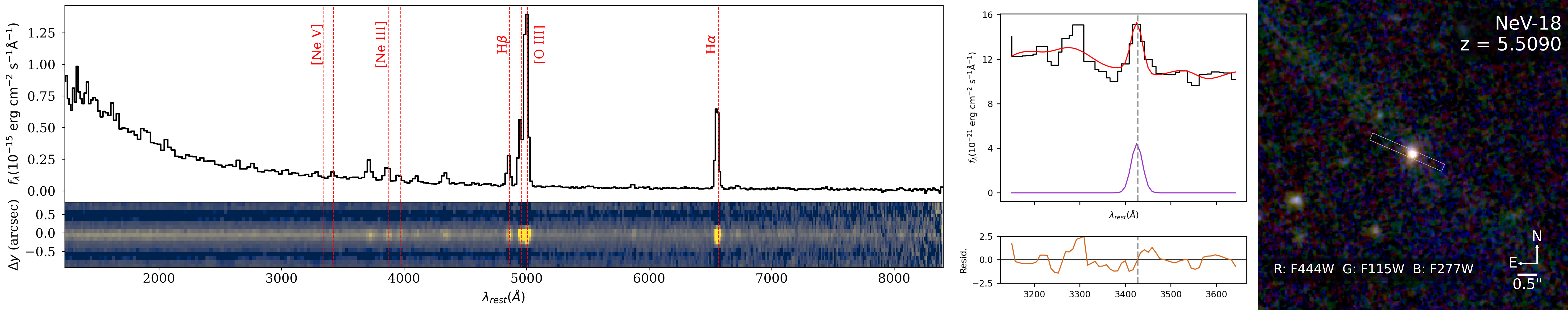}\par\smallskip
    \includegraphics[width=0.99\textwidth]
    {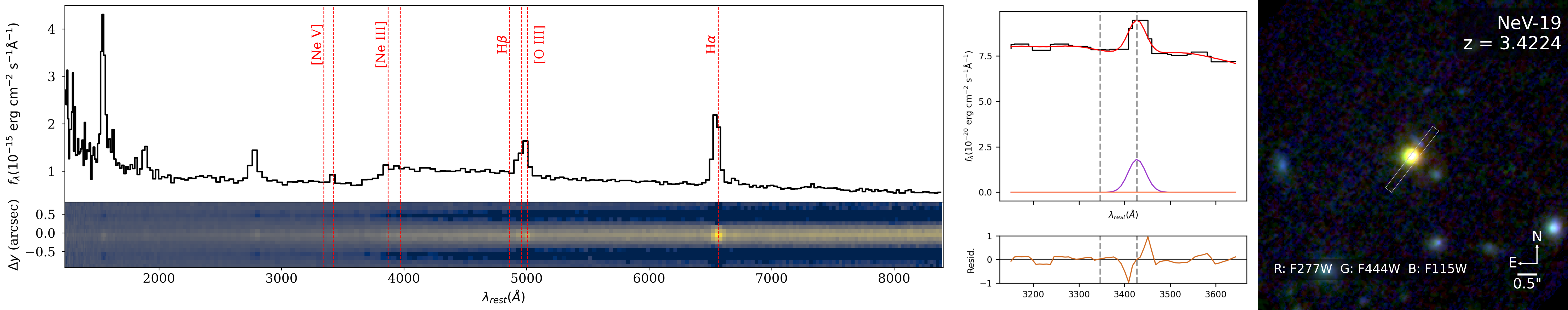}

    \caption{Spectral and imaging atlas for NeV-17, NeV-18, and
    NeV-19, shown from top to bottom. The plotting conventions are the
    same as in Figures~\ref{fig:nev_representative_spectra} and~
    \ref{fig:nev_atlas_1}.}
    \label{fig:nev_atlas_4}
\end{figure*}

\begin{figure*}
    \centering
    \includegraphics[width=0.99\textwidth]
    {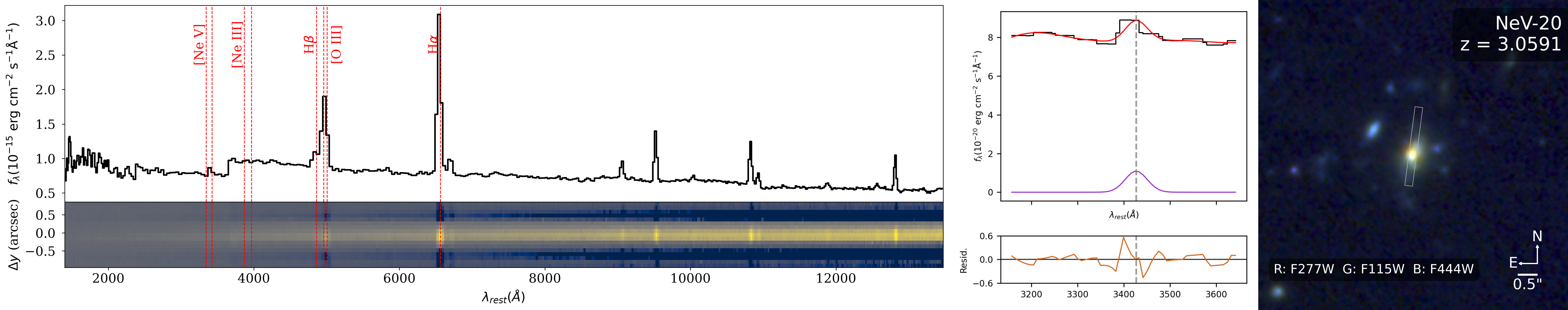}\par\smallskip
    \includegraphics[width=0.99\textwidth]
    {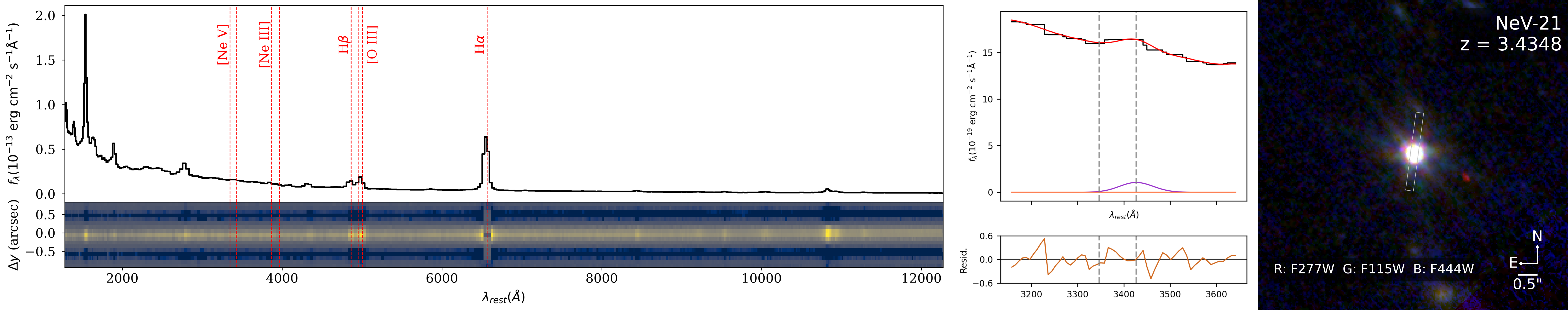}\par\smallskip
    \includegraphics[width=0.99\textwidth]
    {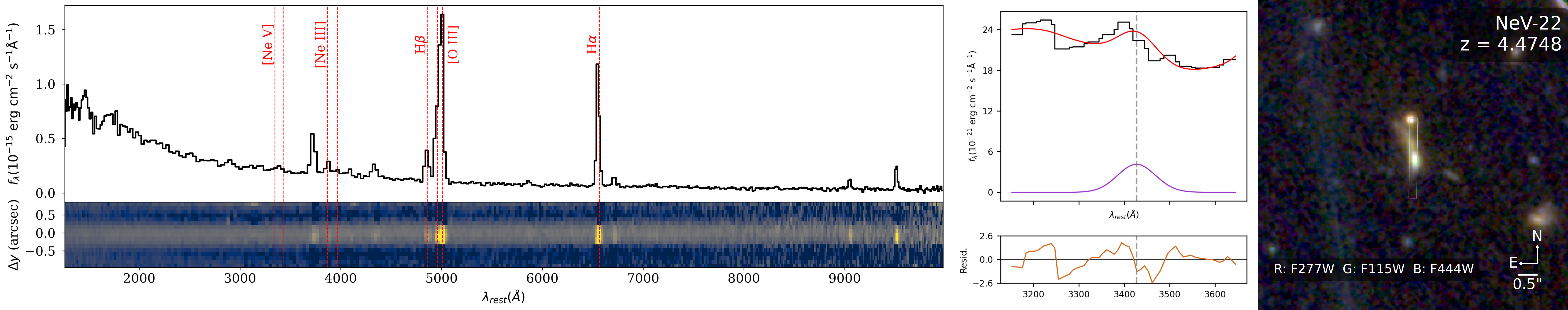}

    \caption{Spectral and imaging atlas for NeV-20, NeV-21, and
    NeV-22, shown from top to bottom. The plotting conventions are the
    same as in Figures~\ref{fig:nev_representative_spectra} and~
    \ref{fig:nev_atlas_1}.}
    \label{fig:nev_atlas_5}
\end{figure*}

\begin{figure*}
    \centering
    \includegraphics[width=0.99\textwidth]
    {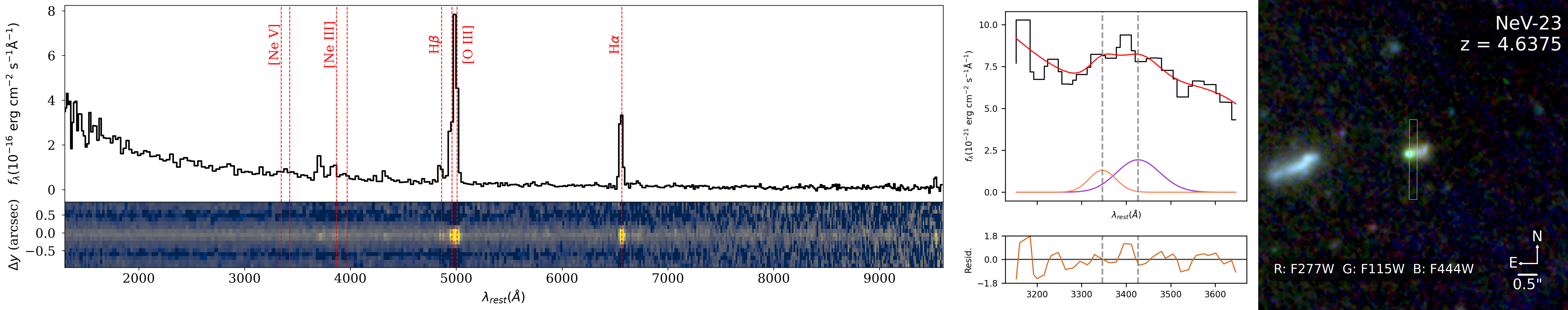}\par\smallskip
    \includegraphics[width=0.99\textwidth]
    {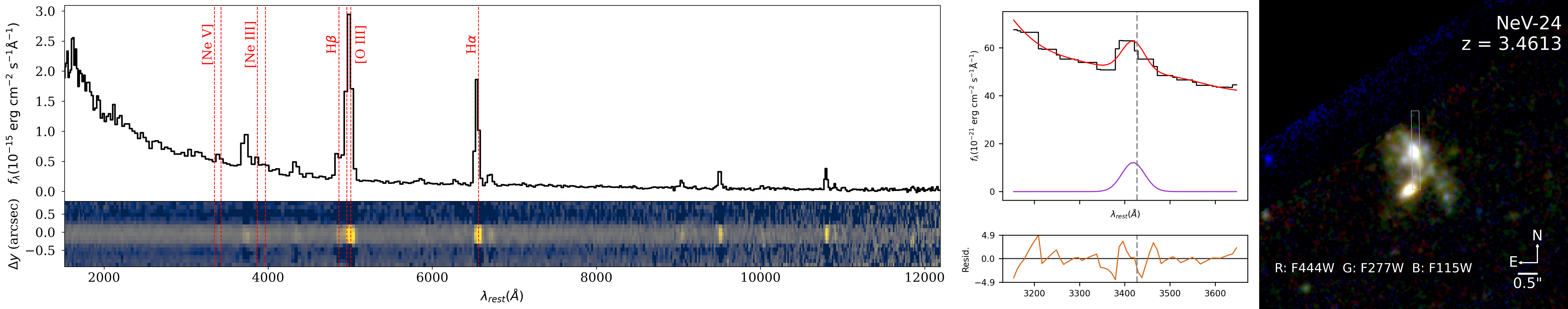}\par\smallskip
    \includegraphics[width=0.99\textwidth]
    {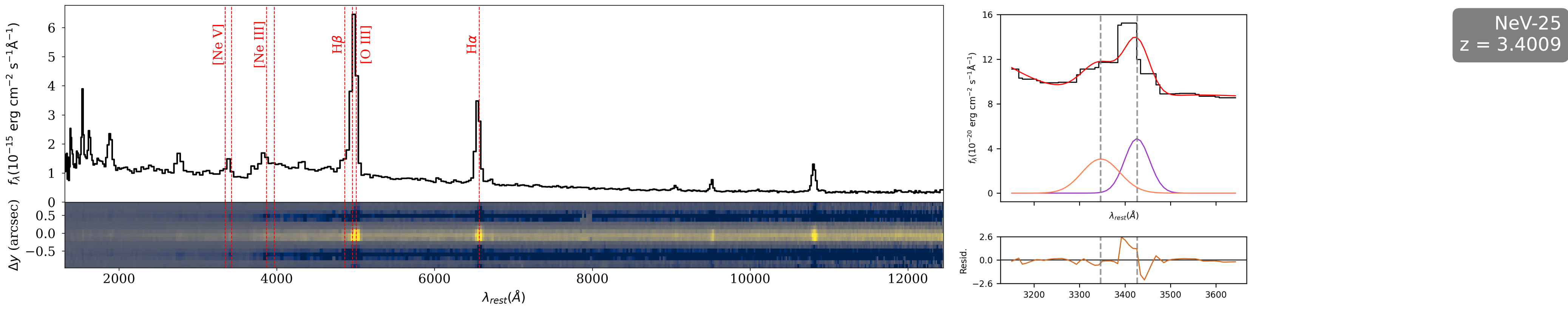}

    \caption{Spectral and imaging atlas for NeV-23, NeV-24, and
    NeV-25, shown from top to bottom. The plotting conventions are the
    same as in Figures~\ref{fig:nev_representative_spectra} and~
    \ref{fig:nev_atlas_1}.}
    \label{fig:nev_atlas_6}
\end{figure*}

Figures~\ref{fig:nev_atlas_1}--\ref{fig:nev_atlas_6} show the PRISM
spectra, detailed [\ion{Ne}{5}] fits, and NIRCam image cutouts for the
21 SHINE-[NeV] sources not included in
Figure~\ref{fig:nev_representative_spectra}. The sources are ordered by
publication identifier. The spectral plotting conventions and the
F115W--F277W--F444W RGB mapping are the same as in the main-text
figure. The cutouts are included for visual comparison and are not
used as quantitative morphology measurements.

\section{Chandra Catalog Matching and X-ray Constraints}
\label{app:xray_catalogs}

\subsection{X-ray Catalog Searches and Counterpart Identification}

We searched for X-ray counterparts to the 25 SHINE-[NeV] galaxies using
the published \textit{Chandra} point-source catalog appropriate to each field:
the 7~Ms CDF-S catalog for GOODS-S, the improved 2~Ms CDF-N catalog for
GOODS-N, the AEGIS-XD catalog for the Extended Groth Strip, and the X-UDS
catalog for the UDS field
\citep{Luo2017,Xue2016,Nandra2015,Kocevski2018}. We also searched the Chandra
Source Catalog Release~2 series (CSC~2.1) at all source positions
\citep{Evans2024}.

We first identified candidate catalog entries within $3\arcsec$ of each
NIRSpec position. Candidate counterparts were then evaluated using the
separation from the X-ray centroid or, where available, the cataloged
multiwavelength counterpart position, together with the catalog positional
uncertainty. The adopted field-specific criteria are summarized in
Table~\ref{tab:xray_catalog_matching}. When more than one catalog measurement
was available, we adopted the measurement from the deeper field-specific
catalog and retained the relevant quality and confusion flags.

\begin{deluxetable*}{llclll}
\tabletypesize{\scriptsize}
\tablewidth{0pt}
\tablecaption{Adopted X-ray Catalogs and Matching Criteria
\label{tab:xray_catalog_matching}}
\tablehead{
\colhead{Field or purpose} &
\colhead{Catalog} &
\colhead{Band (keV)} &
\colhead{Coordinate used} &
\colhead{Adopted criterion} &
\colhead{Reference}
}
\startdata
GOODS-S / CDF-S & 7~Ms CDF-S & 0.5--7 & MW counterpart &
$\Delta\theta<0\farcs5$ & \citealt{Luo2017} \\
GOODS-N / CDF-N & Improved 2~Ms CDF-N & 0.5--7 & X-ray centroid &
$\Delta\theta<0\farcs5$ and $\Delta\theta<2\sigma_{\rm pos}$ &
\citealt{Xue2016} \\
EGS / AEGIS-XD & AEGIS-XD v4.2 & 0.5--10 & MW counterpart &
$\Delta\theta<0\farcs5$ & \citealt{Nandra2015} \\
UDS / X-UDS & X-UDS & 0.5--10 & X-ray centroid &
$\Delta\theta<1\arcsec$ and $\Delta\theta<2\sigma_{\rm pos}$ &
\citealt{Kocevski2018} \\
Catalog nondetections & CSC~2.1 & 0.5--7 & NIRSpec position &
TRUE-class coordinate detection threshold & \citealt{Evans2024} \\
\enddata
\tablecomments{
Candidate catalog entries were initially collected within $3\arcsec$ of each
NIRSpec position. ``MW counterpart'' denotes the multiwavelength counterpart
coordinate provided by the field catalog. When multiple catalog measurements
were available, the deeper field-specific catalog measurement was adopted.
The CSC TRUE-class value is a coordinate-specific catalog detection threshold,
not a statistical source-flux upper limit.
}
\end{deluxetable*}

\subsection{Constraints for Sources without Catalog Counterparts}

For sources without a secure X-ray counterpart, we used the CSC~2.1
TRUE-class broad-band source-detection threshold evaluated at the NIRSpec
position. A usable coordinate-specific threshold was available for 18
galaxies. NeV-15 had no usable source-specific X-ray constraint.

These values describe the local sensitivity of the source catalog. They
estimate the flux at which a source at the specified coordinate would meet the
catalog detection criterion. They are not measurements of the galaxy flux and
are not statistical confidence upper limits on an undetected source. We did
not perform forced X-ray photometry or construct a new field-specific
sensitivity map. The threshold-derived luminosities are therefore shown as
downward arrows and are not treated as measured source luminosities.

\subsection{X-ray Luminosities}

We converted the adopted catalog fluxes and coordinate-specific detection
thresholds to observed-equivalent rest-frame 2--10~keV luminosities. For a
catalog flux measured over the observed band $E_1$--$E_2$, we assumed a
power-law spectrum with photon index $\Gamma=1.8$ . The catalog flux convention was adopted, and we applied no additional
Galactic or intrinsic absorption correction. The resulting luminosities are
therefore observed-equivalent quantities rather than intrinsic,
absorption-corrected luminosities. The [Ne~V] luminosities are the apparent
BADASS values adopted in Table~\ref{tab:nev_sample} and are not uniformly corrected for
extinction or gravitational lensing. The reported
$L_{2-10}/L_{\rm [Ne\,V]}$ values should therefore be interpreted as observed
ratios.

Flux uncertainties were retained when suitable catalog uncertainties were
available. Because their definitions and availability differ among the
adopted catalogs, a common uncertainty is not assigned to every counterpart.
The complete machine-readable X-ray catalog records the available asymmetric
uncertainties and identifies cases for which a directly comparable catalog
uncertainty was unavailable.

Table~\ref{tab:xray_constraints_all25} lists the adopted X-ray state for all
25 emitters. For threshold rows, the tabulated flux, luminosity, and ratio are
detection-threshold quantities and are not formal upper limits.

\begin{deluxetable*}{lllcccccl}
\tabletypesize{\scriptsize}
\rotate
\tablewidth{0pt}
\tablecaption{X-ray Detections and Catalog Detection Thresholds for the
SHINE-[NeV] Sample
\label{tab:xray_constraints_all25}}
\tablehead{
\colhead{NeV ID} &
\colhead{Field} &
\colhead{X-ray state} &
\colhead{Catalog} &
\colhead{Band} &
\colhead{$f_{\rm X}$ or $f_{\rm det}$} &
\colhead{$\log L_{2-10}$ or $\log L_{\rm det}$} &
\colhead{$L_{2-10}/L_{\rm [Ne\,V]}$ or $L_{\rm det}/L_{\rm [Ne\,V]}$} &
\colhead{Balmer status}
}
\startdata
NeV-01 & A2744 & Threshold & CSC 2.1 & 0.5-7 & 2.540 & 44.191 & 67.1 & Broad \\
NeV-02 & A2744 & Threshold & CSC 2.1 & 0.5-7 & 3.113 & 44.851 & 1231.9 & Undetermined \\
NeV-03 & A2744 & Threshold & CSC 2.1 & 0.5-7 & 3.690 & 45.199 & 4201.2 & Undetermined \\
NeV-04 & UDS & Threshold & CSC 2.1 & 0.5-7 & 2.410 & 44.484 & 221.3 & Broad \\
NeV-05\tablenotemark{a} & UDS & Detection & X-UDS & 0.5-10 & 10.030 & 44.840 & 979.5 & Broad \\
NeV-06 & UDS & Threshold & CSC 2.1 & 0.5-7 & 1.448 & 44.505 & 1898.0 & Undetermined \\
NeV-07 & CDF-S & Threshold & CSC 2.1 & 0.5-7 & 0.163 & 43.186 & 228.9 & Undetermined \\
NeV-08 & CDF-S & Threshold & CSC 2.1 & 0.5-7 & 0.439 & 43.432 & 107.4 & Undetermined \\
NeV-09 & CDF-S & Threshold & CSC 2.1 & 0.5-7 & 0.154 & 42.978 & 189.6 & Undetermined \\
NeV-10 & CDF-S & Threshold & CSC 2.1 & 0.5-7 & 0.162 & 43.935 & 1184.8 & No broad comp. \\
NeV-11 & CDF-S & Threshold & CSC 2.1 & 0.5-7 & 0.194 & 43.586 & 601.2 & No broad comp. \\
NeV-12 & CDF-S & Detection & 7 Ms CDF-S & 0.5-7 & 2.160 & 44.190 & 21.2 & Undetermined \\
NeV-13 & CDF-S & Threshold & CSC 2.1 & 0.5-7 & 0.160 & 42.984 & 62.6 & Undetermined \\
NeV-14 & CDF-S & Detection & 7 Ms CDF-S & 0.5-7 & 0.187 & 43.060 & 15.9 & Undetermined \\
NeV-15 & Other & No constraint & \nodata & \nodata & \nodata & \nodata & \nodata & Undetermined \\
NeV-16 & COSMOS & Threshold & CSC 2.1 & 0.5-7 & 1.942 & 44.334 & 653.8 & Undetermined \\
NeV-17 & COSMOS & Threshold & CSC 2.1 & 0.5-7 & 2.389 & 44.064 & 537.1 & Broad \\
NeV-18 & COSMOS & Threshold & CSC 2.1 & 0.5-7 & 2.072 & 44.544 & 1585.4 & Undetermined \\
NeV-19 & CDF-N & Detection & Improved 2 Ms CDF-N & 0.5-7 & 1.000 & 43.774 & 161.9 & Undetermined \\
NeV-20 & AEGIS-XD & Threshold & CSC 2.1 & 0.5-7 & 1.085 & 43.701 & 282.8 & Undetermined \\
NeV-21 & AEGIS-XD & Detection & AEGIS-XD v4.2 & 0.5-10 & 9.122 & 44.666 & 113.7 & Broad \\
NeV-22 & AEGIS-XD & Threshold & CSC 2.1 & 0.5-7 & 0.584 & 43.798 & 218.6 & Undetermined \\
NeV-23 & AEGIS-XD & Threshold & CSC 2.1 & 0.5-7 & 0.642 & 43.873 & 227.7 & No broad comp. \\
NeV-24 & AEGIS-XD & Threshold & CSC 2.1 & 0.5-7 & 1.060 & 43.810 & 193.6 & No broad comp. \\
NeV-25 & AEGIS-XD & Detection & AEGIS-XD v4.2 & 0.5-10 & 1.310 & 43.813 & 44.3 & No broad comp. \\
\enddata
\tablecomments{
The observed band is in keV. Fluxes and detection thresholds are in units of
$10^{-15}\ \mathrm{erg\,s^{-1}\,cm^{-2}}$, and luminosities are in
$\mathrm{erg\,s^{-1}}$. Rows labeled ``Detection'' contain an adopted catalog
flux and observed-equivalent rest-frame 2--10~keV luminosity. Rows labeled
``Threshold'' contain the CSC~2.1 TRUE-class coordinate-specific catalog
detection threshold, its luminosity equivalent, and its ratio to the apparent
[Ne~V] luminosity. These threshold values are not direct source measurements
or formal confidence upper limits, and no inequality sign is assigned to
them. ``No broad comp.'' means that adequate higher-resolution Balmer-line
data did not require a broad component. ``Undetermined'' means that the
presence or absence of broad Balmer emission could not be determined from the
available data.
}
\tablenotetext{a}{
The X-UDS counterpart is retained as secure, but the corresponding CSC entry
carries a confusion flag.
}
\end{deluxetable*}

\section{Additional Emission-Line Incidence Diagnostics}
\label{app:neiii_incidence}

We used [\ion{Ne}{3}]~$\lambda3869$ as an additional diagnostic of
redshift-dependent line detectability. Because [\ion{Ne}{3}] lies
close in wavelength to [\ion{Ne}{5}]~$\lambda3426$, the two lines are
affected by broadly similar wavelength-dependent PRISM coverage and
instrumental sensitivity. We define robust [\ion{Ne}{3}] emitters as
sources with
$F_{\rm [Ne\,III]}/\sigma_{F_{\rm [Ne\,III]}}>4$.

\begin{figure*}[t]
    \centering
    \includegraphics[width=\textwidth]
    {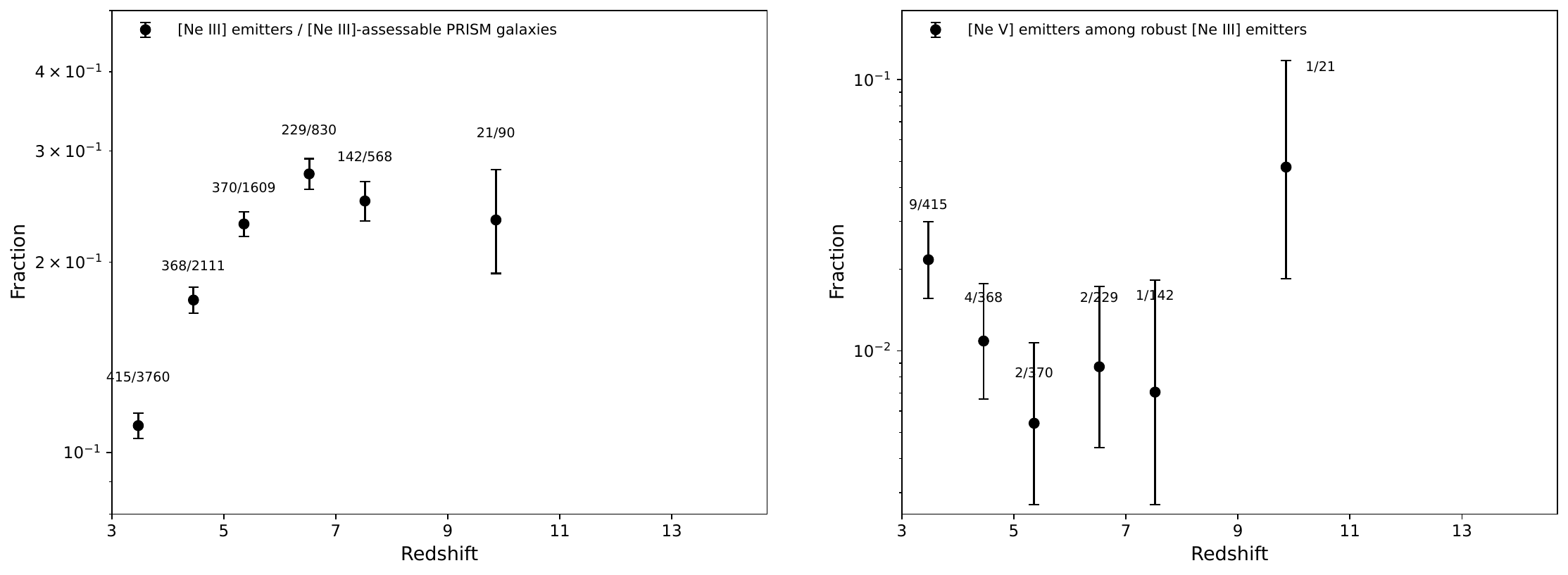}
    \caption{Diagnostic fractions involving [\ion{Ne}{3}] as a
    function of redshift. \textit{Left:} black points show the
    fraction of unique NIRSpec/PRISM galaxies with robust catalog
    [\ion{Ne}{3}] emission, defined by
    $F_{\rm [Ne\,III]}/\sigma_{F_{\rm [Ne\,III]}}>4$.
    \textit{Right:} black points show the fraction of robust
    [\ion{Ne}{3}] emitters that are also [\ion{Ne}{5}] emitters.
    Vertical error bars indicate 68\% Wilson score confidence
    intervals, and labels give the exact numerator and denominator in
    each redshift bin. The [\ion{Ne}{3}] fraction is not defined above
    a common intrinsic line-strength threshold and may therefore
    reflect both astrophysical differences and heterogeneous survey
    sensitivity and targeting. The quantity in the right panel is a
    conditional population fraction, not a source-level
    [\ion{Ne}{5}]/[\ion{Ne}{3}] line ratio. Because the
    [\ion{Ne}{3}]-selected denominator evolves with redshift and
    excludes [\ion{Ne}{5}] emitters without robust [\ion{Ne}{3}]
    detections, both panels should be interpreted as selection
    diagnostics rather than direct measurements of intrinsic AGN
    evolution.}
    \label{fig:neiii_incidence_diagnostics}
\end{figure*}

The left panel of
Figure~\ref{fig:neiii_incidence_diagnostics} shows the fraction of
[\ion{Ne}{3}]-assessable PRISM galaxies satisfying this criterion.
The right panel shows the fraction of robust [\ion{Ne}{3}] emitters
that are also [\ion{Ne}{5}] emitters. The latter is a conditional
population fraction and should not be confused with the source-level
[\ion{Ne}{5}]/[\ion{Ne}{3}] emission-line ratio.

These calculations are intended as selection diagnostics rather than
measurements of intrinsic redshift evolution. The [\ion{Ne}{3}]
fraction is not measured above a common line-flux, luminosity, or
equivalent-width threshold and may reflect both astrophysical
evolution and variations in survey targeting and sensitivity.
Moreover, the [\ion{Ne}{3}]-selected denominator itself changes with
redshift and excludes six [\ion{Ne}{5}] emitters that do not satisfy
the adopted robust [\ion{Ne}{3}] criterion.

\section{Acknowledgments}

We gratefully acknowledge the dedicated teams and broader community whose work made JWST and the development of its data-reduction pipelines possible.

This work is based in part on observations made with the NASA/ESA/CSA JWST and obtained from the Mikulski Archive for Space Telescopes (MAST) at the Space Telescope Science Institute, which is operated by the Association of Universities for Research in Astronomy, Inc., under NASA contract NAS5-03127 for JWST. This work also makes use of data products from the DAWN JWST Archive (DJA), an initiative of the Cosmic Dawn Center (DAWN), which is funded by the Danish National Research Foundation under grant DNRF140.

This work was supported by the National Science Foundation under Cooperative Agreement 2421782 and the Simons Foundation grant MPS-AI-00010515 awarded to the NSF-Simons AI Institute for Cosmic Origins---CosmicAI (\url{https://www.cosmicai.org/}). The work of S.D. was supported by the National Science Foundation Graduate Research Fellowship Program under Grant Number DGE-2236637. Any opinions, findings, and conclusions or recommendations expressed in this material are those of the authors and do not necessarily reflect the views of the National Science Foundation.

This publication makes use of data products from the Wide-field Infrared Survey Explorer, which is a joint project of the University of California, Los Angeles, and the Jet Propulsion Laboratory/California Institute of Technology, funded by the National Aeronautics and Space Administration. This research has made use of the NASA/IPAC Infrared Science Archive, which is funded by the National Aeronautics and Space Administration and operated by the California Institute of Technology.

This research made use of Astropy,\footnote{\url{http://www.astropy.org}} a community-developed core Python package for astronomy \citep{2013A&A...558A..33A}, as well as \textsc{topcat} \citep{2005ASPC..347...29T}. Pipeline processing, spectral fitting, and Cloudy simulations were performed on ARGO and HOPPER, research computing clusters provided by the Office of Research Computing at George Mason University, VA.
(\url{http://orc.gmu.edu})

\bibliographystyle{yahapj}
\bibliography{bib}

\end{document}